\documentclass{article}
\usepackage{setspace}
\usepackage{amssymb}
\usepackage{amsmath}
\usepackage{amsfonts}
\usepackage{amssymb}
\usepackage{setspace}
\usepackage{amsthm}
\usepackage{textcomp}
\usepackage{graphicx}
\usepackage{url}
\usepackage{color}
\usepackage[dvipsnames]{xcolor}
\definecolor{cuteBlue}{rgb}{0.258, 0.387, 0.574}
\definecolor{cuteGreen}{rgb}{0, 0.3, 0}
\usepackage{cancel}
\usepackage{comment}
\usepackage[framemethod=TikZ]{mdframed}
\usepackage{enumitem}
\usepackage{wasysym}
\usepackage{listings}
\usepackage{float}
\usepackage{booktabs}
\usepackage{threeparttable}
\usepackage{titling}
\usepackage{zref-base}
\usepackage{makecell}
\usepackage{array}
\usepackage{hhline}
\usepackage{titlesec}

\usepackage[colorlinks=true, urlcolor=cuteBlue, citecolor=cuteGreen, linkcolor=black]{hyperref}

\usepackage[labelfont=bf, font=small]{caption}

\usepackage[normalem]{ulem}

\usepackage[utf8]{inputenc}

\usepackage{caption}

\usepackage{authblk}

\reversemarginpar
\allowdisplaybreaks

\date{\vspace{-5em}}  %

\titlespacing\section{0pt}{12pt plus 4pt minus 2pt}{-2pt plus 2pt minus 2pt}
\titlespacing\subsection{0pt}{12pt plus 4pt minus 2pt}{-2pt plus 2pt minus 2pt}
\titlespacing\subsubsection{0pt}{12pt plus 4pt minus 2pt}{-2pt plus 2pt minus 2pt}

\setlist{itemsep=0pt, topsep=0pt}

\setitemize{listparindent=\parindent}
\setenumerate{listparindent=\parindent}

\newcolumntype{M}[1]{>{\centering\arraybackslash}m{#1}}
\newcolumntype{N}{@{}m{0pt}@{}}

\usepackage[
	backend=bibtex,
	style=phys, 
	sorting=none,				%
	url=false, 					%
	doi=false, 					%
	isbn=false, 				%
	eprint=false, 			%
	maxbibnames=9, 			%
	giveninits=true,
]{biblatex}
\usepackage[
	aboveskip=1pt,
	labelfont=bf,
	labelsep=period,
	justification=raggedright,
	singlelinecheck=off
]{caption}

\usepackage{titling}

\usepackage{ifthen}
\newboolean{maintext}
\newboolean{sitext}

\usepackage{xparse}

\newcommand{\eref}[1]{Eq.~\ref{#1}}
\newcommand{\fref}[1]{Fig.~\ref{#1}}
\newcommand{\tref}[1]{Table~\ref{#1}}
\newcommand{\secref}[1]{Section~\ref{#1}}
\newcommand{\siref}[1]{Appendix~\ref{#1}}

\newcommand{\beginsupplement}{
				\setcounter{section}{0} %
        \renewcommand{\thesection}{S\arabic{section}}%
        \setcounter{table}{0} %
        \renewcommand{\thetable}{S\arabic{table}}%
        \setcounter{figure}{0} %
        \renewcommand{\thefigure}{S\arabic{figure}}%
        \setcounter{equation}{0} %
        \renewcommand{\theequation}{S\arabic{equation}}%
     }

\newcommand{\kpon}{k^{(p)}_{\text{on}}}
\newcommand{\kpoff}{k^{(p)}_{\text{off}}}
\newcommand{\kron}{k^{(r)}_{\text{on}}}
\newcommand{\kroff}{k^{(r)}_{\text{off}}}
\newcommand{\gm}{\gamma _m}
\newcommand{\gp}{\gamma _p}
\newcommand{\Km}{\bb{K}}
\newcommand{\Rm}{\bb{R}_m}
\newcommand{\Gm}{\bb{\Gamma}_m}
\newcommand{\Rp}{\bb{R}_p}
\newcommand{\Gp}{\bb{\Gamma}_p}
\newcommand{\PP}{\bb{P}}
\newcommand{\ee}[1]{\left\langle #1 \right\rangle}
\newcommand{\bb}[1]{\mathbf{#1}}
\newcommand{\dt}[1]{{d{#1} \over dt}}

\newcommand{\smp}{\sum_m \sum_p}
\newcommand{\bmu}{\boldsymbol{\mu}^{\bb{(x, y)}}}
\newcommand{\foldchange}{\text{fold-change}}
\newcommand{\Nns}{N_\text{NS}}
\newcommand{\eR}{\Delta\varepsilon_r}
\newcommand{\eP}{\Delta\varepsilon_p}
\newcommand{\eAI}{\Delta\varepsilon_{AI}}
\newcommand{\pbound}{p_{_\text{bound}}}

\renewcommand{\th}{^\text{th}}

\title{\textbf{First-principles prediction of the information processing
capacity of a simple genetic circuit}}
\author[1]{Manuel Razo-Mejia}
\author[2]{Sarah Marzen}
\author[1]{Griffin Chure}
\author[2]{Rachel Taubman}
\author[3]{Muir Morrison}
\author[1, 3, *]{Rob Phillips}

\affil[1]{Division of Biology and Biological Engineering, California Institute
of Technology, Pasadena, CA 91125, USA}
\affil[2]{Department of Physics, W. M. Keck Science Department}
\affil[3]{Department of Physics, California Institute of Technology, Pasadena,
CA 91125, USA}
\affil[*]{Correspondence: phillips@pboc.caltech.edu}

\begin{document}
\addtocontents{toc}{\protect\setcounter{tocdepth}{-1}}

	\begin{refsegment}

		\defbibfilter{notother}{not segment=\therefsegment}
		\setboolean{maintext}{true}
		\ifthenelse{\boolean{maintext}}{
		\maketitle %

\begin{abstract}
  Given the stochastic nature of gene expression, genetically identical cells
  exposed to the same environmental inputs will produce different outputs. This
  heterogeneity has been hypothesized to have consequences for how cells are
  able to survive in changing environments. Recent work has explored the use of
  information theory as a framework to understand the accuracy with which cells
  can ascertain the state of their surroundings. Yet the predictive power of
  these approaches is limited and has not been rigorously tested using precision
  measurements. To that end, we generate a minimal model for a simple genetic
  circuit in which all parameter values for the model come from independently
  published data sets. We then predict the information processing capacity of
  the genetic circuit for a suite of biophysical parameters such as protein copy
  number and protein-DNA affinity. We compare these parameter-free predictions
  with an experimental determination of protein expression distributions and the
  resulting information processing capacity of {\it E. coli} cells. We find that
  our minimal model captures the scaling of the cell-to-cell variability in the
  data and the inferred information processing capacity of our simple genetic
  circuit up to a systematic deviation.
\end{abstract}

As living organisms thrive in a given environment, they are faced with
constant changes in their surroundings. From abiotic conditions such as
temperature fluctuations or changes in osmotic pressure, to biological
interactions such as cell-to-cell communication in a tissue or in a bacterial
biofilm, living organisms of all types sense and respond to external signals.
\fref{fig1_intro}(A) shows a schematic of this process for a bacterial cell
sensing a concentration of an extracellular chemical. At the molecular level
where signal transduction unfolds mechanistically, there are physical
constraints on the accuracy and precision of these responses given by intrinsic
stochastic fluctuations \cite{Nemenman2010}. This means that two genetically
identical cells exposed to the same stimulus will not have identical responses
\cite{Eldar2010}.

One implication of this noise in biological systems is that cells do not have
an infinite resolution to distinguish signals and, as a consequence, there is a
one-to-many mapping between inputs and outputs. Furthermore, given the limited
number of possible outputs, there are overlapping responses between different
inputs. This scenario can be map to a Bayesian inference problem where cells
try to infer the state of the environment from their phenotypic response, as
schematized in \fref{fig1_intro}(B). The question then becomes this: how can
one analyze this probabilistic, rather than deterministic, relationship between
inputs and outputs? The abstract answer to this question was worked out in 1948
by Claude Shannon who, in his seminal work, founded the field of information
theory \cite{Shannon1948}. Shannon developed a general framework for how to
analyze information transmission through noisy communication channels. In his
work, Shannon showed that the only quantity that satisfies three reasonable
axioms for a measure of uncertainty was of the same functional form as the
thermodynamic entropy -- thereby christening his metric the information entropy
\cite{MacKay2003}. He also gave a definition, based on this information
entropy, for the relationship between inputs and outputs known as the mutual
information. The mutual information $I$ between input $c$ and output $p$, given
by
\begin{equation}
  I = \sum_c P(c) \sum_p P(p \mid c) \log_2 {P(p \mid c) \over P(p)},
	\label{eq_mutual_info}
\end{equation}
quantifies how much we learn about the state of the input $c$ given that we get
to observe the output $p$. In other words, the mutual information can be
thought of as a generalized correlation coefficient that quantifies the degree
to which the uncertainty about a random event decreases given the knowledge of
the average outcome of another random event \cite{Kinney2010}.

It is natural to conceive of scenarios in which living organisms that can
better resolve signals might have an evolutionary benefit, making it more
likely that their offspring will have a fitness advantage \cite{Taylor2007}. In
recent years there has been a growing interest in understanding the theoretical
limits on cellular information processing \cite{Bialek2005, Gregor2007}, and in
quantifying how close evolution has pushed cellular signaling pathways to these
theoretical limits \cite{Tkacik2008, Dubuis2013, Petkova2019}. While these
studies have treated the signaling pathway as a ``black box,'' explicitly
ignoring all the molecular interactions taking place in them, other studies
have explored the role that molecular players and regulatory architectures have
on these information processing tasks \cite{Rieckh2014, Ziv2007, Voliotis2014a,
Tostevin2009, Tkacik2011, Tkacik2008a, Tabbaa2014}. Despite the great advances
in our understanding of the information processing capabilities of molecular
mechanisms, the field still lacks a rigorous experimental test of these
detailed models with precision measurements on a simple system in which
physical parameters can be perturbed. In this work we approach this task with a
system that is both theoretically and experimentally tractable in which
molecular parameters can be varied in a controlled manner.

Over the last decade the dialogue between theory and experiments in gene
regulation has led to predictive power of models not only over the mean level
of gene expression, but the noise as a function of relevant parameters such as
regulatory protein copy numbers, affinity of these proteins to the DNA
promoter, as well as the extracellular concentrations of inducer molecules
\cite{Golding2005, Garcia2011c, Vilar2013, Xu2015}. These models based on
equilibrium and non-equilibrium statistical physics have reached a predictive
accuracy level such that, for simple cases, it is now possible to design
input-output functions \cite{Brewster2012, Barnes2019}. This opens the
opportunity to exploit these predictive models to tackle the question of how
much information genetic circuits can process. This question lies at the heart
of understanding the precision of the cellular response to environmental
signals. \fref{fig1_intro}(C) schematizes a scenario in which two bacterial
strains respond with different levels of precision to three possible
environmental states, i.e., inducer concentrations. The overlap between the
three different responses is what precisely determines the resolution with
which cells can distinguish different inputs. This is analogous to how the
point spread function limits the ability to resolve two light emitting point
sources.

In this work we follow the same philosophy of theory-experiment dialogue used
to determine model parameters to predict from first principles the effect that
biophysical parameters such as transcription factor copy number and protein-DNA
affinity have on the information processing capacity of a simple genetic
circuit. Specifically, to predict the mutual information between an
extracellular chemical signal (input $c$) and the corresponding cellular
response in the form of protein expression (output $p$), we must compute the
input-output function $P(p \mid c)$. To do so, we use a master-equation-based
model to construct the protein copy number distribution as a function of an
extracellular inducer concentration for different combinations of transcription
factor copy numbers and binding sites. Having these input-output distributions
allows us to compute the mutual information $I$ between inputs and outputs for
any arbitrary input distribution $P(c)$. We opt to compute the channel capacity,
i.e., the maximum information that can be processed by this gene regulatory
architecture, defined as \eref{eq_mutual_info} maximized over all possible input
distributions $P(c)$. By doing so we examine the physical limits of what cells
can do in terms of information processing by harboring these genetic circuits.
Nevertheless, given the generality of the input-output function $P(p \mid c)$ we
derive, the model presented here can be used to compute the mutual information
for any arbitrary input distribution $P(c)$. All parameters used for our model
were inferred from a series of studies that span several experimental techniques
\cite{Garcia2011c, Jones2014a, Brewster2014, Razo-Mejia2018}, allowing us to
make parameter-free predictions of this information processing capacity
\cite{Phillips2019}.

These predictions are then contrasted with experimental data, where the channel
capacity is inferred from single-cell fluorescence distributions taken at
different concentrations of inducer for cells with previously characterized
biophysical parameters \cite{Garcia2011c, Razo-Mejia2018}. We find that our
parameter-free predictions quantitatively track the experimental data up to a
systematic deviation. The lack of numerical agreement between our model and the
experimental data poses new challenges towards having a foundational,
first-principles understanding of the physics of cellular decision-making.

The reminder of the paper is organized as follows. In \secref{sec_model} we
define the minimal theoretical model and parameter inference for a simple
repression genetic circuit. \secref{sec_param} discusses how all parameters for
the minimal model are determined from published datasets that explore different
aspects of the simple repression motif. \secref{sec_moments} computes the
moments of the mRNA and protein distributions from this minimal model. In
\secref{sec_cell_cycle} we explore the consequences of variability in gene copy
number during the cell cycle. In this section we compare experimental and
theoretical quantities related to the moments of the distribution, specifically
the predictions for the fold-change in gene expression (mean expression
relative to an unregulated promoter) and the gene expression noise (standard
deviation over mean). \secref{sec_maxent} follows with reconstruction of the
full mRNA and protein distribution from the moments using the maximum entropy
principle. Finally \secref{sec_channcap} uses the distributions from
\secref{sec_maxent} to compute the maximum amount of information that the
genetic circuit can process. Here we again contrast our zero-parameter fit
predictions with experimental inferences of the channel capacity.

\begin{figure}[h!]
	\centering \includegraphics
  {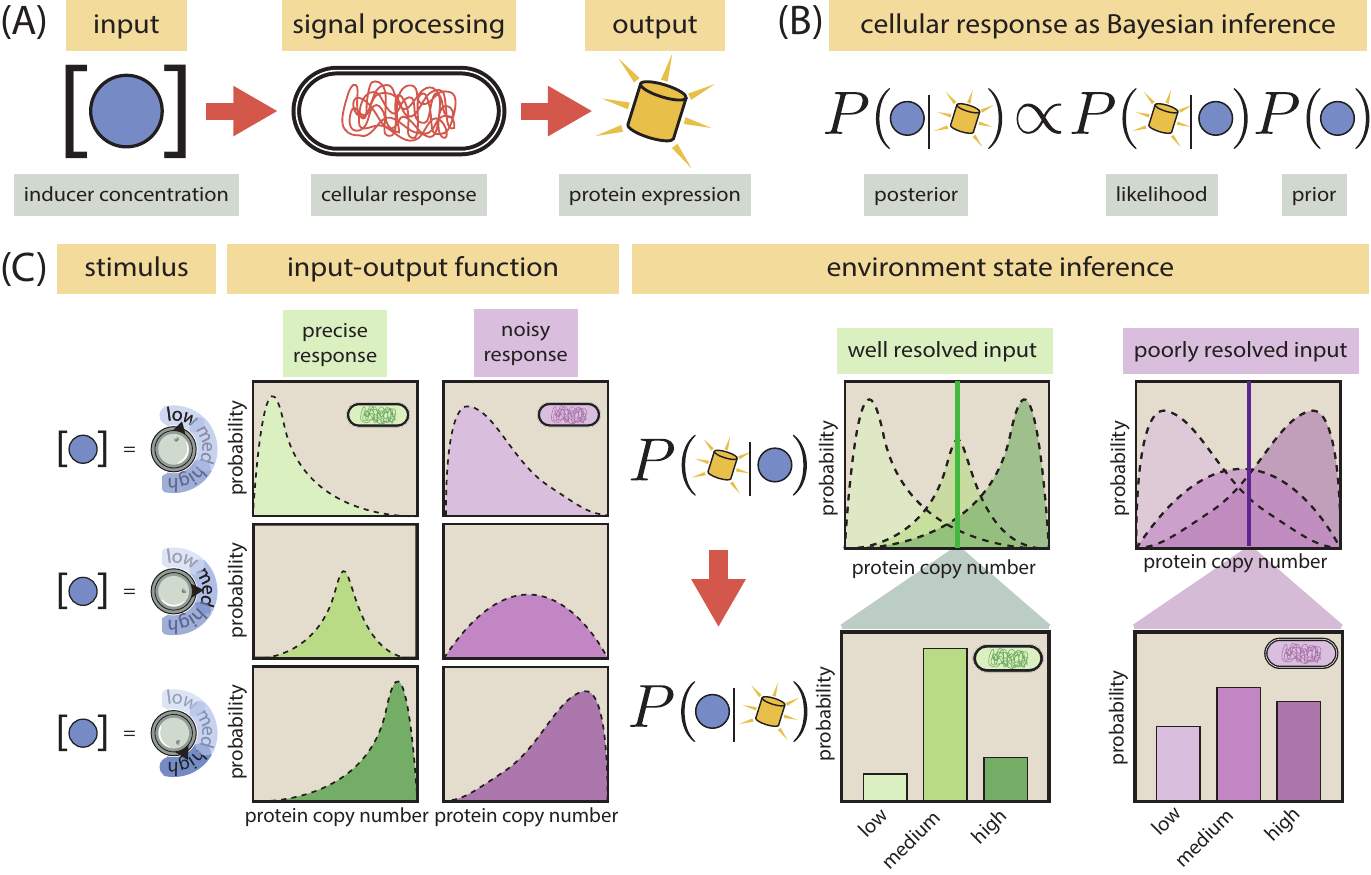}
	\caption{\textbf{Cellular signaling systems sense the environment with
	different degrees of precision}. (A) Schematic representation of a cell as
	a noisy communication channel. From an environmental input (inducer
	molecule concentration) to a phenotypic output (protein expression level),
	cellular signaling systems can be modeled as noisy communication channels.
	(B) We treat cellular response to an external stimulus as a Bayesian
	inference  of the state of the environment. As the phenotype (protein
	level) serves as the internal representation of the environmental state
	(inducer concentration), the probability of a cell being in a specific
	environment given this internal representation $P(c \mid p)$ is a function
	of the probability of the response given that environmental state $P(p \mid
	c)$. (C) The precision of the inference of the environmental state depends
	on how well can cells resolve different inputs. For three different levels
	of input (left panel) the green strain responds more precisely than the
	purple strain since the output distributions overlap less (middle panel).
	This allows the green strain to make a more precise inference of the
	environmental state given a phenotypic response (right panel).}
  \label{fig1_intro}
\end{figure}

\section{Results}

\subsection{Minimal model of transcriptional regulation}\label{sec_model}

As a tractable circuit for which we have control over the parameters both
theoretically and experimentally, we chose the so-called simple repression
motif, a common regulatory scheme among prokaryotes \cite{Rydenfelt2014}. This
circuit consists of a single promoter with an RNA-polymerase (RNAP) binding
site and a single binding site for a transcriptional repressor
\cite{Garcia2011c}. The regulation due to the repressor occurs via exclusion of
the RNAP from its binding site when the repressor is bound, decreasing the
likelihood of having a transcription event. As with many important
macromolecules, we consider the repressor to be allosteric, meaning that it can
exist in two conformations, one in which the repressor is able to bind to the
specific binding site (active state) and one in which it cannot bind the
specific binding site (inactive state). The environmental signaling occurs via
passive import of an extracellular inducer that binds the repressor, shifting
the equilibrium between the two conformations of the repressor
\cite{Razo-Mejia2018}. In previous work we have extensively characterized the
mean response of this circuit under different conditions using equilibrium
based models \cite{Phillips2019}. Here we build upon these models to
characterize the full distribution of gene expression with parameters such as
repressor copy number and its affinity for the DNA being systematically varied.

As the copy number of molecular species is a discrete quantity, chemical master
equations have emerged as a useful tool to model their inherent probability
distribution \cite{Sanchez2013}. In \fref{fig2_minimal_model}(A) we show the
minimal model and the necessary set of parameters needed to compute the full
distribution of mRNA and its protein gene product. Specifically, we assume a
three-state model where the promoter can be found in a 1) transcriptionally
active state  ($A$ state), 2) a transcriptionally inactive state without the
repressor bound ($I$ state) and 3) a transcriptionally inactive state with the
repressor bound ($R$ state). We do not assume that the transition between the
active state $A$ and the inactive state $I$ occurs due to RNAP binding to the
promoter as the transcription initiation kinetics involve several more steps
than simple binding \cite{Browning2004}. We coarse-grain all these steps into
effective ``on" and ``off" states for the promoter, consistent with experiments
demonstrating the bursty nature of gene expression in {\it E. coli}
\cite{Golding2005}. These three states generate a system of coupled
differential equations for each of the three state distributions $P_A(m, p;
t)$, $P_I(m, p; t)$ and $P_R(m, p; t)$, where $m$ and $p$ are the mRNA and
protein count per cell, respectively and $t$ is time. Given the rates depicted
in \fref{fig2_minimal_model}(A) we define the system of ODEs for a specific $m$
and $p$. For the transcriptionally active state, we have
\begin{equation}
  \begin{aligned}
    \dt{P_A(m, p)} &=
    - \overbrace{\kpoff P_A(m, p)}^{A \rightarrow I} %
    + \overbrace{\kpon P_I(m, p)}^{I \rightarrow A}\\ %
    &+ \overbrace{r_m P_A(m-1, p)}^{m-1 \rightarrow m} %
    - \overbrace{r_m P_A(m, p)}^{m \rightarrow m+1}%
    + \overbrace{\gm (m + 1) P_A(m+1 , p)}^{m+1 \rightarrow m} %
    - \overbrace{\gm m P_A(m , p)}^{m \rightarrow m-1}\\ %
    &+ \overbrace{r_p m P_A(m, p - 1)}^{p-1 \rightarrow p} %
    - \overbrace{r_p m P_A(m, p)}^{p \rightarrow p+1} %
    + \overbrace{\gp (p + 1) P_A(m, p + 1)}^{p + 1 \rightarrow p} %
    - \overbrace{\gp p P_A(m, p)}^{p \rightarrow p-1}, %
  \end{aligned}
\end{equation}
where the state transitions for each term are labeled by overbraces. For the
transcriptionally inactive state $I$, we have
\begin{equation}
  \begin{aligned}
    \dt{P_I(m, p)} &=
    \overbrace{\kpoff P_A(m, p)}^{A \rightarrow I} %
    - \overbrace{\kpon P_I(m, p)}^{I \rightarrow A} %
    + \overbrace{\kroff P_R(m, p)}^{R \rightarrow I} %
    - \overbrace{\kron P_I(m, p)}^{I \rightarrow R}\\ %
    &+ \overbrace{\gm (m + 1) P_I(m+1 , p)}^{m+1 \rightarrow m} %
    - \overbrace{\gm m P_I(m , p)}^{m \rightarrow m-1}\\ %
    &+ \overbrace{r_p m P_I(m, p - 1)}^{p-1 \rightarrow p} %
    - \overbrace{r_p m P_I(m, p)}^{p \rightarrow p+1} %
    + \overbrace{\gp (p + 1) P_I(m, p + 1)}^{p + 1 \rightarrow p} %
    - \overbrace{\gp p P_I(m, p)}^{p \rightarrow p-1}. %
  \end{aligned}
\end{equation}
And finally, for the repressor bound state $R$,
\begin{equation}
  \begin{aligned}
    \dt{P_R(m, p)} &=
    - \overbrace{\kroff P_R(m, p)}^{R \rightarrow I} %
    + \overbrace{\kron P_I(m, p)}^{I \rightarrow R}\\ %
    &+ \overbrace{\gm (m + 1) P_R(m+1 , p)}^{m+1 \rightarrow m} %
    - \overbrace{\gm m P_R(m , p)}^{m \rightarrow m-1}\\ %
    &+ \overbrace{r_p m P_R(m, p - 1)}^{p-1 \rightarrow p} %
    - \overbrace{r_p m P_R(m, p)}^{p \rightarrow p+1} %
    + \overbrace{\gp (p + 1) P_R(m, p + 1)}^{p + 1 \rightarrow p} %
    - \overbrace{\gp p P_R(m, p)}^{p \rightarrow p-1}. %
  \end{aligned}
\end{equation}
As we will discuss later in \secref{sec_cell_cycle} the protein degradation
term $\gp$ is set to zero since active protein degradation is slow compared to
the cell cycle of exponentially growing bacteria, but rather we explicitly
implement binomial partitioning of the proteins into daughter cells upon
division \cite{Maurizi1992}.

It is convenient to rewrite these equations in a compact matrix notation
\cite{Sanchez2013}. For this we define the vector $\PP(m, p)$ as
\begin{equation}
  \PP(m, p) = (P_A(m, p), P_I(m, p), P_R(m, p))^T,
\end{equation}
where $^T$ is the transpose. By defining the matrices $\Km$ to contain the
promoter state transitions, $\Rm$ and $\Gm$ to contain the mRNA production and
degradation terms, respectively, and $\Rp$ and $\Gp$ to contain the protein
production and degradation terms, respectively, the system of ODEs can then be
written as (See \siref{supp_model} for full definition of these matrices)
\begin{equation}
  \begin{aligned}
    \dt{\PP(m, p)} &= \left( \Km -\Rm -m\Gm -m\Rp -p\Gp \right) \PP(m, p)\\
    &+ \Rm \PP(m-1, p)
    + (m + 1) \Gm \PP(m + 1, p)\\
    &+ m \Rp \PP(m, p - 1)
    + (p + 1) \Gp \PP(m, p + 1).
  \end{aligned}
  \label{eq_cme_matrix}
\end{equation}
Having defined the gene expression dynamics we now proceed to determine all 
rate parameters in \eref{eq_cme_matrix}.

\subsection{Inferring parameters from published data sets}
\label{sec_param}

A decade of research in our group has characterized the simple repression motif
with an ever expanding array of predictions and corresponding experiments to
uncover the physics of this genetic circuit \cite{Phillips2019}. In doing so we
have come to understand the mean response of a single promoter in the presence
of varying levels of repressor copy numbers and repressor-DNA affinities
\cite{Garcia2011c}, due to the effect that competing binding sites and multiple
promoter copies impose \cite{Brewster2014}, and in recent work, assisted by the
Monod-Wyman-Changeux (MWC) model, we expanded the scope to the allosteric
nature of the repressor \cite{Razo-Mejia2018}. All of these studies have
exploited the simplicity and predictive power of equilibrium approximations to
these non-equilibrium systems \cite{Buchler2003}. We have also used a similar
kinetic model to that depicted in \fref{fig2_minimal_model}(A) to study the
noise in mRNA copy number \cite{Jones2014a}. As a test case of the depth of our
theoretical understanding of this simple transcriptional regulation system we
combine all of the studies mentioned above to inform the parameter values of
the model presented in \fref{fig2_minimal_model}(A).
\fref{fig2_minimal_model}(B) schematizes the data sets and experimental
techniques used to measure gene expression along with the parameters that can
be inferred from them.

\siref{supp_param_inference} expands on the details of how the inference was
performed for each of the parameters. Briefly, the promoter activation and
inactivation rates $\kpon$ and $\kpoff$, as well as the transcription rate
$r_m$ were obtained in units of the mRNA degradation rate $\gm$ by fitting a
two-state promoter model (no state $R$ from \fref{fig2_minimal_model}(A))
\cite{Peccoud1995} to mRNA FISH data of an unregulated promoter (no  repressor
present in the cell) \cite{Jones2014a}. The repressor on rate is assumed to be
of the form $\kron = k_o [R]$ where $k_o$ is a diffusion-limited on rate and
$[R]$ is the concentration of active repressor in the cell \cite{Jones2014a}.
This concentration of active repressor is at the same time determined by the
repressor copy number in the cell, and the fraction of these repressors  that
are in the active state, i.e. able to bind DNA. Existing estimates of the
transition rates between conformations of allosteric molecules set them at the
microsecond scale \cite{Cui2008}. By considering this to be representative for
our repressor of interest, the separation of time-scales between the rapid
conformational changes of the repressor and the slower downstream processes
such as the open-complex formation processes allow us to model the probability
of the repressor being in the active state as an equilibrium MWC process. The
parameters of the MWC model $K_A$, $K_I$ and $\eAI$ were previously
characterized from video-microscopy and flow-cytometry data
\cite{Razo-Mejia2018}. For the repressor off rate, $\kroff$, we take advantage
of the fact that the mean mRNA copy number as derived from the model in
\fref{fig2_minimal_model}(A) cast in the language of rates is of the same
functional form as the equilibrium model cast in the language of binding
energies \cite{Phillips2015}. Therefore the value of the repressor-DNA binding
energy $\eR$ constrains the value of the repressor off rate $\kroff$. These
constraints on the rates allow us to make self-consistent predictions under
both the equilibrium and the kinetic framework. Having all parameters in hand,
we can now proceed to solve he gene expression dynamics.

\begin{figure}[h!]
	\centering \includegraphics
  {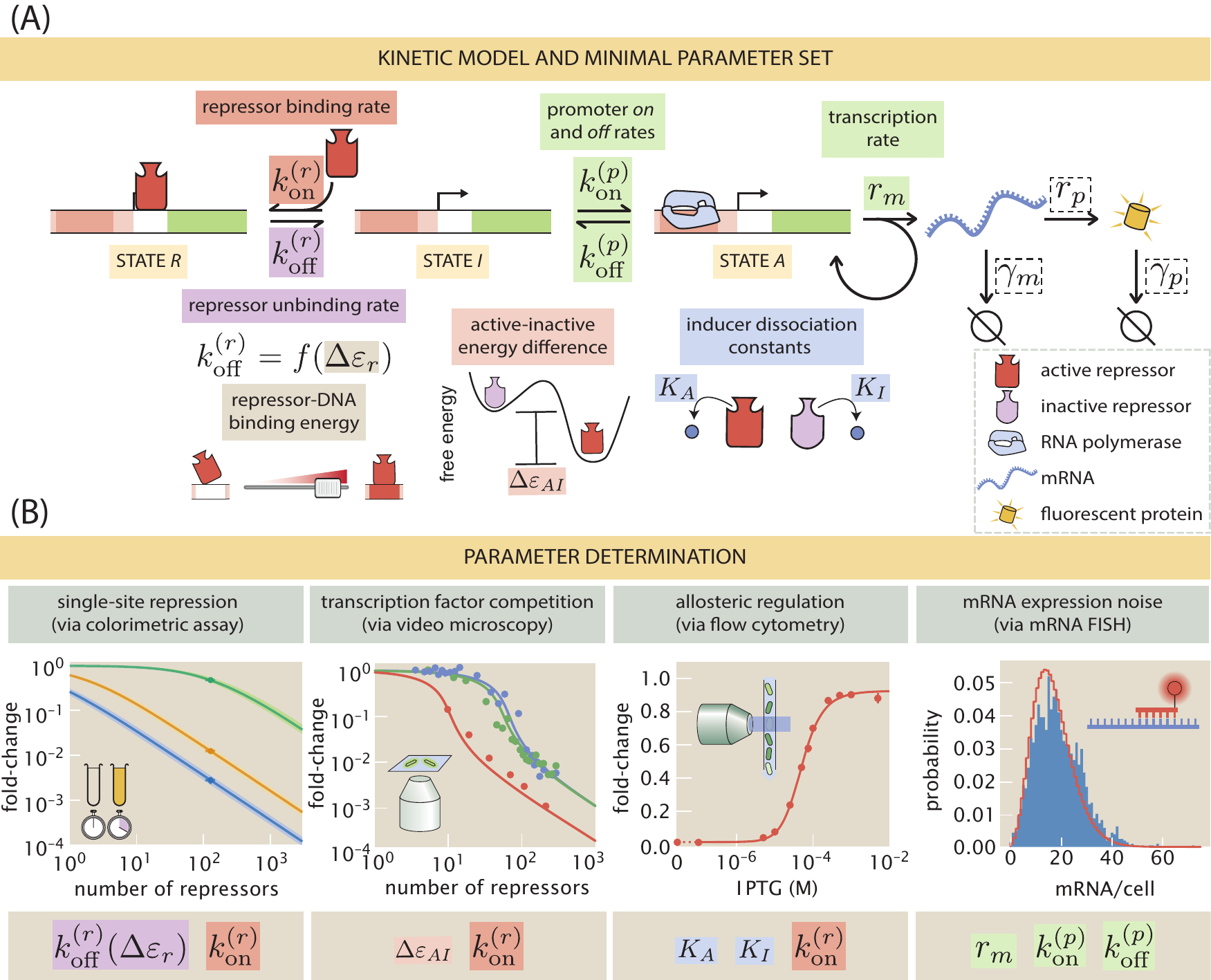}
	\caption{\textbf{Minimal kinetic model of transcriptional regulation for a
	simple repression architecture.} (A) Three-state promoter stochastic model of
	transcriptional regulation by a repressor. The regulation by the repressor
	occurs via exclusion of the transcription initiation machinery, not allowing
	the promoter to transition to the transcriptionally active state. All
	parameters highlighted with colored boxes were determined from published
	datasets based on the same genetic circuit. Parameters in dashed boxes were
	taken directly from values reported in the literature or adjusted to satisfy
	known biological restrictions. (B) Data sets used to infer the parameter
	values. From left to right Garcia \& Phillips \cite{Garcia2011c} is used to
	determine $\kroff$ and $\kron$, Brewster et al. \cite{Brewster2014} is used
	to determine $\eAI$ and $\kron$, Razo-Mejia et al. \cite{Razo-Mejia2018} is
	used to determine $K_A$, $K_I$, and $\kron$ and Jones et al.
	\cite{Jones2014a} is used to determine $r_m$, $\kpon$, and $\kpoff$.}
  \label{fig2_minimal_model}
\end{figure}

\subsection{Computing the moments of the mRNA and protein distributions}
\label{sec_moments}

Finding analytical solutions to chemical master equations is often fraught with
difficulty. An alternative approach is to to approximate the distribution. One
such scheme of approximation, the maximum entropy principle, makes use of the
moments of the distribution to approximate the full distribution. In this
section we will demonstrate an iterative algorithm to compute the mRNA and
protein distribution moments.

The kinetic model for the simple repression motif depicted in
\fref{fig2_minimal_model}(A) consists of an infinite system of ODEs for each
possible pair of mRNA and protein copy number, $(m, p)$. To compute any moment
of the distribution, we define a vector
\begin{equation}
	\ee{\bb{m^x p^y}} \equiv (\ee{m^x p^y}_A, \ee{m^x p^y}_I, \ee{m^x p^y}_R)^T,
\end{equation}
where $\ee{m^x p^y}_S$ is the expected value of $m^x p^y$ in state $S \in \{A,
I, R\}$ for $x, y \in \mathbb{N}$. In other words, just as we defined the
vector $\PP(m, p)$, here we define a vector to collect the expected value of
each of the promoter states. By definition, any of these moments $\ee{m^x
p^y}_S$ can be computed as
\begin{equation}
  \ee{m^x p^y}_S \equiv \sum_{m=0}^\infty \sum_{p=0}^\infty m^x p^y P_S(m, p).
  \label{eq_mom_def}
\end{equation}
Summing over all possible values for $m$ and $p$ in \eref{eq_cme_matrix}
results in an ODE for any moment of the distribution of the form (See
\siref{supp_moments} for full derivation)
\begin{equation}
  \begin{aligned}
    \dt{\bb{\ee{m^x p^y}}} &=
    \Km \bb{\ee{m^x p^y}}\\
    &+ \Rm \bb{\ee{p^y \left[ (m + 1)^x -m^x \right]}}
     + \Gm \bb{\ee{m p^y \left[ (m - 1)^x - m^x \right]}}\\
    &+ \Rp \bb{\ee{m^{(x + 1)} \left[ (p + 1)^y - p^y \right]}}
     + \Gp \bb{\ee{m^x p \left[ (p - 1)^y - p^y \right]}}.
    \label{eq_gral_mom}
  \end{aligned}
\end{equation}

Given that all transitions in our stochastic model are first order reactions,
\eref{eq_gral_mom} has no moment-closure problem \cite{Voliotis2014a}. This
means that the dynamical equation for a given moment only depends on lower
moments (See \siref{supp_moments} for full proof). This feature of our model
implies, for example, that the second moment of the protein distribution
$\ee{p^2}$ depends only on the first two moments of the mRNA distribution
$\ee{m}$ and $\ee{m^2}$, the first protein moment $\ee{p}$, and the
cross-correlation term $\ee{mp}$. We can therefore define $\bmu$ to be a vector
containing all moments up to $\bb{\ee{m^x p^y}}$ for all promoter states,
\begin{equation}
\bmu = \left[ \bb{\ee{m^0 p^0}},
							\bb{\ee{m^1 p^0}},
							\ldots, \bb{\ee{m^x p^y}} \right]^T.
\end{equation}
Explicitly for the three-state promoter model depicted in
\fref{fig2_minimal_model}(A) this vector takes the form
\begin{equation}
	\bmu = \left[ \ee{m^0 p^0}_A, \ee{m^0 p^0}_I, \ee{m^0 p^0}_R,
	\ldots, \ee{m^x p^y}_A, \ee{m^x p^y}_I, \ee{m^x p^y}_R \right]^T.
\end{equation}

Given this definition we can compute the general moment dynamics as
\begin{equation}
\dt{\bmu} = \bb{A} \bmu,
\label{eq_mom_dynamics}
\end{equation}
where $\bb{A}$ is a square matrix that contains all the numerical coefficients
that relate each of the moments. We can then use \eref{eq_gral_mom} to build
matrix $\bb{A}$ by iteratively substituting values for the exponents $x$ and
$y$ up to a specified value. In the next section, we will use
\eref{eq_mom_dynamics} to numerically integrate the dynamical equations for our
moments of interest as cells progress through the cell cycle. We will then use
the value of the moments of the distribution to approximate the full gene
expression distribution. This method is computationally more efficient than
trying to numerically integrate the infinite set of equations describing the
full probability distribution $\bb{P}(m, p)$, or using a stochastic algorithm 
to sample from the distribution.

\subsection{Accounting for cell-cycle dependent variability in gene dosage}
\label{sec_cell_cycle}

As cells progress through the cell cycle, the genome has to be replicated to
guarantee that each daughter cell receives a copy of the genetic material. As
replication of the genome can take longer than the total cell cycle, this
implies that cells spend part of the cell cycle with multiple copies of each
gene depending on the cellular growth rate and the relative position of the
gene with respect to the replication origin \cite{Bremer1996}. Genes closer to
the replication origin spend a larger fraction of the cell cycle with multiple
copies compared to genes closer to the replication termination site
\cite{Bremer1996}. \fref{fig3_cell_cycle}(A) depicts a schematic of this
process where the replication origin ({\it oriC}) and the relevant locus for
our experimental measurements ({\it galK}) are highlighted.

Since this change in gene copy number has been shown to have an effect on
cell-to-cell variability in gene expression \cite{Jones2014a, Peterson2015}, we
now extend our minimal model to account for these changes in gene copy number
during the cell cycle.  We reason that the only difference between the
single-copy state and the two-copy state of the promoter is a doubling of the
mRNA production rate $r_m$. In particular, the promoter activation and
inactivation rates $\kpon$ and $\kpoff$ and the mRNA production rate $r_m$
inferred in \secref{sec_model} assume that cells spend a fraction $f$ of the
cell cycle  with one copy of the promoter (mRNA production rate $r_m$) and a
fraction $(1-f)$ of the cell cycle with two copies of the promoter (mRNA
production rate $2 r_m$). This inference was performed considering that at each
cell state the mRNA level immediately reaches the steady state value for the
corresponding mRNA production rate. This assumption is justified since the
timescale to reach this steady state depends only on the degradation rate
$\gm$, which for the mRNA  is much shorter ($\approx 3$ min) than the length of
the cell cycle ($\approx$ 60 min for our experimental conditions)
\cite{Dong1995}. \siref{supp_param_inference} shows that a model accounting for
this gene copy number variability is able to capture data from single molecule
mRNA counts of an unregulated (constitutively expressed) promoter.

Given that the protein degradation rate $\gp$ in our model is set by the cell
division time, we do not expect that the protein count will reach the
corresponding steady state value for each stage in the cell cycle. In other
words, cells do not spend long enough with two copies of the promoter for the
protein level to reach the steady state value corresponding to a transcription
rate of $2 r_m$. We therefore use the dynamical equations developed in
\secref{sec_moments} to numerically integrate the time trajectory of the
moments of the distribution with the corresponding parameters for each phase of
the cell cycle. \fref{fig3_cell_cycle}(B) shows an example corresponding to the
mean mRNA level (upper panel) and the mean protein level (lower panel) for the
case of the unregulated promoter. Given that we inferred the promoter rate
parameters considering that mRNA reaches steady state in each stage, we see
that the numerical integration of the equations is consistent with the
assumption of having the mRNA reach a stable value in each stage (See
\fref{fig3_cell_cycle}(B) upper panel). On the other hand, the mean protein
level does not reach a steady state at either of the cellular stages.
Nevertheless it is notable that after several cell cycles the trajectory from
cycle to cycle follows a repetitive pattern (See \fref{fig3_cell_cycle}(B)
lower panel). Previously we have experimentally observed this repetitive
pattern by tracking the expression level over time with video microscopy as
observed in Fig. 18 of \cite{Phillips2019}.

To test the effects of including this gene copy number variability in our model
we now compare the predictions of the model with experimental data. As detailed
in the Methods section, we obtained single-cell fluorescence values of
different {\it E. coli} strains carrying a YFP gene under the control of the
LacI repressor. Each strain was exposed to twelve different inducer
concentrations for $\approx 8$ generations for cells to adapt to the media. The
strains imaged spanned three orders of magnitude in repressor copy number and
three distinct repressor-DNA affinities. Since growth was asynchronous, we
reason that cells were randomly sampled at all stages of the cell cycle.
Therefore, when computing statistics from the data such as the mean fluorescence
value, in reality we are averaging over the cell cycle. In other words, as
depicted in \fref{fig3_cell_cycle}(B), quantities such as the mean protein copy
number change over time, i.e. $\ee{p} \equiv \ee{p(t)}$. This means that
computing the mean of a population of unsynchronized cells is equivalent to
averaging this time-dependent mean protein copy number over the span of the cell
cycle. Mathematically this is expressed as
\begin{equation}
	\ee{p}_c = \int_{t_o}^{t_d} \ee{p(t)} P(t) dt,
	\label{eq_time_avg}
\end{equation}
where $\ee{p(t)}$ represents the first moment of the protein distribution as
computed from \eref{eq_gral_mom}, $\ee{p}_c$ represents the average protein
copy number over a cell cycle, $t_o$ represents the start of the cell cycle,
$t_d$ represents the time of cell division, and $P(t)$ represents the
probability of any cell being at time $t \in [t_o, t_d]$ of their cell cycle.
We do not consider cells uniformly distributed along the cell cycle since it is
known that cells age is exponentially distributed, having more younger than
older cells at any point in time \cite{Powell1956} (See
\siref{supp_cell_age_dist} for further details). All computations hereafter are
therefore done by applying an average like that in \eref{eq_time_avg} for the
span of a cell cycle. We remind the reader that these time averages are done
under a fixed environmental state. It is the trajectory of cells over cell
cycles under a constant environment that we need to account for. It is through
this averaging over the span of a cell cycle that we turn a periodic process as
the one shown in \fref{fig3_cell_cycle}(B) into a stationary process that we can
compare with experimental data and, as we will see later, use to reconstruct the
steady state gene expression distribution.

\fref{fig3_cell_cycle}(C) compares zero-parameter fit predictions (lines) with
experimentally determined quantities (points). The upper row shows the
non-dimensional quantity known as the fold-change in gene expression
\cite{Garcia2011c}. This fold-change is defined as the relative mean gene
expression level with respect to an unregulated promoter. For protein this is
\begin{equation}
	\foldchange = {\ee{p(R > 0)}_c \over \ee{p(R = 0)}_c},
\end{equation}
where $\ee{p(R > 0)}_c$ represents the mean protein count for cells with
non-zero repressor copy number count $R$ over the entire cell cycle, and
$\ee{p(R = 0)}_c$ represents the equivalent for a strain with no repressors
present. The experimental points were determined from the YFP fluorescent
intensities of cells with varying repressor copy number and a $\Delta lacI$
strain with no repressor gene present (See Methods for further details). The
fold-change in gene expression has previously served as a metric to test the
validity of equilibrium-based models \cite{Phillips2015}. We note that the
curves shown in the upper panel of \fref{fig3_cell_cycle}(C) are consistent
with the predictions from equilibrium models \cite{Razo-Mejia2018} despite
being generated from a clearly non-equilibrium process as shown in
\fref{fig3_cell_cycle}(B). The kinetic model from \fref{fig2_minimal_model}(A)
goes beyond the equilibrium picture to generate predictions for moments of the
distribution other than the mean mRNA or mean protein count. To test this
extended predictive power the lower row of \fref{fig3_cell_cycle}(C) shows the
noise in gene expression defined as the standard deviation over the mean
protein count, accounting for the changes in gene dosage during the cell cycle.
Although our model systematically underestimates the noise in gene expression,
the zero-parameter fits capture the scaling of this noise. Possible origins of
this systematic discrepancy could be the intrinsic cell-to-cell variability of
rate parameters given the variability in the molecular components of the central
dogma machinery \cite{Jones2014a}, or noise generated by irreversible
non-equilibrium reactions not explicitly taken into account in our minimal model
\cite{Grah2020}.  The large errors for the highly repressed strains (lower left
panel in \fref{fig3_cell_cycle}(C)) are a result of having a small number in the
denominator - mean fluorescence level - when computing the noise. Although the
model is still highly informative about the physical nature of how cells
regulate their gene expression, the lack of exact numerical agreement between
theory and data opens an opportunity to gain new insights into the biophysical
origin of cell-to-cell variability. In \siref{supp_empirical} we explore
empirical ways to account for this systematic deviation. We direct the reader to
\siref{supp_multi_gene} where equivalent predictions are done ignoring the
changes in gene dosage due to the replication of the genome.

\begin{figure}[h!]
	\centering \includegraphics
  {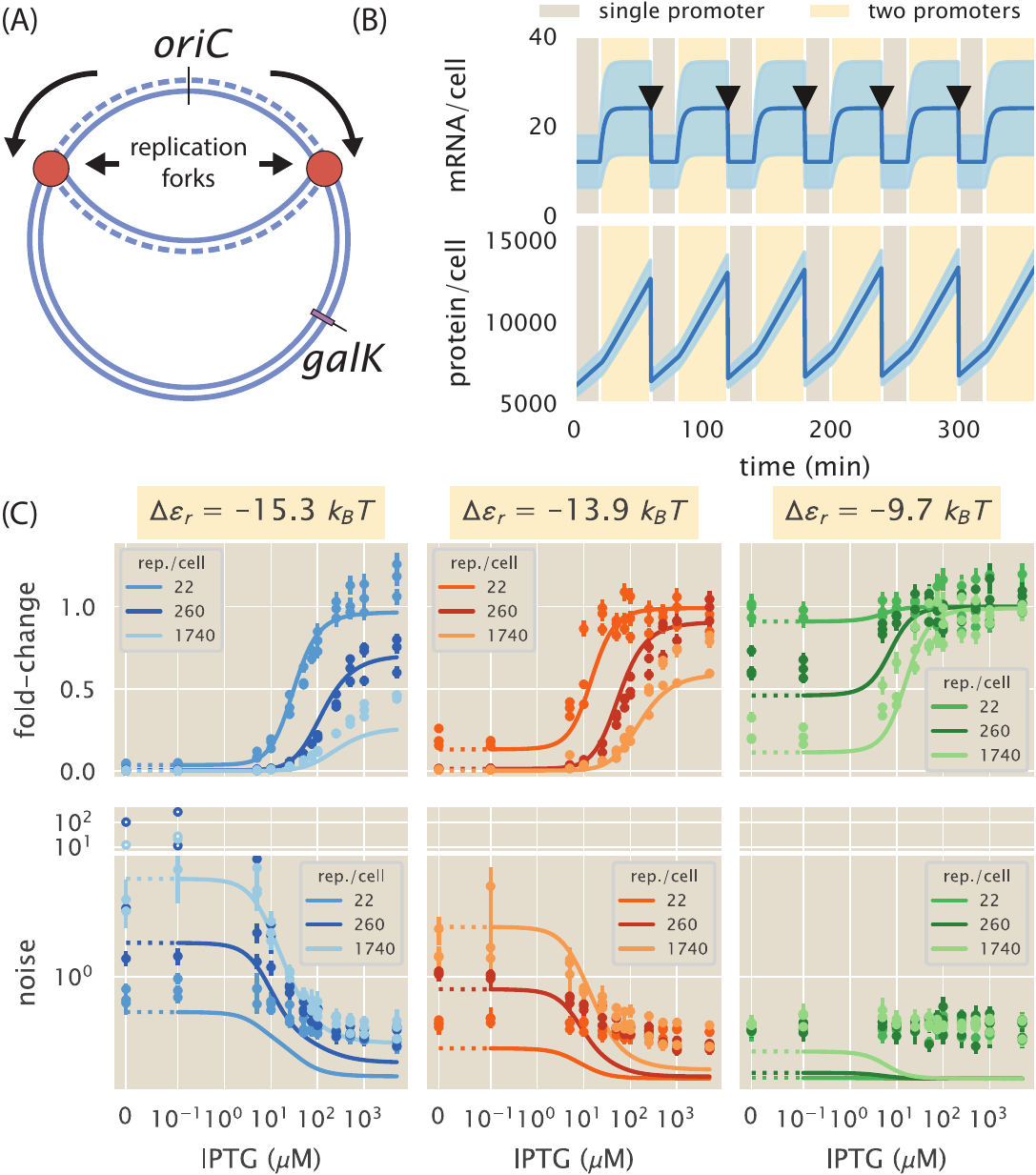}
	\caption{\textbf{Accounting for gene copy number variability during the
	cell cycle.} (A) Schematic of a replicating bacterial genome. As cells
	progress through the cell cycle the genome is replicated, duplicating gene
	copies for a fraction of the cell cycle before the cell divides. {\it oriC}
	indicates the replication origin, and {\it galK} indicates the locus at
	which the YFP reporter construct was integrated. (B) mean (solid line)
	$\pm$ standard deviation (shaded region) for the mRNA (upper panel) and
	protein (lower panel) dynamics. Cells spend a fraction of the cell cycle
	with a single copy of the promoter (light brown) and the rest of the cell
	cycle with two copies (light yellow). Black arrows indicate time of cell
	division. (C) Zero parameter-fit predictions (lines) and experimental data
	(circles) of the gene expression fold-change (upper row) and noise (lower
	row) for repressor binding sites with different affinities (different
	columns) and different repressor copy numbers per cell (different lines on
	each panel). Error bars in data represent the 95\% confidence interval on
	the quantities as computed from 10,000 bootstrap estimates generated from
	$> 500$ single-cell fluorescence measurements. In the theory curves, dotted
	lines indicate plot in linear scale to include zero, while solid lines
	indicate logarithmic scale. For visual clarity, data points in the noise
	panel with exceptionally large values coming from highly repressed strains
	are plotted on a separate panel.}
	\label{fig3_cell_cycle}
\end{figure}

\subsection{Maximum Entropy approximation}\label{sec_maxent}

Having numerically computed the moments of the mRNA and protein distributions
as cells progress through the cell cycle, we now proceed to make an
approximate reconstruction of the full distributions given this limited
information. As hinted in \secref{sec_moments} the maximum entropy principle,
first proposed by E.T. Jaynes in 1957 \cite{Jaynes1957}, approximates the
entire distribution by maximizing the Shannon entropy subject to constraints
given by the values of the moments of the distribution \cite{Jaynes1957}. This
procedure leads to a probability distribution of the form (See
\siref{supp_maxent} for full derivation)
\begin{equation}
  P(m, p) = {1 \over \mathcal{Z}}
              \exp \left( - \sum_{(x,y)} \lambda_{(x,y)} m^x p^y \right),
  \label{eq_maxEnt_joint}
\end{equation}
where $\lambda_{(x,y)}$ is the Lagrange multiplier associated with the
constraint set by the moment $\ee{m^x p^y}$, and $\mathcal{Z}$ is a
normalization constant. The more moments $\ee{m^x p^y}$ included as
constraints, the more accurate the approximation resulting from
\eref{eq_maxEnt_joint} becomes.

The computational challenge then becomes an optimization routine in which the
values for the Lagrange multipliers $\lambda_{(x,y)}$ that are consistent with
the constraints set by the moment values $\ee{m^x p^y}$ need to be found. This
is computationally more efficient than sampling directly out of the master
equation with a stochastic algorithm (see \siref{supp_gillespie} for further
comparison between maximum entropy estimates and the Gillespie algorithm).
\siref{supp_maxent} details our implementation of a robust algorithm to find
the values of the Lagrange multipliers. \fref{fig4_maxent}(A) shows example
predicted protein distributions reconstructed using the first six moments of
the protein distribution for a suite of different biophysical parameters and
environmental inducer concentrations. As repressor-DNA binding affinity
(columns in \fref{fig4_maxent}(A)) and repressor copy number (rows in
\fref{fig4_maxent}(A)) are varied, the responses to different signals, i.e.
inducer concentrations, overlap to varying degrees. For example, the upper
right corner frame with a weak binding site ($\eR = -9.7 \; k_BT$) and a low
repressor copy number (22 repressors per cell) have virtually identical
distributions regardless of the input inducer concentration. This means that
cells with this set of parameters cannot resolve any difference in the
concentration of the signal. As the number of repressors is increased, the
degree of overlap between distributions decreases, allowing cells to better
resolve the value of the signal input. On the opposite extreme the lower left
panel shows a strong binding site ($\eR = -15.3 \; k_BT$) and a high repressor
copy number (1740 repressors per cell). This parameter combination shows
overlap between distributions since the high degree of repression centers all
distributions towards lower copy numbers, again giving little ability for the
cells to resolve the inputs. In \fref{fig4_maxent}(B) and \siref{supp_maxent}
we show the comparison of these predicted cumulative distributions with the
experimental single-cell fluorescence distributions. Given the systematic
deviation of our predictions for the protein copy number noise highlighted in
\fref{fig3_cell_cycle}(C), the theoretical distributions (dashed lines)
underestimate the width of the experimental data. We again direct the reader to
\siref{supp_empirical} for an exploration of empirical changes to the moments
that improve the agreement of the predictions. In the following section we
formalize the notion of how well cells can resolve different inputs from an
information theoretic perspective via the channel capacity.

\begin{figure}[h!]
	\centering \includegraphics
  {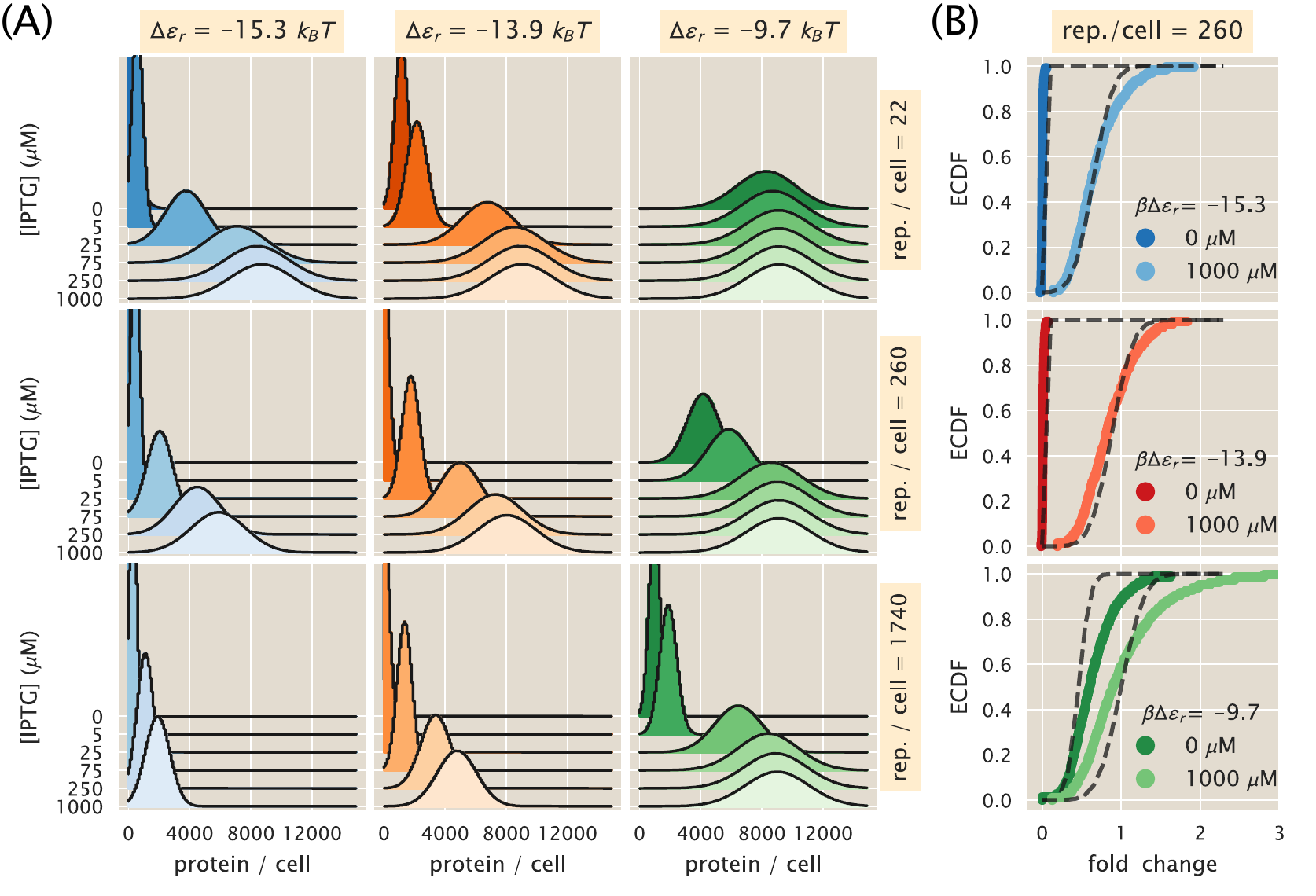}
	\caption{\textbf{Maximum entropy protein distributions for varying physical
	parameters.} (A) Predicted protein distributions under different inducer
	(IPTG) concentrations for different combinations of repressor-DNA
	affinities (columns) and repressor copy numbers (rows). The first six
	moments of the protein distribution used to constrain the maximum entropy
	approximation were computed by integrating \eref{eq_gral_mom} as cells
	progressed through the cell cycle as described in \secref{sec_cell_cycle}.
	(B) Theory-experiment comparison of predicted fold-change empirical
	cumulative distribution functions (ECDF). Each panel shows two example
	concentrations of inducer (colored curves) with their corresponding
	theoretical predictions (dashed lines). Distributions were normalized to the
	mean expression value of the unregulated strain in order to compare
	theoretical predictions in discrete protein counts with experimental
	fluorescent measurements in arbitrary units.}
  \label{fig4_maxent}
\end{figure}

\subsection{Theoretical prediction of the channel capacity}
\label{sec_channcap}

We now turn our focus to the channel capacity, which is a metric by which we
can quantify the degree to which cells can measure the environmental state (in
this context, the inducer concentration). The channel capacity is defined as
the mutual information $I$ between input and output (\eref{eq_mutual_info}),
maximized over all possible input distributions $P(c)$. If used as a metric of
how reliably a signaling system can infer the state of the external signal, the
channel capacity, when measured in bits, is commonly interpreted as the
logarithm of the number of states that the signaling system can properly
resolve. For example, a signaling system with a channel capacity of $C$ bits is
interpreted as being able to resolve $2^C$ states, though channel capacities
with fractional values are allowed. We therefore prefer the Bayesian
interpretation that the mutual information quantifies the improvement in the
inference of the input when considering the output compared to just using the
prior distribution of the input by itself for prediction \cite{Voliotis2014a,
Bowsher2014}. Under this interpretation a channel capacity of a fractional bit
still quantifies an improvement in the ability of the signaling system to infer
the value of the extracellular signal compared to having no sensing system at
all.

Computing the channel capacity implies optimizing over an infinite space of
possible distributions $P(c)$. For special cases in which the noise is small
compared to the dynamic range, approximate analytical equations have been
derived \cite{Tkacik2008a}. But given the high cell-to-cell variability that
our model predicts, the conditions of the so-called small noise approximation
are not satisfied. We therefore appeal to a numerical solution known as the
Blahut-Arimoto algorithm \cite{Blahut1972} (See \siref{supp_channcap} for
further details). \fref{fig5_channcap}(A) shows zero-parameter fit predictions
of the channel capacity as a function of the number of repressors for different
repressor-DNA affinities (solid lines). These predictions are contrasted with
experimental determinations of the channel capacity as inferred from
single-cell fluorescence intensity distributions taken over 12 different
concentrations of inducer. Briefly, from single-cell fluorescence measurements
we can approximate the input-output distribution $P(p \mid c)$. Once these
conditional distributions are fixed, the task of finding the input distribution
at channel capacity becomes a computational optimization routine that can be
undertaken using conjugate gradient or similar algorithms. For the particular
case of the channel capacity on a system with a discrete number of inputs and
outputs the Blahut-Arimoto algorithm is built in such a way that it guarantees
the convergence towards the optimal input distribution (See
\siref{supp_channcap} for further details). \fref{fig5_channcap}(B) shows
example input-output functions for different values of the channel capacity.
This illustrates that having access to no information (zero channel capacity)
is a consequence of having overlapping input-output functions (lower panel). On
the other hand, the more separated the input-output distributions are (upper
panel) the higher the channel capacity can be.

All theoretical predictions in \fref{fig5_channcap}(A) are systematically above
the experimental data. Although our theoretical predictions in
\fref{fig5_channcap}(A) do not numerically match the experimental inference of
the channel capacity, the model does capture interesting qualitative features of
the data that are worth highlighting. On one extreme, for cells with no
transcription factors, there is no information processing potential as this
simple genetic circuit would be constitutively expressed regardless of the
environmental state. As cells increase the transcription factor copy number, the
channel capacity increases until it reaches a maximum before falling back down
at high repressor copy number since the promoter would be permanently repressed.
The steepness of the increment in channel capacity as well as the height of the
maximum expression is highly dependent on the repressor-DNA affinity. For strong
binding sites (blue curve in \fref{fig5_channcap}(A)) there is a rapid increment
in the channel capacity, but the maximum value reached is smaller compared to a
weaker binding site (orange curve in \fref{fig5_channcap}(A)). In
\siref{supp_empirical} we show using the small noise approximation
\cite{Tkacik2008, Tkacik2008a} that if the systematic deviation of our
predictions on the cell-to-cell variability was explained with a multiplicative
constant, i.e. all noise predictions can be corrected by multiplying them by a
single constant, we would expect the channel capacity to be off by a constant
additive factor. This factor of $\approx 0.43$ bits can recover the agreement
between the model and the experimental data.

\begin{figure}[h!]
	\centering \includegraphics
  {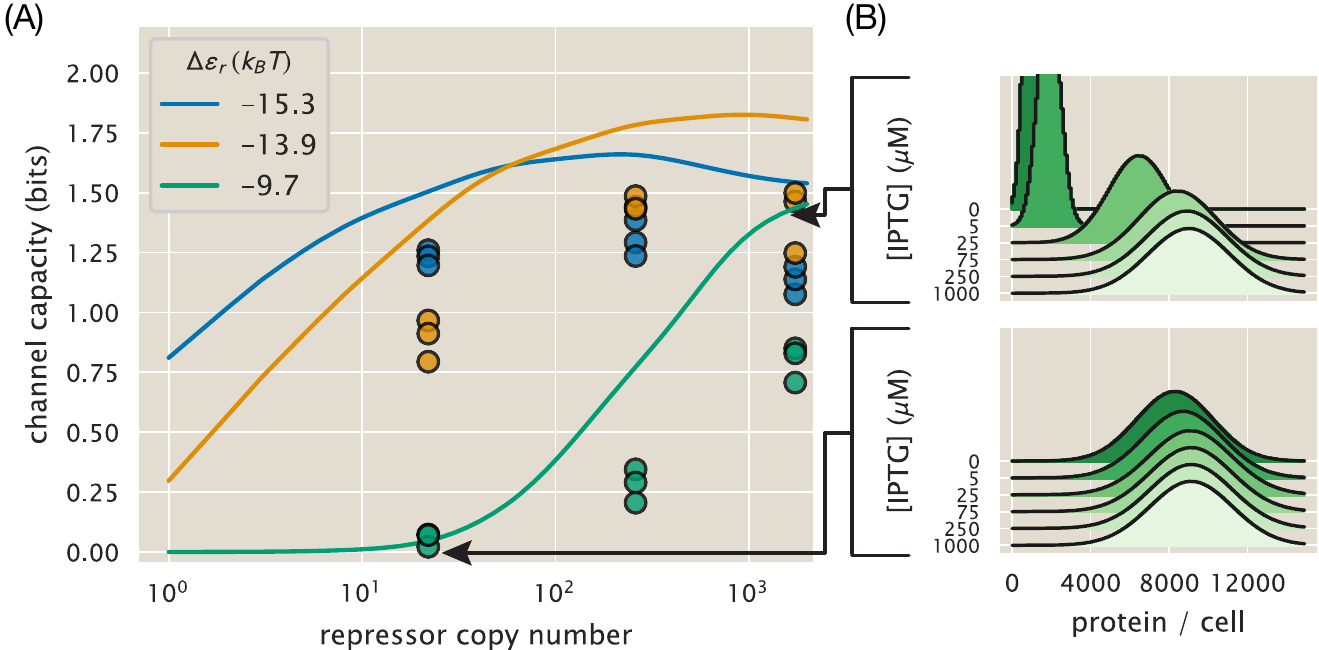}
	\caption{\textbf{Comparison of theoretical and experimental channel
	capacity.} (A) Channel capacity as inferred using the Blahut-Arimoto
	algorithm \cite{Blahut1972} for varying number of repressors and
	repressor-DNA affinities. All inferences were performed using 12 IPTG
	concentrations as detailed in the Methods. Curves represent zero-parameter
	fit predictions made with the maximum entropy distributions as shown in
	\fref{fig4_maxent}. Points represent inferences made from single cell
	fluorescence distributions (See \siref{supp_channcap} for further details).
	Theoretical curves were smoothed using a Gaussian kernel to remove
	numerical precision errors. (B) Example input-output functions in opposite
	limits of channel capacity. Lower panel illustrates that zero channel
	capacity indicates that all distributions overlap. Upper panel illustrates
	that as the channel capacity increases, the separation between
	distributions increases as well. Arrows point to the corresponding channel
	capacity computed from the predicted distributions.}
  \label{fig5_channcap}
\end{figure}

\section{Discussion}

Building on Shannon's formulation of information theory, there have been
significant efforts using this theoretical framework to understand the
information processing capabilities of biological systems, and the evolutionary
consequences for organisms harboring signal transduction systems
\cite{Bergstrom2004, Taylor2007, Tkacik2008, Polani2009, Nemenman2010,
Rivoire2011}. Recently, with the mechanistic dissection of molecular signaling
pathways, significant progress has been made on the question of the physical
limits of cellular detection and the role that features such as feedback loops
play in this task \cite{Bialek2005, Libby2007, Tkacik2011, Rhee2012a,
Voliotis2014a}. But the field still lacks a rigorous experimental test of these
ideas with precision measurements on a system that is tractable both
experimentally and theoretically.

In this paper we take advantage of the recent progress on the quantitative
modeling of input-output functions of genetic circuits to build a minimal model
of the simple repression motif \cite{Phillips2019}. By combining a series of
studies on this circuit spanning diverse experimental methods for measuring
gene expression under a myriad of different conditions, we possess complete a
priori parametric knowledge -- allowing us to generate parameter-free
predictions for processes related to information processing. Some of the model
parameters for our kinetic formulation of the input-output function are
informed by inferences made from equilibrium models. We use the fact that if
both kinetic and thermodynamic languages describe the same system, the
predictions must be self-consistent. In other words, if the equilibrium model
can only make statements about the mean mRNA and mean protein copy number
because of the way these models are constructed, those predictions must be
equivalent to what the kinetic model has to say about these same quantities.
This condition therefore constrains the values that the kinetic rates in the
model can take. To test whether or not the equilibrium picture can reproduce
the predictions made by the kinetic model we compare the experimental and
theoretical fold-change in protein copy number for a suite of biophysical
parameters and environmental conditions (\fref{fig3_cell_cycle}(C) upper row).
The agreement between theory and experiment demonstrates that these two
frameworks can indeed make consistent predictions.

The kinetic treatment of the system brings with it increasing predictive power
compared to the equilibrium picture. Under the kinetic formulation, the
predictions are not limited only to the mean but to any of the moments of the
mRNA and protein distributions. We first test these novel predictions by
comparing the noise in protein copy number (standard deviation / mean) with
experimental data. Our minimal model predicts the noise up to a systematic
deviation. The physical or biological origins of this discrepancy remain an
open question. In that way the work presented here exposes the status quo of
our understanding of gene regulation in bacteria, posing new questions to be
answered with future refinements of the model. We then extend our analysis to
infer entire protein distributions at different input signal concentrations by
using the maximum entropy principle. What this means is that we compute moments
of the protein distribution, and then use these moments to build an
approximation to the full distribution. These predicted distributions are then
compared with experimental single-cell distributions as shown in
\fref{fig4_maxent}(B) and \siref{supp_maxent}. Again, here although our minimal
model systematically underestimates the width of the distributions, it informs
how changes in parameters such as protein copy number or protein-DNA binding
affinity will affect the full probabilistic input-output function of the genetic
circuit, up to a multiplicative constant. We then use our model to predict the
information processing capacity.

By maximizing the mutual information between input signal concentration and
output protein distribution over all possible input distributions, we predict
the channel capacity of the system over a suite of biophysical parameters such
as varying repressor protein copy number and repressor-DNA binding affinity.
Although there is no reason to assume the the simplified synthetic circuit we
used as an experimental model operates optimally given the distribution of
inputs, the relevance of the channel capacity comes from its interpretation as
a metric of the physical limit of how precise an inference cells can
make about what the state of the environment is. Our model, despite the
systematic deviations, makes non-trivial predictions such as the existence of
an optimal repressor copy number for a given repressor-DNA binding energy,
predicting the channel capacity up to an additive constant (See
\fref{fig5_channcap}). The origin of this optimal combination of repressor copy
number and binding energy differs from previous publications in which an extra
term associated with the cost of producing protein was included in the model
\cite{Tkacik2011}. This optimal parameter combination is a direct consequence
of the fact that the LacI repressor cannot be fully deactivated
\cite{Razo-Mejia2018}. This implies that as the number of repressors increases,
a significant number of them are still able to bind to the promoter even at
saturating concentrations of inducer. This causes all of the input-output
functions to be shift towards low expression levels, regardless of the inducer
concentration, decreasing the amount of information that the circuit is able to
process.

We consider it important to highlight the limitations of the work presented
here. The previously discussed systematic deviation for the noise and skewness
of the predicted distributions (See \siref{supp_multi_gene}), and therefore of
the predicted distributions and channel capacity, remains an unresolved question
that deserves to be addressed in further iterations of our minimal model. Also,
as first reported in \cite{Razo-Mejia2018}, our model fails to capture the
steepness of the fold-change induction curve for the weakest repressor binding
site (See \fref{fig3_cell_cycle}(B)). Furthermore the minimal model in
\fref{fig2_minimal_model}(A), despite being widely used, is an
oversimplification of the physical picture of how the transcriptional machinery
works. The coarse-graining of all the kinetic steps involved in transcription
initiation into two effective promoter states -- active and inactive -- ignores
potential kinetic regulatory mechanisms of intermediate states
\cite{Scholes2017}. Moreover it has been argued that despite the fact that the
mRNA count distribution does not follow a Poisson distribution, this effect
could be caused by unknown factors not at the level of transcriptional
regulation \cite{Choubey2018}.

The findings of this work open the opportunity to accurately test intriguing
ideas that connect Shannon's metric of how accurately a signaling system can
infer the state of the environment, with Darwinian fitness \cite{Taylor2007}.
Beautiful work along these lines has been done in the context of the
developmental program of the early {\it Drosophila} embryo \cite{Tkacik2008,
Petkova2019}. These studies demonstrated that the input-output function of the
pair-rule genes works at channel capacity, suggesting that selection has acted
on these signaling pathways, pushing them to operate at the limit of what the
physics of these systems allows. Our system differs from the early embryo in
the sense that we have a tunable circuit with variable amounts of information
processing capabilities. Furthermore, compared with the fly embryo in which the
organism tunes both the input and output distributions over evolutionary time,
we have experimental control of the distribution of inputs that the cells are
exposed to. Consequently this means that instead of seeing the final result of
the evolutionary process, we would be able to set different environmental
challenges, and track over time the evolution of the population. These
experiments could shed light into the suggestive hypothesis of information bits
as a trait on which natural selection acts. We see this exciting direction as
part of the overall effort in quantitative biology of predicting evolution
\cite{Lassig2017}.

\section{Materials and Methods}

\subsection{\textit{E. coli} strains}

All strains used in this study were originally made for \cite{Razo-Mejia2018}.
We chose a subset of three repressor copy numbers that span two orders of
magnitude. We refer the reader to \cite{Razo-Mejia2018} for details on the
construction of these strains. Briefly, the strains have a construct consisting
of the \textit{lacUV5} promoter and one of three possible binding sites for the
\textit{lac} repressor (O1, O2, and O3) controlling the expression of a YFP
reporter gene. This construct is integrated into the genome at the
\textit{galK} locus. The number of repressors per cell is varied by changing
the ribosomal binding site controlling the translation of the \textit{lac}
repressor gene. The repressor constructs were integrated in the \textit{ybcN}
locus. Finally, all strains used in this work constitutively express an mCherry
reporter from a low copy number plasmid. This serves as a volume marker that
facilitates the segmentation of cells when processing microscopy images.

\subsection{Growth conditions}

For all experiments, cultures were initiated from a 50\% glycerol frozen stock
at -80$^\circ$C. Three strains - autofluorescence ($auto$), $\Delta lacI$
($\Delta$), and a strain with a known binding site and repressor copy number
($R$) - were inoculated into individual tubes with 2 mL of Lysogeny Broth (LB
Miller Powder, BD Medical) with 20 $\mu$g/mL of chloramphenicol and 30
$\mu$g/mL of kanamycin. These cultures were grown overnight at 37$^\circ$C with 
rapid agitation to reach saturation. The saturated cultures were diluted 1:1000
into 500 $\mu$L of M9 minimal media (M9 5X Salts, Sigma-Aldrich M6030; 2 mM
magnesium sulfate, Mallinckrodt Chemicals 6066-04; 100 mM calcium chloride,
Fisher Chemicals C79-500) supplemented with 0.5\% (w/v) glucose on a 2 mL
96-deep-well plate. The $R$ strain was diluted into 12 different wells with
minimal media, each with a different IPTG  concentration (0 $\mu$M, 0.1 $\mu$M,
5 $\mu$M, 10 $\mu$M, 25 $\mu$M, 50 $\mu$M, 75 $\mu$M, 100 $\mu$M, 250 $\mu$M,
500 $\mu M$, 1000 $\mu$M, 5000 $\mu$M) while  the $auto$ and $\Delta$ strains
were diluted into two wells (0 $\mu$M, 5000 $\mu$M). Each of the IPTG
concentrations came from a single preparation stock kept in 100-fold
concentrated aliquots. The 96 well plate was then incubated at 37$^\circ$C with
rapid agitation for 8 hours before imaging.

\subsection{Microscopy imaging procedure}

The microscopy pipeline used for this work exactly followed the steps from
\cite{Razo-Mejia2018}. Briefly, twelve 2\% agarose (Life Technologies UltraPure
Agarose, Cat.No. 16500100) gels were made out of M9 media (or PBS buffer) with
the corresponding IPTG concentration (see growth conditions) and placed between
two glass coverslips for them to solidify after microwaving. After the 8 hour
incubation in minimal media, 1 $\mu$L of a 1:10 dilution of the cultures into
fresh media or PBS buffer was placed into small squares (roughly 10 mm $\times$
10 mm) of the different agarose gels. A total of 16 agarose squares - 12
concentrations of IPTG for the $R$ strain, 2 concentrations for the $\Delta$ and
2 for the $auto$ strain - were mounted into a single glass-bottom dish (Ted
Pella Wilco Dish, Cat. No. 14027-20) that was sealed with parafilm.

All imaging was done on an inverted fluorescent microscope (Nikon Ti-Eclipse)
with custom-built laser illumination system. The YFP fluorescence (quantitative
reporter) was imaged with a CrystaLaser 514 nm excitation laser coupled with a
laser-optimized (Semrock Cat. No. LF514-C-000) emission filter. All strains,
including the $auto$ strain, included a constitutively expressed mCherry protein
to aid the segmentation. Therefore, for each image three channels (YFP, On
average 30 images with roughly 20 cells per condition were taken. 25 images of a
fluorescent slide and 25 images of the camera background noise were taken every
imaging session in order to flatten the illumination. The image processing
pipeline for this work is exactly the same as in \cite{Razo-Mejia2018}.

\subsection{Data and Code Availability}

All data and custom scripts were collected and stored using Git version
control. Code for raw data processing, theoretical analysis, and figure
generation is available on the GitHub repository
(\url{https://github.com/RPGroup-PBoC/chann_cap}). The code can also be
accessed via the paper website
(\url{https://www.rpgroup.caltech.edu/chann_cap/}). Raw microscopy data are
stored on the CaltechDATA data repository and can be accessed via DOI
\url{https://doi.org/10.22002/d1.1184}. Bootstrap estimates of experimental
channel capacity are also available on the CaltechDATA data repository via
\url{https://doi.org/10.22002/D1.1185}.

\section{Acknowledgements}

We would like to also thank Nathan Belliveau, Michael Betancourt, William
Bialek, Justin Bois, Emanuel Flores, Hernan Garcia, Alejandro Granados,
Porfirio Quintero,  Catherine Triandafillou, and Ned Wingreen for useful advice
and discussion. We would especially like to thank Alvaro Sanchez, Gasper
Tkacik, and Jane Kondev for critical observations on the manuscript. We thank
Rob Brewster for providing the raw mRNA FISH data for inferences, and David
Drabold for advice on the maximum entropy inferences. We are grateful to Heun
Jin Lee for his key support with the quantitative microscopy. This work was
supported by La Fondation Pierre-Gilles de Gennes, the Rosen Center at Caltech,
and the NIH 1R35 GM118043 (MIRA). M.R.M. was supported by the Caldwell CEMI
fellowship. 
		}{}%
		\printbibliography[segment=\therefsegment]
	\end{refsegment}

\clearpage

\title{\textbf{Supplemental Information for: First-principles prediction of the
information processing capacity of a simple genetic circuit}}

\setboolean{sitext}{true}
\ifthenelse{\boolean{sitext}}{
\maketitle

\addtocontents{toc}{\protect\setcounter{tocdepth}{2}}

	\begin{refsegment}
		\beginsupplement
		\tableofcontents

\section{Three-state promoter model for simple repression}\label{supp_model}

In order to tackle the question of how much information the simple repression
motif can process we require the joint probability distribution of mRNA and
protein $P(m, p; t)$. To obtain this distribution we use the chemical master
equation formalism as described in \secref{sec_model}. Specifically, we assume
a three-state model where the promoter can be found 1) in a transcriptionally
active state  ($A$ state), 2) in a transcriptionally inactive state without the
repressor bound ($I$ state) and 3) with the repressor bound ($R$ state). (See
\fref{fig2_minimal_model}(A)). These three states generate a system of coupled
differential equations for each of the three state distributions $P_A(m, p)$,
$P_I(m, p)$ and $P_R(m, p)$. Given the rates shown in
\fref{fig2_minimal_model}(A) let us define the system of ODEs. For the
transcriptionally active state we have
\begin{equation}
	\begin{aligned}
	\dt{P_A(m, p)} &=
	- \overbrace{\kpoff P_A(m, p)}^{A \rightarrow I} %
	+ \overbrace{\kpon P_I(m, p)}^{I \rightarrow A}\\ %
	&+ \overbrace{r_m P_A(m-1, p)}^{m-1 \rightarrow m} %
	- \overbrace{r_m P_A(m, p)}^{m \rightarrow m+1}%
	+ \overbrace{\gm (m + 1) P_A(m+1 , p)}^{m+1 \rightarrow m} %
	- \overbrace{\gm m P_A(m , p)}^{m \rightarrow m-1}\\ %
	&+ \overbrace{r_p m P_A(m, p - 1)}^{p-1 \rightarrow p} %
	- \overbrace{r_p m P_A(m, p)}^{p \rightarrow p+1} %
	+ \overbrace{\gp (p + 1) P_A(m, p + 1)}^{p + 1 \rightarrow p} %
	- \overbrace{\gp p P_A(m, p)}^{p \rightarrow p-1}. %
	\end{aligned}
\end{equation}
For the inactive promoter state $I$ we have
\begin{equation}
	\begin{aligned}
	\dt{P_I(m, p)} &=
	\overbrace{\kpoff P_A(m, p)}^{A \rightarrow I} %
	- \overbrace{\kpon P_I(m, p)}^{I \rightarrow A} %
	+ \overbrace{\kroff P_R(m, p)}^{R \rightarrow I} %
	- \overbrace{\kron P_I(m, p)}^{I \rightarrow R}\\ %
	&+ \overbrace{\gm (m + 1) P_I(m+1 , p)}^{m+1 \rightarrow m} %
	- \overbrace{\gm m P_I(m , p)}^{m \rightarrow m-1}\\ %
	&+ \overbrace{r_p m P_I(m, p - 1)}^{p-1 \rightarrow p} %
	- \overbrace{r_p m P_I(m, p)}^{p \rightarrow p+1} %
	+ \overbrace{\gp (p + 1) P_I(m, p + 1)}^{p + 1 \rightarrow p} %
	- \overbrace{\gp p P_I(m, p)}^{p \rightarrow p-1}. %
	\end{aligned}
\end{equation}
And finally for the repressor bound state $R$ we have
\begin{equation}
	\begin{aligned}
	\dt{P_R(m, p)} &=
	- \overbrace{\kroff P_R(m, p)}^{R \rightarrow I} %
	+ \overbrace{\kron P_I(m, p)}^{I \rightarrow R}\\ %
	&+ \overbrace{\gm (m + 1) P_R(m+1 , p)}^{m+1 \rightarrow m} %
	- \overbrace{\gm m P_R(m , p)}^{m \rightarrow m-1}\\ %
	&+ \overbrace{r_p m P_R(m, p - 1)}^{p-1 \rightarrow p} %
	- \overbrace{r_p m P_R(m, p)}^{p \rightarrow p+1} %
	+ \overbrace{\gp (p + 1) P_R(m, p + 1)}^{p + 1 \rightarrow p} %
	- \overbrace{\gp p P_R(m, p)}^{p \rightarrow p-1}. %
	\end{aligned}
\end{equation}

For an unregulated promoter, i.e. a promoter in a cell that has no repressors
present, and therefore constitutively expresses the gene, we use a two-state
model in which the state $R$ is not allowed. All the terms in the system of ODEs
containing $\kron$ or $\kroff$ are then set to zero.

As detailed in \secref{sec_model} it is convenient to express this system using
matrix notation \cite{Sanchez2013}. For this we define $\PP(m, p) = (P_A(m, p),
P_I(m, p), P_R(m, p))^T$. Then the system of ODEs can be expressed as
\begin{equation}
	\begin{aligned}
	\dt{\PP(m, p)} &= \Km \PP(m, p)
	- \Rm \PP(m, p) + \Rm \PP(m-1, p)
	- m \Gm \PP(m, p) + (m + 1) \Gm \PP(m + 1, p)\\
	&- m \Rp \PP(m, p) + m \Rp \PP(m, p - 1)
	- p \Gp \PP(m, p) + (p + 1) \Gp \PP(m, p + 1),
	\end{aligned}
\end{equation}
where we defined matrices representing the promoter state transition $\Km$,
\begin{align}
	\Km \equiv
	\begin{bmatrix}
	-\kpoff   & \kpon         & 0\\
	\kpoff    & -\kpon -\kron  & \kroff\\
	0         & \kron         & -\kroff
	\end{bmatrix},
\end{align}
mRNA production, $\Rm$, and degradation, $\Gm$, as
\begin{equation}
	\Rm \equiv
	\begin{bmatrix}
	r_m   & 0 & 0\\
	0     & 0 & 0\\
	0     & 0 & 0\\
	\end{bmatrix},
\end{equation}
and
\begin{equation}
	\Gm \equiv
	\begin{bmatrix}
	\gm   & 0   & 0\\
	0     & \gm & 0\\
	0     & 0   & \gm\\
	\end{bmatrix}.
\end{equation}
For the protein we also define production $\Rp$ and degradation $\Gp$ matrices
as
\begin{equation}
	\Rp \equiv
	\begin{bmatrix}
	r_p   & 0   & 0\\
	0     & r_p & 0\\
	0     & 0   & r_p\\
	\end{bmatrix}
\end{equation}
and
\begin{equation}
	\Gp \equiv
	\begin{bmatrix}
	\gp   & 0   & 0\\
	0     & \gp & 0\\
	0     & 0   & \gp\\
	\end{bmatrix}.
\end{equation}

The corresponding equation for the unregulated two-state promoter takes the
exact same form with the definition of the matrices following the same scheme
without including the third row and third column, and setting $\kron$ and
$\kroff$ to zero.

A closed-form solution for this master equation might not even exist. The
approximate solution of chemical master equations of this kind is an active
area of research. As we will see in \siref{supp_param_inference} the two-state
promoter master equation has been analytically solved for the mRNA
\cite{Peccoud1995} and protein distributions \cite{Shahrezaei2008}. For our
purposes, in \siref{supp_maxent} we will detail how to use the Maximum Entropy
principle to approximate the full distribution for the two- and three-state
promoter.

\section{Parameter inference}\label{supp_param_inference}

(Note: The Python code used for the calculations presented in this section can
be found in the
\href{https://www.rpgroup.caltech.edu//chann_cap/software/chemical_master_mRNA_FISH_mcmc.html}{following
link} as an annotated Jupyter notebook)

With the objective of generating falsifiable predictions with meaningful
parameters, we infer the kinetic rates for this three-state promoter model
using different data sets generated in our lab over the last decade concerning
different aspects of the regulation of the simple repression motif. For
example, for the unregulated promoter transition rates $\kpon$ and $\kpoff$ and
the mRNA production rate $r_m$, we use single-molecule mRNA FISH counts from an
unregulated promoter \cite{Jones2014a}. Once these parameters are fixed, we use
the values to constrain the repressor rates $\kron$ and $\kroff$. These
repressor rates are obtained using information from mean gene expression
measurements from bulk LacZ colorimetric assays \cite{Garcia2011c}. We also
expand our model to include the allosteric nature of the repressor protein,
taking advantage of video microscopy measurements done in the context of
multiple promoter copies \cite{Brewster2014} and flow-cytometry measurements of
the mean response of the system to different levels of induction
\cite{Razo-Mejia2018}. In what follows of this section we detail the steps
taken to infer the parameter values. At each step the values of the parameters
inferred in previous steps constrain the values of the parameters that are not
yet determined, building in this way a self-consistent model informed by work
that spans several experimental techniques.

\subsection{Unregulated promoter rates}

We begin our parameter inference problem with the promoter on and off rates
$\kpon$ and $\kpoff$, as well as the mRNA production rate $r_m$. In this case
there are only two states available to the promoter -- the inactive state $I$
and the transcriptionally active  state $A$. That means that the third ODE for
$P_R(m, p)$ is removed from the system. The mRNA steady state distribution for
this particular two-state promoter model was solved analytically by Peccoud and
Ycart \cite{Peccoud1995}. This distribution $P(m) \equiv P_I(m) + P_A(m)$ is of
the form
\begin{equation}
	\small
	P(m \mid \kpon, \kpoff, r_m, \gm) =
	{\Gamma \left( \frac{\kpon}{\gm} + m \right) \over
	\Gamma(m + 1) \Gamma\left( \frac{\kpoff+\kpon}{\gm} + m \right)}
	{\Gamma\left( \frac{\kpoff+\kpon}{\gm} \right) \over
	\Gamma\left( \frac{\kpon}{\gm} \right)}
	\left( {r_m \over \gm} \right)^m
	F_1^1 \left( {\kpon \over \gm} + m,
	{\kpoff + \kpon \over \gm} + m,
	-{r_m \over \gm} \right),
	\label{seq_two_state_mRNA}
\end{equation}
where $\Gamma(\cdot)$ is the gamma function, and $F_1^1$ is the confluent
hypergeometric function of the first kind. This rather complicated expression
will aid us to find parameter values for the rates. The inferred rates $\kpon$,
$\kpoff$ and $r_m$ will be expressed in units of the mRNA degradation rate
$\gm$. This is because the model in \eref{seq_two_state_mRNA} is homogeneous in
time, meaning that if we divide all rates by a constant it would be equivalent
to multiplying the characteristic time scale by the same constant. As we will
discuss in the next section, \eref{seq_two_state_mRNA} has degeneracy in the
parameter values. What this means is that a change in one of the parameters,
specifically $r_m$, can be compensated by a change in another parameter,
specifically $\kpoff$, to obtain the exact same distribution. To work around
this intrinsic limitation of the model we will include in our inference prior
information from what we know from equilibrium-based models.

\subsubsection{Bayesian parameter inference of RNAP rates}

In order to make progress at inferring the unregulated promoter state
transition rates, we make use of the single-molecule mRNA FISH data from Jones
et al. \cite{Jones2014a}. \fref{sfig_lacUV5_FISH} shows the distribution of mRNA per cell
for the \textit{lacUV5} promoter used for our inference. This
promoter, being very strong, has a mean copy number of $\ee{m} \approx 18$
mRNA/cell.

\begin{figure}[h!]
	\centering \includegraphics
	{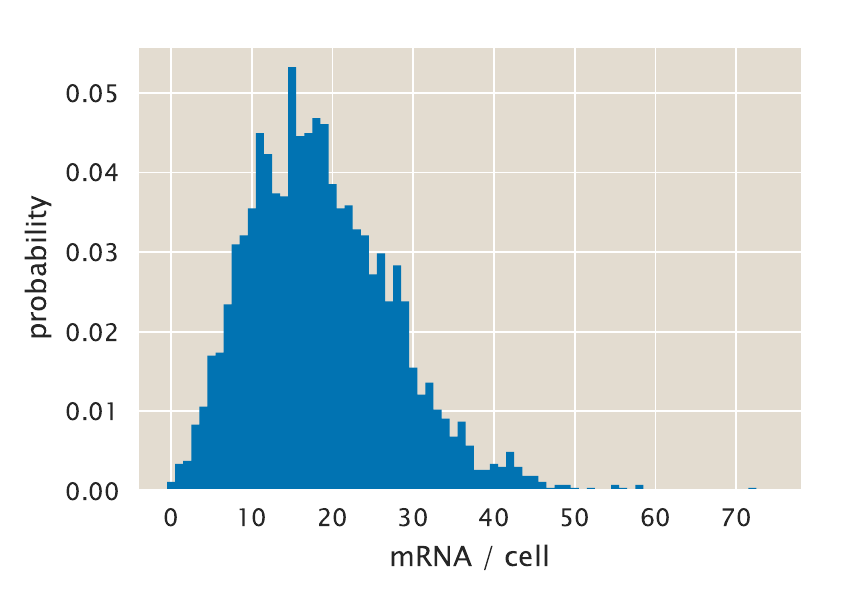}
	\caption{\textbf{\textit{lacUV5} mRNA per cell distribution.} Data from
	\cite{Jones2014a} of the unregulated \textit{lacUV5} promoter as inferred
	from single molecule mRNA FISH.}
	\label{sfig_lacUV5_FISH}
\end{figure}

Having this data in hand we now turn to Bayesian parameter inference. Writing
Bayes theorem we have
\begin{equation}
	P(\kpon, \kpoff, r_m \mid D) = {P(D \mid \kpon, \kpoff, r_m)
	P(\kpon, \kpoff, r_m) \over P(D)},
	\label{seq_bayes_rnap_rates}
\end{equation}
where $D$ represents the data. For this case the data consists of single-cell
mRNA counts $D = \{ m_1, m_2, \ldots, m_N \}$, where $N$ is the number of
cells. We assume that each cell's measurement is independent of the others such
that we can rewrite \eref{seq_bayes_rnap_rates} as
\begin{equation}
	P(\kpon, \kpoff, r_m \mid \{m_i\}) \propto
	\left[\prod_{i=1}^N P(m_i \mid \kpon, \kpoff, r_m) \right]
	P(\kpon, \kpoff, r_m),
	\label{seq_bayes_sample}
\end{equation}
where we ignore the normalization constant $P(D)$. The likelihood term $P(m_i
\mid \kpon, \kpoff, r_m)$ is exactly given by \eref{seq_two_state_mRNA} with
$\gm = 1$. Given that we have this functional form for the distribution, we can
use Markov Chain Monte Carlo (MCMC) sampling to explore the 3D parameter space
in order to fit \eref{seq_two_state_mRNA} to the mRNA-FISH data.

\subsubsection{Constraining the rates given prior thermodynamic knowledge.}

One of the strengths of the Bayesian approach is that we can include all the
prior knowledge on the parameters when performing an inference
\cite{MacKay2003}. Basic features such as the fact that the rates have to be
strictly positive constrain the values that these parameters can take. For
the specific rates analyzed in this section we know more than the simple
constraint of non-negative values. The expression of an unregulated promoter has
been studied from a thermodynamic perspective \cite{Brewster2012}. Given the
underlying assumptions of these equilibrium models, in which the probability of
finding the RNAP bound to the promoter is proportional to the transcription
rate \cite{Bintu2005a}, they can only make statements about the mean expression
level. Nevertheless if both the thermodynamic and the kinetic model describe
the same process, the predictions for the mean gene expression level must
agree. That means that we can use what we know about the mean gene expression,
and how this is related to parameters such as molecule copy numbers and binding
affinities, to constrain the values that the rates in question can take.

In the case of this two-state promoter it can be shown that the mean number of
mRNA is given by \cite{Sanchez2013} (See \siref{supp_moments} for moment
computation)
\begin{equation}
	\ee{m} = {r_m \over \gm} {\kpon \over \kpon + \kpoff}.
	\label{seq_mean_kinetic}
\end{equation}
Another way of expressing this is as ${r_m \over \gm} \times
p_{\text{active}}^{(p)}$, where $p_{\text{active}}^{(p)}$ is the probability of
the promoter being in the transcriptionally active state. The thermodynamic
picture has an equivalent result where the mean number of mRNA is given by
\cite{Brewster2012, Bintu2005a}
\begin{equation}
	\left\langle m \right\rangle = {r_m \over \gm}
	{{P \over \Nns} e^{-\beta\eP}
	\over 1 + {P \over \Nns} e^{-\beta\eP}},
	\label{seq_mean_thermo}
\end{equation}
where $P$ is the number of RNAP per cell, $\Nns$ is the number of non-specific
binding sites, $\eP$ is the RNAP binding energy in $k_BT$ units and $\beta\equiv
{(k_BT)}^{-1}$. Using \eref{seq_mean_kinetic} and \eref{seq_mean_thermo} we can
easily see that if these frameworks are to be equivalent, then it must be true
that
\begin{equation}
	{\kpon \over \kpoff} = {P \over \Nns} e^{-\beta\eP},
\end{equation}
or equivalently
\begin{equation}
	\ln \left({\kpon \over \kpoff}\right) =
	-\beta\eP + \ln P - \ln \Nns.
\end{equation}
To put numerical values into these variables we can use information from the
literature. The RNAP copy number is order $P \approx 1000-3000$ RNAP/cell for a
1 hour doubling time \cite{Klumpp2008}. As for the number of non-specific
binding sites and the binding energy, we have that $\Nns = 4.6\times 10^6$
\cite{Bintu2005a} and $-\beta\eP \approx 5 - 7$ \cite{Brewster2012}. Given
these values we define a Gaussian prior for the log ratio of these two
quantities of the form
\begin{equation}
	P\left(\ln \left({\kpon \over \kpoff}\right) \right) \propto
	\exp \left\{ - {\left(\ln \left({\kpon \over \kpoff}\right) -
	\left(-\beta\eP + \ln P - \ln \Nns \right) \right)^2
	\over 2 \sigma^2} \right\},
	\label{seq_prior_single}
\end{equation}
where $\sigma$ is the variance that accounts for the uncertainty in these
parameters. We include this prior as part of the prior term $P(\kpon, \kpoff,
r_m)$ of \eref{seq_bayes_sample}. We then use MCMC to sample out of the
posterior distribution given by \eref{seq_bayes_sample}. \fref{sfig_mcmc_rnap}
shows the MCMC samples of the posterior distribution. For the case of the
$\kpon$ parameter there is a single symmetric peak. $\kpoff$ and $r_m$ have a
rather long tail towards large values. In fact, the 2D projection of $\kpoff$ vs
$r_m$ shows that the model is sloppy, meaning that the two parameters are highly
correlated. This feature is a common problem for many non-linear systems used in
biophysics and systems biology \cite{Transtrum2015}. What this implies is that
we can change the value of $\kpoff$, and then compensate by a change in $r_m$ in
order to maintain the shape of the mRNA distribution. Therefore it is impossible
from the data and the model themselves to narrow down a single value for the
parameters. Nevertheless since we included the prior information on the rates as
given by the analogous form between the equilibrium and non-equilibrium
expressions for the mean mRNA level, we obtained a more constrained parameter
value for the RNAP rates and the transcription rate that we will take as the
peak of this long-tailed distribution.

\begin{figure}[h!]
	\centering \includegraphics
	{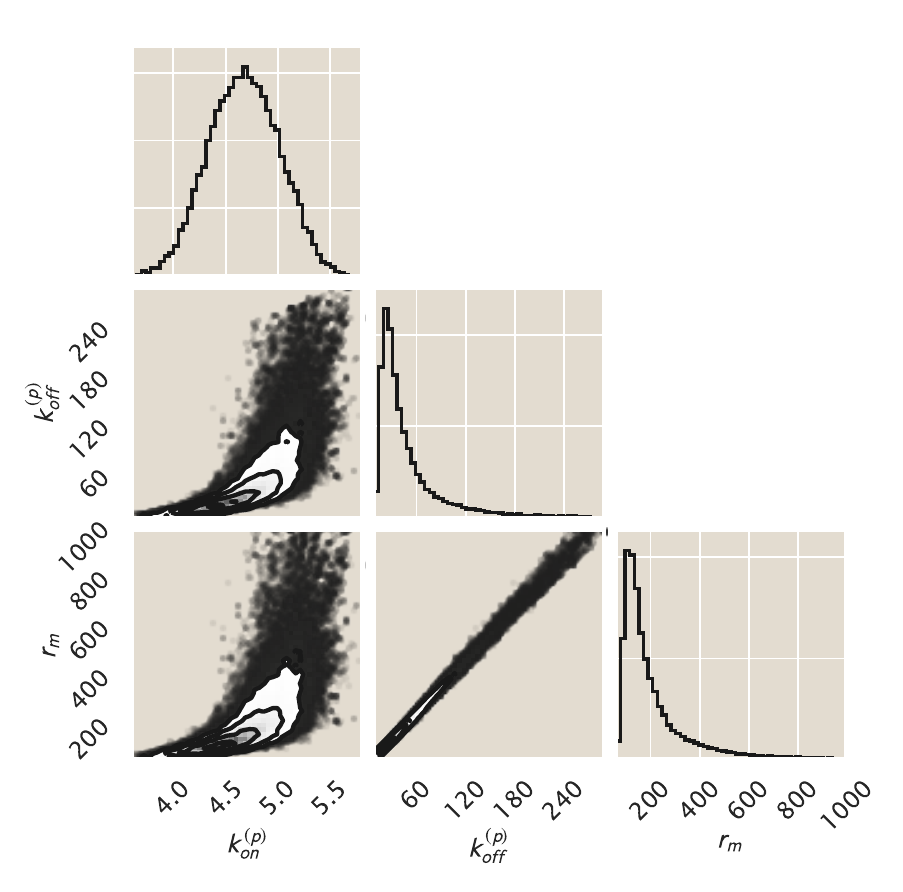}
	\caption{\textbf{MCMC posterior distribution.} Sampling out of
	\eref{seq_bayes_sample} the plot shows 2D and 1D projections of the 3D
	parameter space. The parameter values are (in units of the mRNA degradation
	rate $\gm$) $\kpon = 4.3^{+1}_{-0.3}$, $\kpoff = 18.8^{+120}_{-10}$ and $r_m =
	103.8^{+423}_{-37}$ which are the modes of their respective distributions,
	where the superscripts and subscripts represent the upper and lower bounds of
	the 95$^\text{th}$ percentile of the parameter value distributions}
	\label{sfig_mcmc_rnap}
\end{figure}

The inferred values $\kpon = 4.3^{+1}_{-0.3}$, $\kpoff = 18.8^{+120}_{-10}$
and $r_m = 103.8^{+423}_{-37}$ are given in units of the mRNA degradation
rate $\gm$. Given the asymmetry of the parameter distributions we report the
upper and lower bound of the 95$^\text{th}$ percentile of the posterior
distributions. Assuming a mean life-time for mRNA of $\approx$ 3 min 
we have an mRNA degradation rate of $\gm \approx 2.84 \times 10^{-3} s^{-1}$.
Using this value gives the following values for the inferred rates: $\kpon =
0.024_{-0.002}^{+0.005} s^{-1}$, $\kpoff = {0.11}_ {-0.05}^{+0.66} s^{-1}$, and
$r_m = 0.3_{-0.2}^{+2.3} s^{-1}$.

\fref{sfig_lacUV5_theory_data} compares the experimental data from
\fref{sfig_lacUV5_FISH} with the resulting distribution obtained by substituting
the most likely parameter values into \eref{seq_two_state_mRNA}. As we can see
this two-state model fits the data adequately.

\begin{figure}[h!]
	\centering \includegraphics[width=0.5\columnwidth]
	{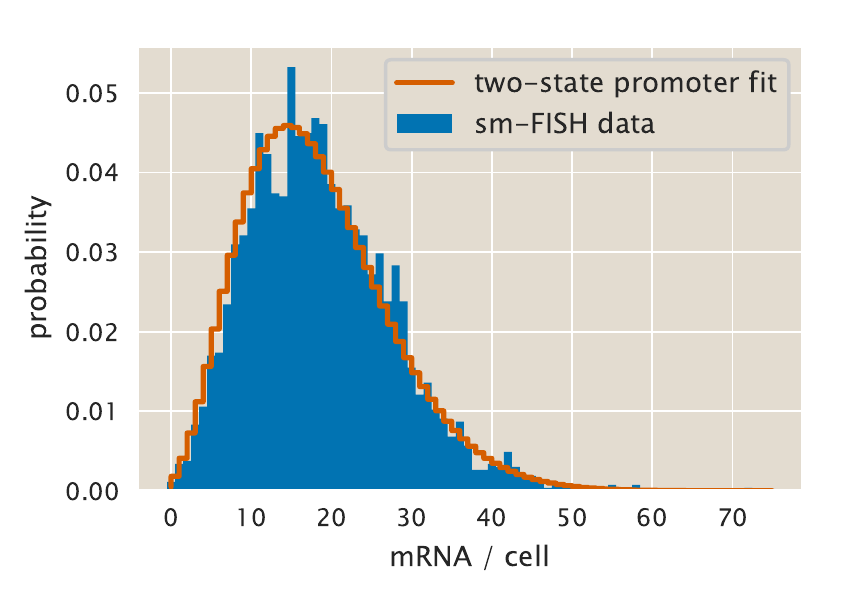}
	\caption{\textbf{Experimental vs. theoretical distribution of mRNA per cell
	using parameters from Bayesian inference.} Dotted line shows the result of
	using \eref{seq_two_state_mRNA} along with the parameters inferred for the
	rates. Blue bars are the same data as \fref{sfig_lacUV5_FISH} obtained from
	\cite{Jones2014a}.}
	\label{sfig_lacUV5_theory_data}
\end{figure}

\subsection{Accounting for variability in the number of promoters}

As discussed in ref. \cite{Jones2014a} and further expanded in
\cite{Peterson2015} an important source of cell-to-cell variability in gene
expression in bacteria is the fact that, depending on the growth rate and the
position relative to the chromosome replication origin, cells can have multiple
copies of any given gene. Genes closer to the replication origin have on
average higher gene copy number compared to genes at the opposite end. For the
locus in which our reporter construct is located (\textit{galK}) and the
doubling time of the mRNA FISH experiments we expect to have $\approx$ 1.66
copies of the gene \cite{Jones2014a, Bremer1996}. This implies that the cells
spend 2/3 of the cell cycle with two copies of the promoter and the rest with a
single copy.

To account for this variability in gene copy we extend the model assuming that
when cells have two copies of the promoter the mRNA production rate is $2 r_m$
compared to the rate $r_m$ for a single promoter copy. The probability of
observing a certain mRNA copy $m$ is therefore given by
\begin{equation}
	P(m) = P(m \mid \text{one promoter}) \cdot P(\text{one promoter}) +
	P(m \mid \text{two promoters}) \cdot P(\text{two promoters}).
	\label{seq_prob_multipromoter}
\end{equation}
Both terms $P(m \mid \text{promoter copy})$ are given by
\eref{seq_two_state_mRNA} with the only difference being the rate $r_m$. It is
important to acknowledge that \eref{seq_prob_multipromoter} assumes that once
the gene is replicated the time scale in which the mRNA count relaxes to the
new steady state is much shorter than the time that the cells spend in this two
promoter copies state. This approximation should be valid for a short lived
mRNA molecule, but the assumption is not applicable for proteins whose
degradation rate is comparable to the cell cycle length as explored in
\secref{sec_cell_cycle}.

In order to repeat the Bayesian inference including this variability in gene
copy number we must split the mRNA count data into two sets -- cells with a
single copy of the promoter and cells with two copies of the promoter. For the
single molecule mRNA FISH data there is no labeling of the locus, making it
impossible to determine the number of copies of the promoter for any given
cell. We therefore follow Jones et al. \cite{Jones2014a} in using the cell area
as a proxy for stage in the cell cycle. In their approach they sorted cells by
area, considering cells below the 33$\th$ percentile having a single promoter
copy and the rest as having two copies. This approach ignores that cells are
not uniformly distributed along the cell cycle. As first derived in
\cite{Powell1956} populations of cells in a log-phase are exponentially
distributed along the cell cycle. This distribution is of the form
\begin{equation}
P(a) = (\ln 2) \cdot 2^{1 - a},
\label{seq_cell_cycle_dist}
\end{equation}
where $a \in [0, 1]$ is the stage of the cell cycle, with $a = 0$ being the
start of the cycle and $a = 1$ being the cell division (See
\siref{supp_cell_age_dist} for a derivation of \eref{seq_cell_cycle_dist}).
\fref{sfig_cell_area} shows the separation of the two groups based on area
where \eref{seq_cell_cycle_dist} was used to weight the distribution along the
cell cycle.

\begin{figure}[h!]
	\centering \includegraphics
	{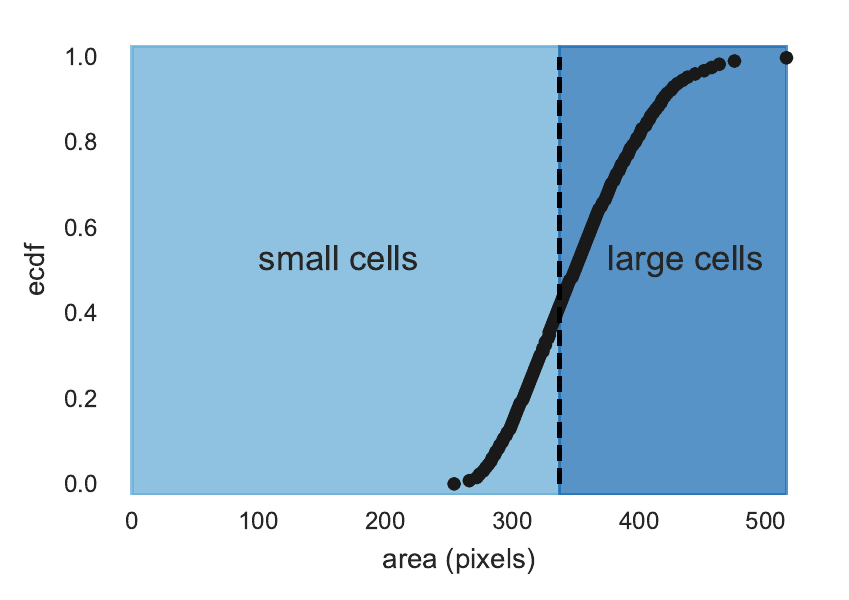}
	\caption{\textbf{Separation of cells based on cell size.} Using the area as
	a proxy for position in the cell cycle, cells can be sorted into two groups --
	small cells (with one promoter copy) and large cells (with two promoter
	copies). The vertical black line delimits the threshold that divides both
	groups as weighted by \eref{seq_cell_cycle_dist}.}
	\label{sfig_cell_area}
\end{figure}

A subtle, but important consequence of \eref{seq_cell_cycle_dist} is that
computing any quantity for a single cell is not equivalent to computing the same
quantity for a population of cells. For example, let us assume that we want to
compute the mean mRNA copy number $\ee{m}$. For a single cell this would be of
the form
\begin{equation}
	\ee{m}_{\text{cell}} = \ee{m}_1 \cdot f + \ee{m}_2 \cdot (1 - f),
\end{equation}
where $\ee{m}_i$ is the mean mRNA copy number with $i$ promoter copies in the
cell, and $f$ is the fraction of the cell cycle that cells spend with a single
copy of the promoter. For a single cell the probability of having a single
promoter copy is equivalent to this fraction $f$. But \eref{seq_cell_cycle_dist}
tells us that if we sample unsynchronized cells we are not sampling uniformly
across the cell cycle. Therefore for a population of cells the mean mRNA is
given by
\begin{equation}
	\ee{m}_{\text{population}} = \ee{m}_1 \cdot \phi + \ee{m}_2 \cdot (1 - \phi)
	\label{seq_mean_m_pop}
\end{equation}
where the probability of sampling a cell with one promoter $\phi$ is given by
\begin{equation}
	\phi = \int_0^f P(a) da,
\end{equation}
where $P(a)$ is given by \eref{seq_age_prob}. What this equation computes is
the probability of sampling a cell during a stage of the cell cycle $< f$ where
the reporter gene hasn't been replicated yet. \fref{sfig_mRNA_by_size} shows the
distribution of both groups. As expected larger cells have a higher mRNA copy
number on average.

\begin{figure}[h!]
	\centering \includegraphics
	{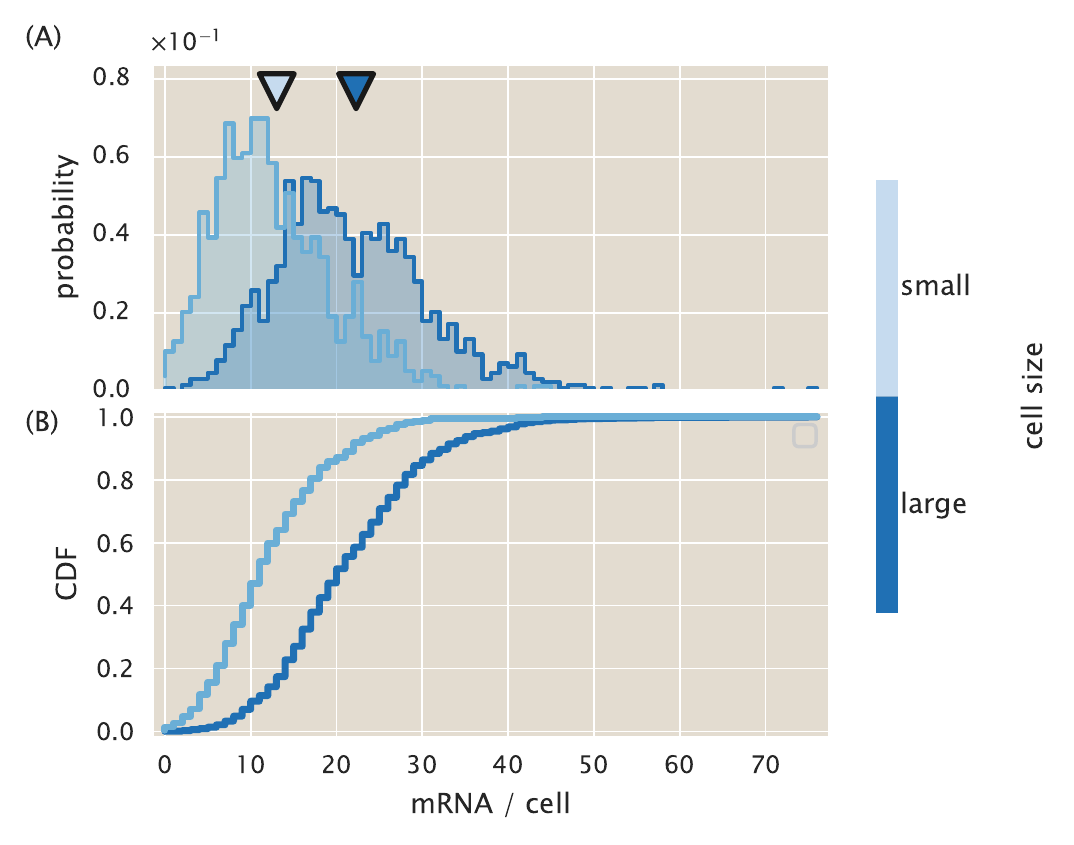}
	\caption{\textbf{mRNA distribution for small and large cells.} (A) histogram
	and (B) cumulative distribution function of the small and large cells as
	determined in \fref{sfig_cell_area}. The triangles above histograms in (A)
	indicate the mean mRNA copy number for each group.}
	\label{sfig_mRNA_by_size}
\end{figure}

We modify \eref{seq_bayes_sample} to account for the two separate groups of
cells. Let $N_s$ be the number of cells in the small size group and $N_l$ the
number of cells in the large size group. Then the posterior distribution for the
parameters is of the form
\begin{equation}
	\small
P(\kpon, \kpoff, r_m \mid \{m_i\}) \propto
	\left[\prod_{i=1}^{N_s} P(m_i \mid \kpon, \kpoff, r_m)\right]
	\left[\prod_{j=1}^{N_l} P(m_j \mid \kpon, \kpoff, 2 r_m)\right]
	P(\kpon, \kpoff, r_m),
	\label{seq_bayes_sample_double}
\end{equation}
where we split the product of small and large cells.

For the two-promoter model the prior shown in \eref{seq_prior_single} requires a
small modification. \eref{seq_mean_m_pop} gives the mean mRNA copy number of a
population of asynchronous cells growing at steady state. Given that we assume
that the only difference between having one vs. two promoter copies is the
change in transcription rate from $r_m$ in the single promoter case to $2 r_m$
in the two-promoter case we can
write \eref{seq_mean_m_pop} as
\begin{equation}
	\ee{m} = \phi \cdot {r_m \over \gm} {\kpon \over \kpon + \kpoff} +
		(1 -\phi) \cdot {2 r_m \over \gm} {\kpon \over \kpon + \kpoff}.
\end{equation}
This can be simplified to
\begin{equation}
	\ee{m} = (2 - \phi) {r_m \over \gm} {\kpon \over \kpon + \kpoff}.
	\label{seq_mean_m_double_rates}
\end{equation}

Equating \eref{seq_mean_m_double_rates} and \eref{seq_mean_thermo} to again
require self-consistent predictions of the mean mRNA from the equilibrium and
kinetic models gives
\begin{equation}
	(2 - \phi) {\kpon \over \kpon + \kpoff} =
	{{P \over \Nns} e^{-\beta\eP}
	\over 1 + {P \over \Nns} e^{-\beta\eP}}.
\end{equation}
Solving for $\kpon \over \kpoff$ results in
\begin{equation}
	\left({\kpon \over \kpoff}\right) =
	{\rho \over \left[ (1 + \rho)(2 - \phi) - \rho \right]},
	\label{seq_kinetic_thermo_equiv}
\end{equation}
where we define $\rho \equiv {P \over \Nns} e^{-\beta\eP}$. To simplify things
further we notice that for the specified values of $P = 1000 - 3000$ per cell,
$\Nns = 4.6 \times 10^6$ bp, and $-\beta\eP = 5 - 7$, we can safely assume that
$\rho \ll 1$. This simplifying assumption has been previously called the weak
promoter approximation \cite{Garcia2011c}. Given this we can simplify
\eref{seq_kinetic_thermo_equiv} as
\begin{equation}
	{\kpon \over \kpoff} = {1 \over 2 - \phi} {P \over \Nns} e^{-\beta\eP}.
\end{equation}
Taking the log of both sides gives
\begin{equation}
	\ln\left({\kpon \over \kpoff}\right) = -\ln (2 - \phi) + \ln P - \ln\Nns
	- \beta\eP.
\end{equation}
With this we can set as before a Gaussian prior to constrain the ratio of the
RNAP rates as
\begin{equation}
	P\left(\ln \left({\kpon \over \kpoff}\right) \right)  \propto
	\exp \left\{ - {\left(\ln \left({\kpon \over \kpoff}\right) -
	\left[-\ln(2 - \phi) -\beta\eP + \ln P - \ln \Nns \right) \right]^2
	\over 2 \sigma^2} \right\}.
	\label{seq_prior_double}
\end{equation}

\fref{sfig_mcmc_rnap_double} shows the result of sampling out of
\eref{seq_bayes_sample_double}. Again we see that the model is highly sloppy
with large credible regions obtained for $\kpoff$ and $r_m$. Nevertheless,
again the use of the prior information allows us to get a parameter values
consistent with the equilibrium picture.

\begin{figure}[h!]
	\centering \includegraphics
	{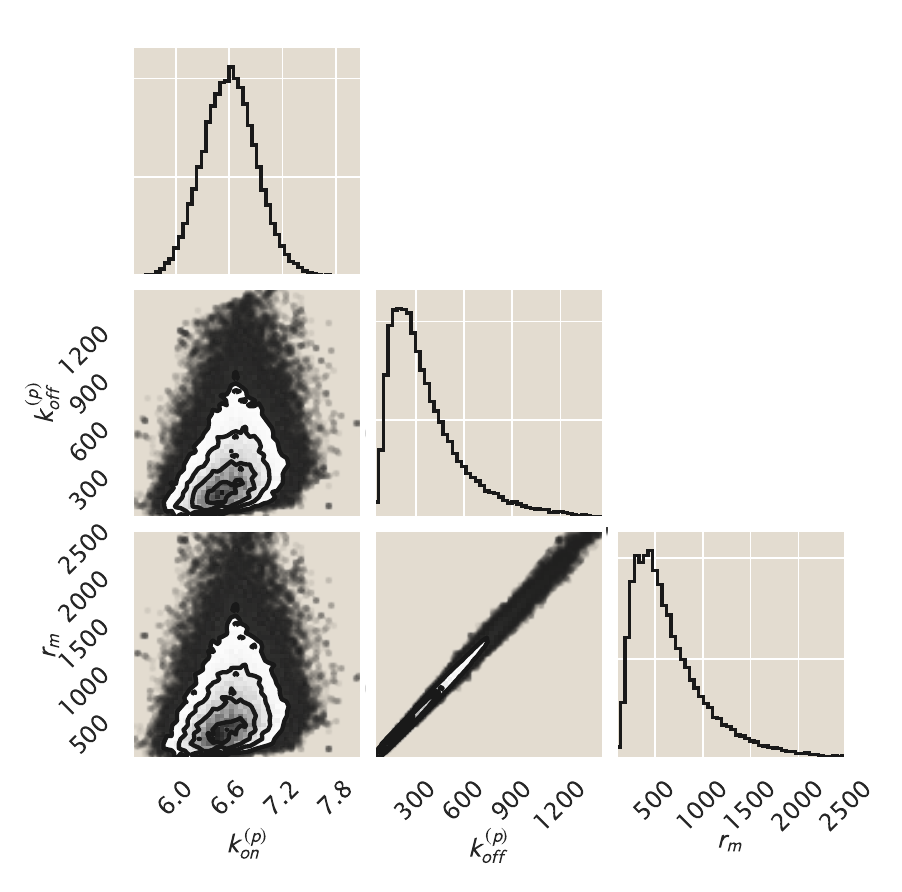}
	\caption{\textbf{MCMC posterior distribution for a multi-promoter model.}
	Sampling out of \eref{seq_bayes_sample_double} the plot shows 2D and 1D
	projections of the 3D parameter space. The parameter values are (in units of
	the mRNA degradation rate $\gm$) $\kpon = 6.4^{+0.8}_{-0.4}$, $\kpoff =
	132^{+737}_{-75}$ and $r_m = 257^{+1307}_{-132}$ which are the modes of
	their respective distributions, where the superscripts and subscripts
	represent the upper and lower bounds of the 95$\th$ percentile of the
	parameter value distributions. The sampling was bounded to values $<$ 1000 for
	numerical stability when computing the confluent hypergeometric function.}
	\label{sfig_mcmc_rnap_double}
\end{figure}

Using again a mRNA mean lifetime of $\approx 3$ min gives the following
values for the parameters: $\kpon = {0.03}_{-0.002}^{+0.004} s^{-1}$, $\kpoff =
{0.7} _{-0.4}^{+4.1} s^{-1}$, and $r_m = {1.4}_{-0.7}^{+7.3} s^{-1}$.
\fref{sfig_lacUV5_theory_data_double} shows the result of applying
\eref{seq_prob_multipromoter} using these parameter values. Specifically
\fref{sfig_lacUV5_theory_data_double}(A) shows the global distribution
including cells with one and two promoters and
\fref{sfig_lacUV5_theory_data_double}(B) splits the distributions within the
two populations. Given that the model adequately describes both populations
independently and pooled together we confirm that using the cell area as a
proxy for stage in the cell cycle and the doubling of the transcription rate
once cells have two promoters are reasonable approximations.

\begin{figure}[h!]
	\centering \includegraphics
	{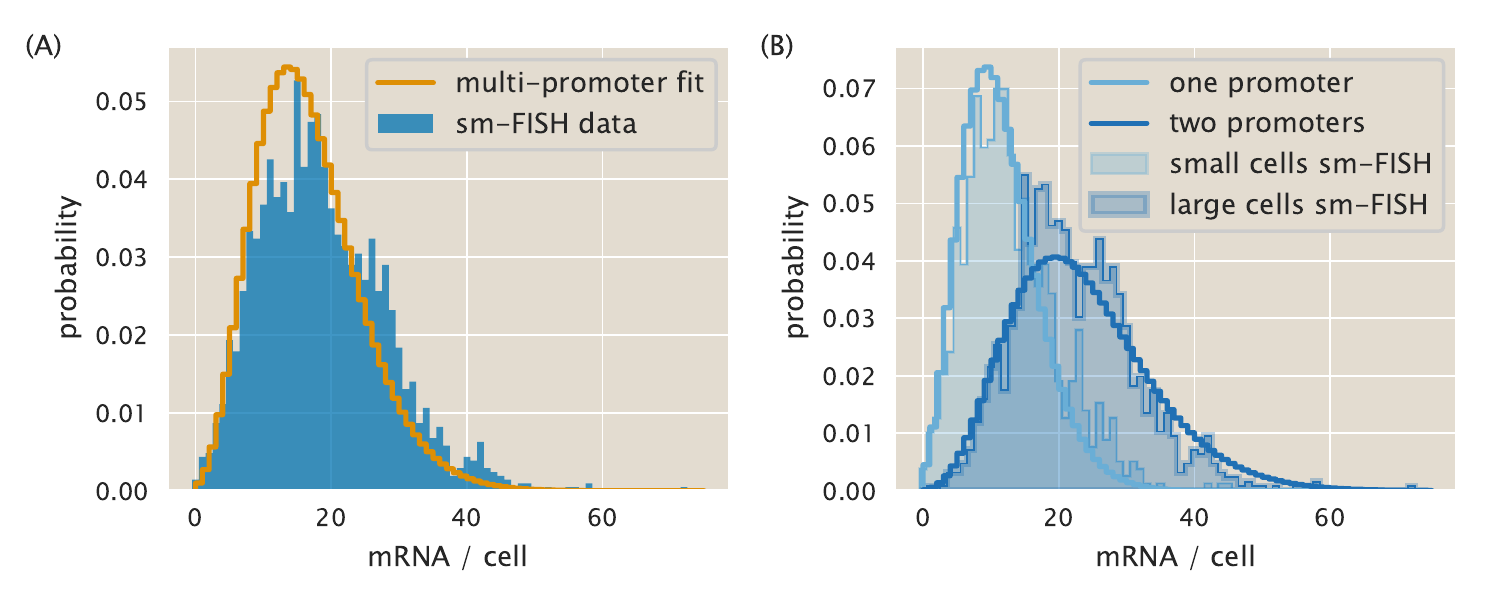}
	\caption{\textbf{Experimental vs. theoretical distribution of mRNA per cell
	using parameters for multi-promoter model.} (A) Solid line shows the result
	of using \eref{seq_prob_multipromoter} with the parameters inferred by
	sampling \eref{seq_bayes_sample_double}. Blue bars are the same data as
	\fref{sfig_lacUV5_FISH} from \cite{Jones2014a}. (B) Split distributions of
	small cells (light blue bars) and large cells (dark blue) with the
	corresponding theoretical predictions with transcription rate $r_m$ (light
	blue line) and transcription rate $2 r_m$ (dark blue line)}
	\label{sfig_lacUV5_theory_data_double}
\end{figure}

It is hard to make comparisons with literature reported values because these
kinetic rates are effective parameters hiding a lot of the complexity of
transcription initiation \cite{Browning2004}. Also the non-identifiability of
the parameters restricts our explicit comparison of the actual numerical values
of the inferred rates. Nevertheless from the model we can see that the mean
burst  size for each transcription event is given by $r_m / \kpoff$. From our
inferred values we obtain then a mean burst size of $\approx 1.9$ transcripts
per cell. This is similar to the reported burst size of 1.15 on a similar
system on \textit{E. coli} \cite{Yu2006}.

\subsection{Repressor rates from three-state regulated promoter.}

Having determined the unregulated promoter transition rates we now proceed to
determine the repressor rates $\kron$ and $\kroff$. The values of these rates
are constrained again by the correspondence between our kinetic picture and
what we know from equilibrium models \cite{Phillips2015}. For this analysis we
again exploit the feature that, at the mean, both the kinetic language and the
thermodynamic language should have equivalent predictions. Over the last decade
there has been great effort in developing equilibrium models for gene
expression regulation \cite{Buchler2003, Vilar2011, Bintu2005a}. In particular
our group has extensively characterized the simple repression motif using this
formalism \cite{Garcia2011c, Brewster2014, Razo-Mejia2018}.

The dialogue between theory and experiments has led to simplified expressions
that capture the phenomenology of the gene expression response as a function of
natural variables such as molecule count and affinities between molecular
players. A particularly interesting quantity for the simple repression motif
used by Garcia \& Phillips \cite{Garcia2011c} is the fold-change in gene
expression, defined as
\begin{equation}
	\foldchange = {\ee{\text{gene expression}(R \neq 0)} \over
					\ee{\text{gene expression}(R = 0)}},
\end{equation}
where $R$ is the number of repressors per cell and $\ee{\cdot}$ is the
population average. The fold-change is simply the mean expression level in the
presence of the repressor relative to the mean expression level in the absence
of regulation. In the language of statistical mechanics this quantity takes the
form
\begin{equation}
	\foldchange = \left( 1 + {R \over \Nns} e^{-\beta\eR} \right)^{-1},
	\label{seq_fc_thermo}
\end{equation}
where $\eR$ is the repressor-DNA binding energy, and as before $\Nns$ is the
number of non-specific binding sites where the repressor can bind
\cite{Garcia2011c}.

To compute the fold-change in the chemical master equation language we compute
the first moment of the steady sate mRNA distribution $\ee{m}$ for both the
three-state promoter ($R \neq 0$) and the two-state promoter case ($R=0$) (See
\siref{supp_moments} for moment derivation). The unregulated (two-state)
promoter mean mRNA copy number is given by \eref{seq_mean_m_double_rates}. For
the regulated (three-state) promoter we have an equivalent expression of the
form
\begin{equation}
	\ee{m (R \neq 0)} = (2 - \phi){r_m \over \gm} {\kroff\kpon
	\over \kpoff\kroff + \kpoff\kron + \kroff\kpon}.
\end{equation}
Computing the fold-change then gives
\begin{equation}
	\foldchange = {\ee{m (R \neq 0)} \over \ee{m (R = 0)}} =
	{\kroff \left( \kpoff + \kpon \right) \over
	\kpoff\kron + \kroff \left( \kpoff + \kpon \right)},
	\label{seq_fold_change_cme}
\end{equation}
where the factor $(2 - \phi)$ due to the multiple promoter copies, the
transcription rate $r_m$ and the mRNA degradation rate $\gm$ cancel out.

Given that the number of repressors per cell $R$ is an experimental variable
that we can control, we assume that the rate at which the promoter transitions
from the transcriptionally inactive state to the repressor bound state,
$\kron$, is given by the concentration of repressors $[R]$ times a diffusion
limited on rate $k_o$. For the diffusion limited constant $k_o$ we use the
value used by Jones et al. \cite{Jones2014a}. With this in hand we can rewrite
\eref{seq_fold_change_cme} as
\begin{equation}
	\foldchange = \left( 1 + {k_0 [R] \over \kroff}
				{\kpoff \over \kpon + \kpoff} \right)^{-1}.
	\label{seq_fc_kinetic}
\end{equation}

We note that both \eref{seq_fc_thermo} and \eref{seq_fc_kinetic} have the same
functional form. Therefore if both languages predict the same output for the
mean gene expression level, it must be true that
\begin{equation}
	{k_o [R] \over \kroff}{\kpoff \over \kpon + \kpoff} =
	{R \over \Nns} e^{-\beta\eR}.
\end{equation}
Solving for $\kroff$ gives
\begin{equation}
	\kroff = {k_o [R] \Nns e^{\beta\eR} \over R}{\kpoff \over \kpon + \kpoff}.
	\label{seq_kroff_complete}
\end{equation}

Since the reported value of $k_o$ is given in units of nM$^{-1}$s$^{-1}$ in
order for the units to cancel properly the repressor concentration has to be
given in nM rather than absolute count. If we consider that the repressor
concentration is equal to
\begin{equation}
[R] = \frac{R}{V_{cell}}\cdot \frac{1}{N_A},
\end{equation}
where $R$ is the absolute repressor copy number per cell, $V_{cell}$ is the cell
volume and $N_A$ is Avogadro's number. The \textit{E. coli} cell volume is 2.1
fL \cite{Radzikowski2016}, and Avogadro's number is $6.022 \times 10^{23}$. If
we further include the conversion factor to turn M into nM we find that
\begin{equation}
[R] = {R \over 2.1 \times 10^{-15} L} \cdot {1 \over 6.022 \times 10^{23}}
\cdot {10^9 \text{ nmol} \over 1 \text{ mol}} \approx 0.8 \times R.
\end{equation}
Using this we simplify \eref{seq_kroff_complete} as
\begin{equation}
	\kroff \approx 0.8 \cdot k_o \cdot \Nns e^{\beta\eR}
	\cdot {\kpoff \over \kpon + \kpoff}.
	\label{seq_kroff}
\end{equation}
What \eref{seq_kroff} shows is the direct relationship that must be satisfied if
the equilibrium model is set to be consistent with the non-equilibrium kinetic
picture. \tref{stab_koff} summarizes the values obtained for the three operator
sequences used throughout this work. To compute these numbers the number of
non-specific binding sites $\Nns$ was taken to be $4.6 \times 10^6$ bp, i.e. the
size of the {\it E. coli} K12 genome.

\begin{table}[]
	\caption{\textbf{Binding sites and corresponding parameters.}}
\begin{tabular}{|c|c|c|}
\hline
	Operator & $\eR\; (k_BT)$ & $\kroff \; (s^{-1})$  \\ \hline
	O1 & -15.3 & 0.002  \\ \hline
	O2 & -13.9 & 0.008  \\ \hline
	O3 & -9.7  & 0.55   \\ \hline
\end{tabular}
\label{stab_koff}
\end{table}

{\it In-vivo} measurements of the Lac repressor off rate have been done with
single-molecule resolution \cite{Hammar2014}. The authors report a mean
residence time of $5.3 \pm 0.2$ minutes for the repressor on an O1 operator. The
corresponding rate is $\kroff \approx 0.003$ $(s^{-1})$, very similar value to
what we inferred from our model. In this same reference the authors determined
that on average the repressor takes $30.9 \pm 0.5$ seconds to bind to the
operator \cite{Hammar2014}. Given the kinetic model presented in
\fref{fig2_minimal_model}(A) this time can be converted to the probability of
not being on the repressor bound state $P_{\text{not }R}$. This is computed as
\begin{equation}
	P_{\text{not }R} = {\tau_{\text{not }R} \over
						\tau_{\text{not }R} + \tau_{R}},
\end{equation}
where $\tau_{\text{not }R}$ is the average time that the operator is not
occupied by the repressor and $\tau_{R}$ is the average time that the repressor
spends bound to the operator. Substituting the numbers from \cite{Hammar2014}
gives $P_{\text{not }R} \approx 0.088$. From our model we can compute the zeroth
moment $\ee{m^0 p^0}$ for each of the three promoter states. This moment is
equivalent to the probability of being on each of the promoter states. Upon
substitution of our inferred rate parameters we can compute $P_{\text{not }R}$
as
\begin{equation}
	P_{\text{not }R} = 1 - P_R \approx 0.046,
\end{equation}
where $P_R$ is the probability of the promoter being bound by the repressor. The
value we obtained is within a factor of two from the one reported in
\cite{Hammar2014}.

\section{Computing moments from the master equation}\label{supp_moments}

In this section we will compute the moment equations for the distribution  $P(m,
p)$. Without lost of generality here we will focus on the three-state
regulated promoter. The computation of the moments for the two-state promoter
follows the exact same procedure, changing only the definition of the matrices 
in the master equation.

\subsection{Computing moments of a distribution}

(Note: The Python code used for the calculations presented in this section can
be found in the
\href{https://www.rpgroup.caltech.edu//chann_cap/software/moment_dynamics_system.html}{following
link} as an annotated Jupyter notebook)

To compute any moment of our chemical master equation (\eref{eq_cme_matrix})
let us define a vector
\begin{equation}
  \ee{\bb{m^x p^y}} \equiv (\ee{m^x p^y}_A, \ee{m^x p^y}_I, \ee{m^x p^y}_R)^T,
\end{equation}
where $\ee{m^x p^y}_S$ is the expected value of $m^x p^y$ in state $S \in \{A,
I, R\}$ with $x, y \in \mathbb{N}$. In other words, just as we defined the
vector $\PP(m, p)$, here we define a vector to collect the expected value of
each of the promoter states. By definition, these moments $\ee{m^x p^y}_S$ are
computed as
\begin{equation}
  \ee{m^x p^y}_S \equiv \sum_{m=0}^\infty \sum_{p=0}^\infty m^x p^y P_S(m, p).
  \label{seq_mom_def}
\end{equation}
To simplify the notation, let $\sum_x \equiv \sum_{x=0}^\infty$. Since we are
working with a system of three ODEs, one for each state, let us define the
following operation:
\begin{equation}
  \ee{\bb{m^x p^y}} =
  \smp m^x p^y \PP(m, p) \equiv
  \begin{bmatrix}
    \smp m^x p^y P_A(m, p)\\
    \smp m^x p^y P_I(m, p)\\
    \smp m^x p^y P_R(m, p)\\
  \end{bmatrix}.
\end{equation}

With this in hand we can then apply this sum over $m$ and $p$ to
\eref{eq_cme_matrix}. For the left-hand side we have
\begin{equation}
  \smp m^x p^y \dt{\PP(m, p)} = \dt{}\left[ \smp m^x p^y \PP(m, p) \right],
  \label{seq_sum_mom}
\end{equation}
where we made use of the linearity property of the derivative to switch the
order between the sum and the derivative. Notice that the right-hand side of
\eref{seq_sum_mom} contains the definition of a moment from \eref{seq_mom_def}.
That means that we can rewrite it as
\begin{equation}
  \dt{}\left[ \smp m^x p^y \PP(m, p) \right] = \dt{\bb{\ee{m^x p^y}}}.
\end{equation}

Distributing the sum on the right-hand side of \eref{eq_cme_matrix} gives
\begin{equation}
  \begin{aligned}
    \dt{\bb{\ee{m^x p^y}}} &=
    \Km \smp m^x p^y \PP(m, p)\\
    &- \Rm \smp m^x p^y \PP(m, p) + \Rm \smp m^x p^y \PP(m-1, p)\\
    &- \Gm \smp (m) m^x p^y \PP(m, p) + \Gm \smp (m + 1) m^x p^y \PP(m + 1, p)\\
    &- \Rp \smp (m) m^x p^y \PP(m, p) + \Rp \smp (m) m^x p^y \PP(m, p - 1)\\
    &- \Gp \smp (p) m^x p^y \PP(m, p) + \Gp \smp (p + 1) m^x p^y \PP(m, p + 1).
  \end{aligned}
  \label{seq_master_sum}
\end{equation}

Let's look at each term on the right-hand side individually. For the terms in
\eref{seq_master_sum} involving $\PP(m, p)$ we can again use \eref{seq_mom_def}
to rewrite them in a more compact form. This means that we can rewrite the
state transition term as
\begin{equation}
  \Km \smp m^x p^y \PP(m, p) = \Km \bb{\ee{m^x p^y}}.
\end{equation}
The mRNA production term involving $\PP(m, p)$ can be rewritten as
\begin{equation}
  \Rm \smp m^x p^y \PP(m, p) = \Rm \bb{\ee{m^x p^y}}.
\end{equation}
In the same way the mRNA degradation term gives
\begin{equation}
  \Gm \smp (m) m^x p^y \PP(m, p) = \Gm \bb{\ee{m^{(x + 1)} p^y}}.
\end{equation}
For the protein production and degradation terms involving $\PP(m, p)$ we have
\begin{equation}
  \Rp \smp (m) m^x p^y \PP(m, p) = \Rp \bb{\ee{m^{(x + 1)} p^y}},
\end{equation}
and
\begin{equation}
  \Gp \smp (p) m^x p^y \PP(m, p) = \Gp \bb{\ee{m^x p^{(y + 1)}}},
\end{equation}
respectively.

For the sums terms in \eref{seq_master_sum} involving $\PP(m \pm 1, p)$ or
$\PP(m, p \pm 1)$ we can reindex the sum to work around this mismatch. To be
more specific let's again look at each term case by case. For the mRNA
production term involving $\PP(m-1, p)$ we define $m' \equiv m - 1$. Using this
we write
\begin{equation}
  \Rm \smp m^x p^y \PP(m-1, p) =
  \Rm \sum_{m' = -1}^\infty \sum_p (m' + 1)^x p^y \PP(m', p).
\end{equation}
Since having negative numbers of mRNA or protein doesn't make physical sense
we have that $\PP(-1, p) = 0$. Therefore we can rewrite the sum starting from 0
rather than from -1, obtaining
\begin{equation}
  \Rm \sum_{m' = -1}^\infty \sum_p (m' + 1)^x p^y \PP(m', p) =
  \Rm \sum_{m'=0}^\infty \sum_p (m' + 1)^x p^y \PP(m', p).
  \label{seq_reindex}
\end{equation}
Recall that our distribution $\PP(m, p)$ takes $m$ and $p$ as numerical inputs 
and returns a probability associated with such a molecule count.  Nevertheless,
$m$ and $p$ themselves are dimensionless quantities that serve as indices of how
many molecules are in the cell. The distribution is the same whether the
variable is called $m$ or $m'$; for a specific number, let's say $m = 5$, or $m'
= 5$, $\PP(5, p)$ will return the same result. This means that the variable name
is arbitrary, and the right-hand side of
\eref{seq_reindex} can be written as
\begin{equation}
  \Rm \sum_{m'=0}^\infty \sum_p (m' + 1)^x p^y \PP(m', p) =
  \Rm \bb{\ee{(m+1)^x p^y}},
\end{equation}
since the left-hand side corresponds to the definition of a moment.

For the mRNA degradation term involving $\PP(m + 1, p)$ we follow a similar
procedure in which we define $m' = m + 1$ to obtain
\begin{equation}
  \Gm \smp (m + 1) m^x p^y \PP(m + 1, p) =
  \Gm \sum_{m' = 1}^\infty \sum_p m' (m' - 1)^x p^y \PP(m', p).
\end{equation}
In this case since the term on the right-hand side of the equation is multiplied
by $m'$, starting the sum over $m'$ from 0 rather than from 1 will not affect
the result since this factor will not contribute to the total sum. Nevertheless
this is useful since our definition of a moment from \eref{seq_mom_def} requires
the sum to start at zero. This means that we can rewrite this term as
\begin{equation}
  \Gm \sum_{m' = 1}^\infty m' \sum_p (m' - 1)^x p^y \PP(m', p) =
  \Gm \sum_{m' = 0}^\infty m' \sum_p (m' - 1)^x p^y \PP(m', p).
\end{equation}
Here again we can change the arbitrary label $m'$ back to $m$ obtaining
\begin{equation}
  \Gm \sum_{m' = 0}^\infty m' \sum_p (m' - 1)^x p^y \PP(m', p) =
  \Gm \bb{\ee{m (m - 1)^x p^y}}.
\end{equation}

The protein production term involving $\PP(m, p - 1)$ can be reindexed by 
defining $p' \equiv p - 1$. This gives
\begin{equation}
  \Rp \smp (m) m^x p^y \PP(m, p - 1) =
  \Rp \sum_m \sum_{p'=-1}^\infty m^{(x + 1)} (p + 1)^y \PP(m, p').
\end{equation}
We again use the fact that negative molecule copy numbers are assigned with
probability zero to begin the sum from 0 rather than -1 and the arbitrary nature
of the label $p'$ to write
\begin{equation}
  \Rp \sum_m \sum_{p'=0}^\infty m^{(x + 1)} (p + 1)^y \PP(m, p') =
  \Rp \bb{\ee{m^{(x + 1)} (p + 1)^y}}.
\end{equation}
Finally, we take care of the protein degradation term involving $\PP(m, p + 1)$.
As before, we define $p' = p + 1$ and substitute this to obtain
\begin{equation}
  \Gp \smp (p + 1) m^x p^y \PP(m, p + 1) =
  \Gp \sum_m \sum_{p'=1}^\infty (p') m^x (p' - 1)^y \PP(m, p').
\end{equation}
Just as with the mRNA degradation term, having a term $p'$  inside the sum
allows us to start the sum over $p'$ from 0 rather than 1. We can therefore
write
\begin{equation}
  \Gp \sum_m \sum_{p'=0}^\infty (p') m^x (p' - 1)^y \PP(m, p') =
  \Gp \bb{\ee{m^x p (p - 1)^y}}.
\end{equation}

Putting all these terms together we can write the general moment ODE. This is
of the form
\begin{equation}
  \begin{aligned}
    \dt{\bb{\ee{m^x p^y}}} &=
    \Km \bb{\ee{m^x p^y}}
    \text{  (promoter state transition)}\\
    &- \Rm \bb{\ee{m^x p^y}} + \Rm \bb{\ee{(m+1)^x p^y}}
    \text{  (mRNA production)}\\
    &- \Gm \bb{\ee{m^{(x + 1)} p^y}} + \Gm \bb{\ee{m (m - 1)^x p^y}}
    \text{  (mRNA degradation)}\\
    &- \Rp \bb{\ee{m^{(x + 1)} p^y}} + \Rp \bb{\ee{m^{(x + 1)} (p + 1)^y}}
    \text{  (protein production)}\\
    &- \Gp \bb{\ee{m^x p^{(y + 1)}}} + \Gp \bb{\ee{m^x p (p - 1)^y}}
    \text{  (protein degradation)}.
  \end{aligned}
  \label{seq_mom_ode}
\end{equation}

\subsection{Moment closure of the simple-repression distribution}

A very interesting and useful feature of \eref{seq_mom_ode} is that for a given
value of $x$ and $y$ the moment $\bb{\ee{m^x p^y}}$ is only a function of lower
moments. Specifically $\bb{\ee{m^x p^y}}$ is a function of moments
$\bb{\ee{m^{x'} p^{y'}}}$ that satisfy two conditions:
\begin{equation}
  \begin{aligned}
    &1) y' \leq y,\\
  &2) x' + y' \leq x + y.
  \end{aligned}
  \label{seq_mom_conditions}
\end{equation}

To prove this we rewrite \eref{seq_mom_ode} as
\begin{equation}
  \begin{aligned}
    \dt{\bb{\ee{m^x p^y}}} &=
    \Km \bb{\ee{m^x p^y}}\\
    &+ \Rm \bb{\ee{p^y \left[ (m + 1)^x -m^x \right]}}\\
    &+ \Gm \bb{\ee{m p^y \left[ (m - 1)^x - m^x \right]}}\\
    &+ \Rp \bb{\ee{m^{(x + 1)} \left[ (p + 1)^y - p^y \right]}}\\
    &+ \Gp \bb{\ee{m^x p \left[ (p - 1)^y - p^y \right]}},
    \label{seq_mom_ode_factorized}
  \end{aligned}
\end{equation}
where the factorization is valid given the linearity of expected values. Now the
objective is to find the highest moment for each term once the relevant
binomial, such as $(m-1)^x$, is expanded. Take, for example, a simple case in
which we want to find the second moment of the mRNA distribution. We then set $x
= 2$ and $y = 0$. \eref{seq_mom_ode_factorized} then becomes
\begin{equation}
  \begin{aligned}
    \dt{\bb{\ee{m^2 p^0}}} &=
    \Km \bb{\ee{m^2 p^0}}\\
    &+ \Rm \bb{\ee{p^0 \left[ (m + 1)^2 - m^2 \right]}}\\
    &+ \Gm \bb{\ee{m p^0 \left[ (m - 1)^2 - m^2 \right]}}\\
    &+ \Rp \bb{\ee{m^{(2 + 1)} \left[ (p + 1)^0 - p^0 \right]}}\\
    &+ \Gp \bb{\ee{m^2 p \left[ (p - 1)^0 - p^0 \right]}}.
  \end{aligned}
\end{equation}
Simplifying this equation gives
\begin{equation}
    \dt{\bb{\ee{m^2}}} =
    \Km \bb{\ee{m^2}}
    + \Rm \bb{\ee{\left[ 2m + 1 \right]}}
    + \Gm \bb{\ee{\left[- 2m^2 + m \right]}}.
    \label{seq_second_mom_mRNA}
\end{equation}

\eref{seq_second_mom_mRNA} satisfies both of our conditions. Since we set $y$ to
be zero, none of the terms depend on any moment that involves the protein number,
therefore $y' \leq y$ is satisfied. Also the highest moment in
\eref{seq_second_mom_mRNA} also satisfies $x' + y' \leq x + y$ since the second
moment of mRNA doesn't depend on any moment higher than $\bb{\ee{m^2}}$. To
demonstrate that this is true for any $x$ and $y$ we now rewrite
\eref{seq_mom_ode_factorized}, making use of the binomial expansion
\begin{equation}
  (z \pm 1)^n = \sum_{k=0}^n {n \choose k} (\pm 1)^{k} z^{n-k}.
\end{equation}
Just as before let's look at each term individually. For the mRNA production
term we have
\begin{equation}
  \Rm \bb{\ee{p^y \left[ (m + 1)^x -m^x \right]}} =
  \Rm \bb{\ee{p^y \left[ \sum_{k=0}^x {x \choose k} m^{x-k} - m^x \right]}}.
\end{equation}
When $k = 0$, the term inside the sum on the right-hand side cancels with the
other $m^x$, so we can simplify to
\begin{equation}
  \Rm \bb{\ee{p^y \left[ (m + 1)^x -m^x \right]}} =
  \Rm \bb{\ee{p^y \left[ \sum_{k=1}^x {x \choose k} m^{x-k} \right]}}.
\end{equation}
Once the sum is expanded we can see that the highest moment in this sum is given
by $\bb{\ee{m^{(x-1)} p^y}}$ which satisfies both of the conditions on
\eref{seq_mom_conditions}.

For the mRNA degradation term we similarly have
\begin{equation}
  \Gm \bb{\ee{m p^y \left[ (m - 1)^x - m^x \right]}} =
  \Gm \bb{\ee{m p^y \left[ \sum_{k=0}^x {x \choose k}(-1)^k m^{x-k} -
                          m^x \right]}}.
\end{equation}
Simplifying terms we obtain
\begin{equation}
  \Gm \bb{\ee{m p^y \left[ \sum_{k=0}^x {x \choose k}(-1)^k m^{x-k} -
                          m^x \right]}} =
  \Gm \bb{\ee{p^y \left[ \sum_{k=1}^x {x \choose k}(-1)^k m^{x+1-k} \right]}}.
\end{equation}
The largest moment in this case is $\bb{\ee{m^x p^y }}$, which again satisfies
the conditions on \eref{seq_mom_conditions}.

The protein production term gives
\begin{equation}
  \Rp \bb{\ee{m^{(x + 1)} \left[ (p + 1)^y - p^y \right]}} =
  \Rp \bb{\ee{m^{(x + 1)} \left[ \sum_{k=0}^y {y \choose k} (-1)^k p^{y-k}
                                - p^y \right]}}.
\end{equation}
Upon simplification we obtain
\begin{equation}
  \Rp \bb{\ee{m^{(x + 1)} \left[ \sum_{k=0}^y {y \choose k} (-1)^k p^{y-k}
                                - p^y \right]}} =
  \Rp \bb{\ee{m^{(x + 1)} \left[ \sum_{k=1}^y {y \choose k}(-1)^k p^{y-k}
  \right]}}.
\end{equation}
Here the largest moment is given by $\bb{\ee{m^{x+1} p^{y-1}}}$, that again
satisfies both of our conditions. For the last term, for protein degradation we
have
\begin{equation}
  Rp \bb{\ee{m^{(x + 1)} \left[ (p + 1)^y - p^y \right]}} =
  Rp \bb{\ee{m^{(x + 1)} \left[ \sum_{k=1}^y {y \choose k} (-1^k) p^{y - k}
  \right]}}.
\end{equation}
The largest moment involved in this term is therefore $\bb{\ee{m^x p^{y-1}}}$.
With this, we show that the four terms involved in our general moment equation
depend only on lower moments that satisfy \eref{seq_mom_conditions}.

As a reminder, what we showed in this section is that the kinetic model
introduced in \fref{fig2_minimal_model}(A) has no moment-closure problem. In
other words, moments of the joint mRNA and protein distribution can be computed
just from knowledge of lower moments. This allows us to cleanly integrate the
moments of the distribution dynamics as cells progress through the cell cycle.

\subsection{Computing single promoter steady-state moments}

(Note: The Python code used for the calculations presented in this section can
be found in the
\href{https://www.rpgroup.caltech.edu//chann_cap/software/chemical_master_steady_state_moments_general.html}{following
link} as an annotated Jupyter notebook)

As discussed in \secref{sec_cell_cycle}, one of the main factors contributing to
cell-to-cell variability in gene expression is the change in gene copy number
during the cell cycle as cells replicate their genome before cell division. Our
minimal model accounts for this variability by considering the time trajectory
of the distribution moments as given by \eref{seq_mom_ode_factorized}. These
predictions will be contrasted with the predictions from a kinetic model that
doesn't account for changes in gene copy number during the cell cycle in
\siref{supp_multi_gene}.

If we do not account for change in gene copy number during the cell cycle or for
the partition of proteins during division, the dynamics of the moments of the
distribution described in this section will reach a steady state. In order to
compute the steady-state moments of the kinetic model with a single gene across
the cell cycle, we use the moment closure property of our master equation. By
equating \eref{seq_mom_ode_factorized} to zero for a given $\bb{x}$ and
$\bb{y}$, we can solve the resulting linear system and obtain a solution for
$\bb{\ee{m^x p^y}}$ at steady state as a function of moments $\bb{\ee{m^{x'}
p^{y'}}}$ that satisfy \eref{seq_mom_conditions}. Then, by solving for the
zero$\th$ moment $\bb{\ee{m^0 p^0}}$ subject to the constraint that the
probability of the promoter being in any state should add up to one, we can
substitute back all of the solutions in terms of moments $\bb{\ee{m^{x'}
p^{y'}}}$ with solutions in terms of the rates shown in
\fref{fig2_minimal_model}. In other words, through an iterative process, we can
get at the value of any moment of the distribution. We start by solving for the
zero$\th$ moment. Since all higher moments, depend on lower moments we can use
the solution of the zero$\th$ moment to compute the first mRNA moment. This
solution is then used for higher moments in a hierarchical iterative process.

\section{Accounting for the variability in gene copy number during the cell
cycle}\label{supp_multi_gene}

(Note: The Python code used for the calculations presented in this section can
be found in the
\href{https://www.rpgroup.caltech.edu/chann_cap/src/theory/html/moment_dynamics_cell_division.html}{following
link} as an annotated Jupyter notebook)

When growing in rich media, bacteria can double every $\approx$ 20 minutes.
With two replication forks each traveling at $\approx$ 1000 bp per second, and
a genome of $\approx$ 5 Mbp for {\it E. coli} \cite{Moran2010}, a cell would
need $\approx$ 40 minutes to replicate its genome. The apparent paradox  of
growth rates faster than one division per 40 minutes is solved by the fact that
cells have multiple replisomes, i.e. molecular machines that replicate the
genome running in parallel. Cells can have up to 8 copies of the genome being
replicated simultaneously depending on the growth rate \cite{Bremer1996}.

This observation implies that during the cell cycle gene copy number varies.
This variation depends on the growth rate and the relative position of the gene
with respect to the replication origin, having genes close to the replication
origin spending more time with multiple copies compare to genes closer to the
replication termination site. This change in gene dosage has a direct effect on
the cell-to-cell variability in gene expression \cite{Jones2014a,
Peterson2015}.

\subsection{Numerical integration of moment equations}

(Note: The Python code used for the calculations presented in this section can
be found in the
\href{https://www.rpgroup.caltech.edu//chann_cap/software/moment_dynamics_cell_division.html}{following
link} as an annotated Jupyter notebook)

For our specific locus ({\it galK}) and a doubling time of $\approx$ 60 min for
our experimental conditions, cells have on average 1.66 copies of the reporter
gene during the cell cycle \cite{Jones2014a}. What this means is that cells
spend 60\% of the time having one copy of the gene and 40\% of the time with
two copies. To account for this variability in gene copy number across the cell
cycle we numerically integrate the moment equations derived in
\siref{supp_moments} for a time $t = [0, t_s]$ with an mRNA production rate
$r_m$, where $t_s$ is the time point at which the replication fork reaches our
specific locus. For the remaining time before the cell division $t = [t_s,
t_d]$ that the cell spends with two promoters, we assume that the only
parameter that changes is the mRNA production rate from $r_m$ to $2 r_m$. This
simplifying assumption ignores potential changes in protein translation rate
$r_p$ or changes in the repressor copy number that would be reflected in
changes on the repressor on rate $\kron$.

\subsubsection{Computing distribution moments after cell division}

(Note: The Python code used for the calculations presented in this section can
be found in the
\href{https://www.rpgroup.caltech.edu//chann_cap/software/binomial_moments.html}{following
link} as an annotated Jupyter notebook)

We have already solved a general form for the dynamics of the moments of the
distribution, i.e. we wrote differential equations for the moments ${d\ee{m^x
p^y}\over dt}$. Given that we know all parameters for our model we can simply
integrate these equations numerically to compute how the moments of the
distribution evolve as cells progress through their cell cycle. Once the cell
reaches a time $t_d$ when is going to divide the mRNA and proteins that we are
interested in undergo a binomial partitioning between the two daughter cells.
In other words, each molecule flips a coin and decides whether to go to either
daughter. The question then becomes given that we have a value for the moment
$\ee{m^x p^y}_{t_d}$ at a time before the cell division, what would the value
of this moment be after the cell division takes place $\ee{m^x p^y}_{t_o}$?

The probability distribution of mRNA and protein after the cell division
$P_{t_o}(m, p)$ must satisfy
\begin{equation}
  P_{t_o}(m, p) = \sum_{m'=m}^\infty \sum_{p'=p}^\infty 
                  P(m, p \mid m', p') P_{t_d}(m', p'),
\label{eq_dist_post_div}
\end{equation}
where we are summing over all the possibilities of having $m'$ mRNA and $p'$
proteins before cell division. Note that the sums start at $m$ and $p$; this is
because for a cell to have these copy numbers before cell division it is a
requirement that the mother cell had at least such copy number since we are not
assuming that there is any production at the instantaneous cell division time.
Since we assume that the partition of mRNA is independent from the partition of
protein, the conditional probability $P(m, p \mid m', p')$ is simply given by a
product of two binomial distributions, one for the mRNA and one for the
protein, i.e.
\begin{equation}
P(m, p \mid m', p') = {m' \choose m} \left( {1 \over 2} \right)^{m'} \cdot
                      {p' \choose p} \left( {1 \over 2} \right)^{p'}.
\label{eq_binom_prod}
\end{equation}
Because of these product of binomial probabilities are allowed to extend the
sum from
\eref{eq_dist_post_div} to start at $m'=0$ and $p'=0$ as
\begin{equation}
  P_{t_o}(m, p) = \sum_{m'=0}^\infty \sum_{p'=0}^\infty 
                  P(m, p \mid m', p') P_{t_d}(m', p'),
\end{equation}
since the product of the binomial distributions in \eref{eq_binom_prod} is zero
for all $m' < m$ and/or $p' < 0$. So from now on in this section we will assume
that a sum of the form $\sum_x \equiv \sum_{x=0}^\infty$ to simplify notation.

We can then compute the distribution moments after the cell division $\ee{m^x
p^y}_{t_o}$ as
\begin{equation}
\ee{m^x p^y}_{t_o} = \sum_m \sum_p m^x p^y P_{t_o}(m, p),
\end{equation}
for all $x, y \in \mathbb{N}$. Substituting \eref{eq_dist_post_div} results in
\begin{equation}
\ee{m^x p^y}_{t_o} = \sum_m \sum_p m^x p^y
\sum_{m'} \sum_{p'} P(m, p \mid m', p') P_{t_d}(m', p').
\end{equation}
We can rearrange the sums to be 
\begin{equation}
\ee{m^x p^y}_{t_o} = \sum_{m'} \sum_{p'} P_{t_d}(m', p')
                     \sum_m \sum_p m^x p^y P(m, p \mid m', p').
\end{equation}
The fact that \eref{eq_binom_prod} is the product of two independent events
allows us to rewrite the joint probability $P(m, p \mid m', p')$ as
\begin{equation}
P(m, p \mid m', p') = P(m \mid m') \cdot P(p \mid p').
\end{equation}
With this we can then write the moment $\ee{m^x p^y}_{t_o}$ as
\begin{equation}
\ee{m^x p^y}_{t_o} = \sum_{m'} \sum_{p'} P_{t_d}(m', p')
                     \sum_m  m^x  P(m \mid m')
                     \sum_p p^y P(p \mid p').
\end{equation}
Notice that both terms summing over $m$ and over $p$ are the conditional
expected values, i.e.
\begin{equation}
\sum_z  z^x  P(z \mid z') \equiv \ee{z^x \mid z'}, \; 
{\text{ for } z\in \{m, p \}}.
\end{equation}
These conditional expected values are the expected values of a binomial random
variable $z \sim \text{Bin}(z', 1/2)$, which can be easily computed as we will
show later in this section. We then rewrite the expected values after the cell
division in terms of these moments of a binomial distribution
\begin{equation}
\ee{m^x p^y}_{t_o} = \sum_{m'} \sum_{p'} \ee{m^x \mid m'} \ee{p^y \mid p'} 
                     P_{t_d}(m', p').
  \label{eq_general_binom_mom}
\end{equation}

To see how this general formula for the moments after the cell division works
let's compute the mean protein per cell after the cell division $\ee{p}_{t_o}$.
That is setting $x = 0$, and $y = 1$. This results in
\begin{equation}
\ee{p}_{t_o} = \sum_{m'} \sum_{p'} \ee{m^0 \mid m'} \ee{p \mid p'} 
               P_{t_d}(m', p').
\end{equation}
The zeroth moment $\ee{m^0 \mid m'}$ by definition must be one since we have
\begin{equation}
\ee{m^0 \mid m'} = \sum_m m^0 P(m \mid m') = \sum_m P(m \mid m') = 1,
\end{equation}
since the probability distribution must be normalized. This leaves us then with
\begin{equation}
\ee{p}_{t_o} = \sum_{m'} \sum_{p'} P_{t_d}(m', p') \ee{p \mid p'}.
\end{equation}
If we take the sum over $m'$ we simply compute the marginal probability
distribution $\sum_{m'} P_{t_d}(m', p') = P_{t_d}(p')$, then we have
\begin{equation}
\ee{p}_{t_o} = \sum_{p'} \ee{p \mid p'} P_{t_d}(p').
\end{equation}
For the particular case of the first moment of the binomial distribution with
parameters $p'$ and $1/2$ we know that
\begin{equation}
\ee{p \mid p'} = {p' \over 2}.
\end{equation}
Therefore the moment after division is equal to
\begin{equation}
\ee{p}_{t_o} = \sum_{p'} {p' \over 2} P_{t_d}(p')
             = {1 \over 2} \sum_{p'} p' P_{t_d}(p').
\end{equation}
Notice that this is just 1/2 of the expected value of $p'$ averaging over the
distribution prior to cell division, i.e.
\begin{equation}
\ee{p}_{t_o} = {\ee{p'}_{t_d} \over 2},
\end{equation}
where $\ee{\cdot}_{t_d}$ highlights that is the moment of the distribution
prior to the cell division. This result makes perfect sense. What this is
saying is that the mean protein copy number right after the cell divides is
half of the mean protein copy number just before the cell division. That is
exactly we would expect. So in principle to know the first moment of either the
mRNA distribution $\ee{m}_{t_o}$ or the protein distribution $\ee{m}_{t_o}$
right after cell division it suffices to multiply the moments before the cell
division $\ee{m}_{t_d}$ or $\ee{p}_{t_d}$ by 1/2. Let's now explore how this
generalizes to any other moment $\ee{m^x p^y}_{t_o}$.

\subsubsection{Computing the moments of a binomial distribution}

The result from last section was dependent on us knowing the functional form of
the first moment of the binomial distribution. For higher moments we need some
systematic way to compute such moments. Luckily for us we can do so by using
the so-called moment generating function (MGF). The MGF of a random variable
$X$ is defined as
\begin{equation}
M_X(t) = \ee{e^{tX}},
\end{equation}
where $t$ is a dummy variable. Once we know the MGF we can obtain any moment of
the distribution by simply computing
\begin{equation}
  \ee{X^n} = \left. {d^n \over dt^n} M_X(t) \right\vert_{t=0},
  \label{eq_mgf_def}
\end{equation}
i.e. taking the $n$-th derivative of the MGF returns the $n$-th moment of the
distribution. For the particular case of the binomial distribution $X \sim
\text{Bin}(N, q)$ it can be shown that the MGF is of the form
\begin{equation}
M_X(t) = \left[ (1 - q) + qe^{t} \right]^N.
\end{equation}
As an example let's compute the first moment of this binomially distributed
variable. For this, the first derivative of the MGF results in
\begin{equation}
  {d M_X(t) \over dt} = N [(1 - q) + qe^t]^{N - 1} q e^t.
\end{equation}
We just need to follow \eref{eq_mgf_def} and set $t = 0$ to obtain the first
moment
\begin{equation}
  \left. {d M_X(t) \over dt} \right\vert_{t=0} = N q,
  \label{eq_mgf_mean}
\end{equation}
which is exactly the expected value of a binomially distributed random
variable.

So according to \eref{eq_general_binom_mom} to compute any moment $\ee{m^x
p^y}$ after cell division we can just take the $x$-th derivative and the $y$-th
derivative of the binomial MGF to obtain $\ee{m^x \mid m'}$ and $\ee{p^y \mid
p'}$, respectively, and take the expected value of the result. Let's follow on
detail the specific case for the moment $\ee{m p}$. When computing the moment
after cell division $\ee{mp}_{t_o}$ which is of
the form
\begin{equation}
\ee{mp}_{t_o} = \sum_{m'} \sum{p'} \ee{m \mid m'} \ee{p \mid p'} 
                P_{t_d}(m', p'),
\end{equation}
the product $\ee{m \mid m'} \ee{p \mid p'}$ is then
\begin{equation}
\ee{m \mid m'} \ee{p \mid p'} = {m' \over 2} \cdot {p' \over 2},
\end{equation}
where we used the result in \eref{eq_mgf_mean}, substituting $m$ and $p$ for
$X$, respectively, and $q$ for 1/2. Substituting this result into the moment
gives
\begin{equation}
\ee{mp}_{t_o} = \sum_{m'} \sum_{p'} {m' p' \over 4} P_{t_d}(m', p') 
              = {\ee{m' p'}_{t_d} \over 4}.
\end{equation}
Therefore to compute the moment after cell division $\ee{mp}_{t_o}$ we simply
have to divide by 4 the corresponding equivalent moment before the cell
division. 

Not all moments after cell division depend only on the equivalent moment before
cell division. For example if we compute the third moment of the protein
distribution $\ee{p^3}_{t_o}$, we find
\begin{equation}
  \ee{p^3}_{t_o} = {\ee{p^3}_{t_d} \over 8} + {3 \ee{p^2}_{t_d} \over 8}.
\end{equation}
So for this particular case the third moment of the protein distribution
depends on the third moment and the second moment before the cell division. In
general all moments after cell division $\ee{m^x p^y}_{t_o}$ linearly depend on
moments before cell division. Furthermore, there is ``moment closure'' for this
specific case in the sense that all moments after cell division depend on lower
moments before cell division. To generalize these results to all the moments
computed in this work let us then define a vector to collect all moments before
the cell division up the $\ee{m^x p^y}_{t_d}$ moment, i.e.
\begin{equation}
\bb{\ee{m^x p^y}}_{t_d} = \left(
\ee{m^0 p^0}_{t_d}, \ee{m^1}_{t_d}, \ldots , \ee{m^x p^y}_{t_d}
\right).
\end{equation}
Then any moment after cell division $\ee{m^{x'} p^{y'}}_{t_o}$ for $x' \leq x$ and $y' \leq y$ can be computed as
$$
\ee{m^{x'} p^{y'}}_{t_o} = \bb{z}_{x'y'} \cdot \bb{\ee{m^x p^y}}_{t_d},
$$
where we define the vector $\bb{z}_{x'y'}$ as the vector containing all the
coefficients that we obtain with the product of the two binomial distributions.
For example for the case of the third protein moment $\ee{p^3}_{t_o}$ the
vector $\bb{z}_{x'y'}$ would have zeros for all entries except for the
corresponding entry for $\ee{p^2}_{t_d}$ and for $\ee{p^3}_{t_d}$, where it
would have $3/8$ and $1/8$ accordingly.

If we want then to compute all the moments after the cell division up to
$\ee{m^x p^y}_{t_o}$ let us define an equivalent vector
\begin{equation}
\bb{\ee{m^x p^y}}_{t_o} = \left(
\ee{m^0 p^0}_{t_o}, \ee{m^1}_{t_o}, \ldots , \ee{m^x p^y}_{t_o}
\right).
\end{equation}
Then we need to build a square matrix $\bb{Z}$ such that each row of the matrix
contains the corresponding vector $\bb{z}_{x' y'}$ for each of the moments.
Having this matrix we would simply compute the moments after the cell division
as
\begin{equation}
\bb{\ee{m^x p^x}}_{t_o} = \bb{Z} \cdot \bb{\ee{m^x p^x}}_{t_d}.
\end{equation}
In other words, matrix $\bb{Z}$ will contain all the coefficients that we need
to multiply by the moments before the cell division in order to obtain the
moments after cell division. Matrix $\bb{Z}$ was then generated automatically
using Python's analytical math library sympy \cite{sympy}.

\fref{sfig_first_mom_cycles} (adapted from \fref{fig3_cell_cycle}(B)) shows how
the first moment of both mRNA and protein changes over several cell cycles. The
mRNA quickly relaxes to the steady state corresponding to the parameters for
both a single and two promoter copies. This is expected since the parameters
for the mRNA production were determined in the first place under this
assumption (See \siref{supp_model}). We note that there is no apparent delay
before reaching steady state of the mean mRNA count after the cell divides.
This is because the mean mRNA count for the two promoters copies  state is
exactly twice the expected mRNA count for the single promoter state (See
\siref{supp_model}). Therefore once the mean mRNA count is halved after the
cell division, it is already at the steady state value for the single promoter
case. On the other hand, given that the relaxation time to steady state is
determined by the degradation rate, the mean protein count does not reach its
corresponding steady state value for either promoter copy number state.
Interestingly once a couple of cell cycles have passed the first moment has a
repetitive trajectory over cell cycles. We have observed this experimentally by
tracking cells as they grow under the microscope. Comparing cells at the
beginning of the cell cycle with the daughter cells that appear after cell
division shown that on average all cells have the same amount of protein at the
beginning of the cell cycle (See Fig. 18 of \cite{Phillips2019}), suggesting
that these dynamical steady state takes place \textit{in vivo}.

\begin{figure}[h!]
	\centering \includegraphics
  {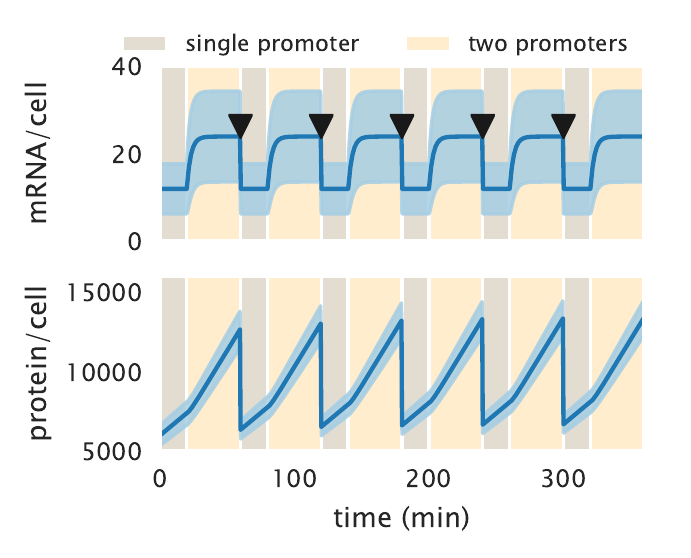}
	\caption{\textbf{First and second moment dynamics over cell the cell cycle.}
	Mean $\pm$ standard deviation mRNA (upper panel) and mean $\pm$ standard
	deviation protein copy number (lower panel) as the cell cycle progresses. The
	dark shaded region delimits the fraction of the cell cycle that cells spend
	with  a single copy of the promoter. The light shaded region delimits the
	fraction of the cell cycle that cells spend with two copies of the promoter.
	For a 100 min doubling time at the {\it galK} locus cells spend 60\% of the
	time with one copy of the promoter and the rest with two copies.}
  \label{sfig_first_mom_cycles}
\end{figure}

In principle when measuring gene expression levels experimentally from an
asynchronous culture, cells are sampled from any time point across their
individual cell cycles. This means that the moments determined experimentally
correspond to an average over the cell cycle. In the following section we
discuss how to account for the fact that cells are not uniformly distributed
across the cell cycle in order to compute these averages.

\subsection{Exponentially distributed ages}

As mentioned in \siref{supp_param_inference}, cells in exponential growth have
exponentially distributed ages across the cell cycle, having more young cells
compared to old ones. Specifically the probability of a cell being at any time
point in the cell cycle is given by \cite{Powell1956}
\begin{equation}
  P(a) = (\ln 2) \cdot 2^{1 - a},
  \label{seq_age_prob}
\end{equation}
where $a \in [0, 1]$ is the stage of the cell cycle, with $a = 0$ being the
start of the cycle and $a = 1$ being the cell division. In
\siref{supp_cell_age_dist} we reproduce this derivation. It is a surprising
result, but can be intuitively thought as follows: If the culture is growing
exponentially, that means that all the time there is an increasing number of
cells. That means for example that if in a time interval $\Delta t$ $N$ ``old''
cells divided, these produced $2N$ ``young'' cells. So at any point there is
always more younger than older cells.

Our numerical integration of the moment equations gave us a time evolution of
the moments as cells progress through the cell cycle. Since experimentally we
sample asynchronous cells that follow \eref{seq_age_prob}, each time point
along the moment dynamic must be weighted by the probability of having sampled
a cell at such specific time point of the cell cycle. Without loss of
generality let's focus on the first mRNA moment $\ee{m(t)}$ (the same can be
applied to all other moments). As mentioned before, in order to calculate the
first moment across the entire cell cycle we must weigh each time point by the
corresponding probability that a cell is found in such point of its cell cycle.
This translates to computing the integral
\begin{equation}
  \ee{m}_c = \int_{\text{beginning cell cycle}}^{\text{end cell cycle}}
                       \ee{m(t)} P(t) dt,
\end{equation}
where $\ee{m}_c$ is the mean mRNA copy number averaged over the entire cell
cycle trajectory, and $P(t)$ is the probability of a cell being at a time $t$ of
its cell cycle.

If we set the time in units of the cell cycle length we can use
\eref{seq_age_prob} and compute instead
\begin{equation}
  \ee{m} = \int_0^1 \ee{m(a)} P(a) da,
  \label{seq_moment_avg}
\end{equation}
where $P(a)$ is given by \eref{seq_age_prob}.

What \eref{seq_moment_avg} implies is that in order to compute the first moment
(or any moment of the distribution) we must weigh each point in the moment
dynamics by the corresponding probability of a cell being at that point along
its cell cycle. That is why when computing a moment we take the time trajectory
of a single cell cycle as the ones shown in \fref{sfig_first_mom_cycles} and
compute the average using \eref{seq_age_prob} to weigh each time point. We
perform this integral numerically for all moments using Simpson's rule.

\subsection{Reproducing the equilibrium picture}

Given the large variability of the first moments depicted in
\fref{sfig_first_mom_cycles} it is worth considering why a simplistic
equilibrium picture has shown to be very successful in predicting the mean
expression level under diverse conditions \cite{Garcia2011c, Brewster2014,
Barnes2019, Razo-Mejia2018}. In this section we compare the simple repression
thermodynamic model with this dynamical picture of the cell cycle. But before
diving into this comparison, it is worth recapping the assumptions that go into
the equilibrium model.

\subsubsection{Steady state under the thermodynamic model}

Given the construction of the thermodynamic model of gene regulation for which
the probability of the promoter microstates rather than the probability of mRNA
or protein counts is accounted for,  we are only allowed to describe the
dynamics of the first moment using this theoretical framework
\cite{Phillips2015}. Again let's only focus on the mRNA first moment $\ee{m}$.
The same principles apply if we consider the protein first moment. We can write
a dynamical system of the form
\begin{equation}
  \dt{\ee{m}} = r_m \cdot \pbound - \gm \ee{m},
\end{equation}
where as before $r_m$ and $\gm$ are the mRNA production and degradation rates
respectively, and $\pbound$ is the probability of finding the RNAP bound to the
promoter \cite{Bintu2005a}. This dynamical system is predicted to have a single
stable fixed point that we can find by computing the steady state. When we
solve for the mean mRNA copy number at steady state $\ee{m}_{ss}$ we find
\begin{equation}
  \ee{m}_{ss} = {r_m \over \gm} \pbound.
\end{equation}

Since we assume that the only effect that the repressor has over the regulation
of the promoter is exclusion of the RNAP from binding to the promoter, we
assume that only $\pbound$ depends on the repressor copy number $R$. Therefore
when computing the fold-change in gene expression we  are left with
\begin{equation}
  \foldchange = {\ee{m (R \neq 0)}_{ss} \over \ee{m (R = 0)}_{ss}}
              = {\pbound (R \neq 0) \over \pbound (R = 0)}.
\end{equation}
As derived in \cite{Garcia2011c} this can be written in the language of
equilibrium statistical mechanics as
\begin{equation}
  \foldchange = \left(1 + {R \over \Nns}e^{-\beta \eR}  \right)^{-1},
  \label{seq_fold_change_thermo}
\end{equation}
where $\beta \equiv (k_BT)^{-1}$, $\eR$ is the repressor-DNA binding energy,
and $\Nns$ is the number of non-specific binding sites where the repressor can
bind.

To arrive at \eref{seq_fold_change_thermo} we ignore the physiological changes
that occur during the cell cycle; one of the most important being the
variability in gene copy number that we are exploring in this section. It is
therefore worth thinking about whether or not the dynamical picture exemplified
in \fref{sfig_first_mom_cycles} can be reconciled with the predictions made by
\eref{seq_fold_change_thermo} both at the mRNA and protein level.

\fref{sfig_lacI_titration} compares the predictions of both theoretical
frameworks for varying repressor copy numbers and repressor-DNA affinities. The
solid lines are directly computed from \eref{seq_fold_change_thermo}. The
hollow triangles and the solid circles, represent the fold-change in mRNA and
protein respectively as computed from the moment dynamics. To compute the
fold-change from the kinetic picture we first numerically integrate the moment
dynamics for both the two- and the three-state promoter (See
\fref{sfig_first_mom_cycles} for the unregulated case) and then average the
time series accounting for the probability of cells being sampled at each stage
of the cell cycle as defined in \eref{seq_moment_avg}. The small systematic
deviations between both models come partly from the simplifying assumption that
the repressor copy number, and therefore the repressor on rate $\kron$ remains
constant during the cell cycle. In principle the gene producing the repressor
protein itself is also subjected to the same duplication during the cell cycle,
changing therefore the mean repressor copy number for both stages.

\begin{figure}[h!]
	\centering \includegraphics
  {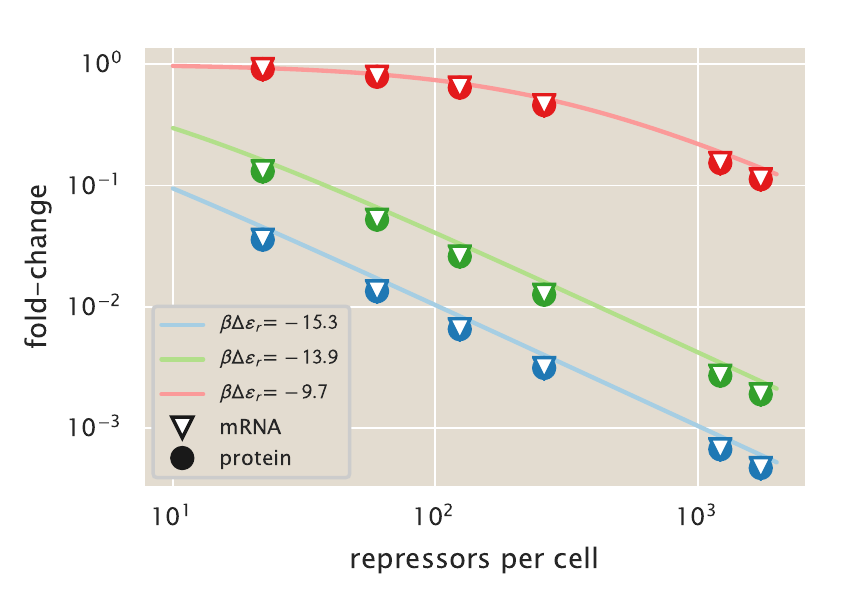}
	\caption{\textbf{Comparison of the equilibrium and kinetic reressor titration
	predictions.} The equilibrium model (solid lines) and the kinetic model with
	variation over the cell cycle (solid circles and white triangles) predictions
	are compared for varying repressor copy numbers and operator binding energy.
	The equilibrium model is directly computed from \eref{seq_fold_change_thermo}
	while the kinetic model is computed by numerically integrating the moment
	equations over several cell cycles, and then averaging over the extent of the
	cell cycle as defined in \eref{seq_moment_avg}.}
  \label{sfig_lacI_titration}
\end{figure}

For completeness \fref{sfig_IPTG_titration} compares the kinetic and
equilibrium models for the extended model of \cite{Razo-Mejia2018} in which the
inducer concentration enters into the equation. The solid line is directly
computed from Eq. 5 of \cite{Razo-Mejia2018}. The hollow triangles and solid
points follow the same procedure as for \fref{sfig_lacI_titration}, where the
only effect that the inducer is assume to have in the kinetics is an effective
change in the number of active repressors, affecting therefore $\kron$.

\begin{figure}[h!]
	\centering \includegraphics
  {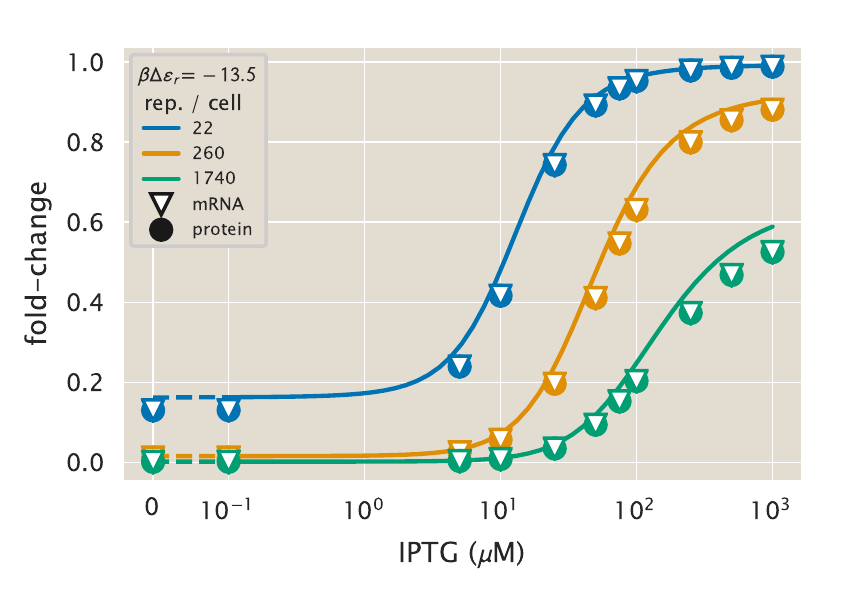}
	\caption{\textbf{Comparison of the equilibrium and kinetic inducer titration
	predictions.} The equilibrium model (solid lines) and the kinetic model with
	variation over the cell cycle (solid circles and white triangles)
	predictions are compared for varying repressor copy numbers and inducer
	concentrations. The equilibrium model is directly computed as Eq. 5 of
	reference \cite{Razo-Mejia2018} with repressor-DNA binding energy $\eR =
	-13.5 \; k_BT$ while the kinetic model is computed by numerically
	integrating the moment dynamics over several cell cycles, and then averaging
	over the extent of a single cell cycle as defined in \eref{seq_moment_avg}.}
  \label{sfig_IPTG_titration}
\end{figure}

\subsection{Comparison between single- and multi-promoter kinetic model}

After these calculations it is worth questioning whether the inclusion of this
change in gene dosage is drastically different with respect to
the simpler picture of a kinetic model that ignores the gene copy number
variability during the cell cycle. To this end we systematically computed the
average moments for varying repressor copy number and repressor-DNA affinities.
We then compare these results with the moments obtained from a single-promoter
model and their corresponding parameters. The derivation of the steady-state
moments of the distribution for the single-promoter model are detailed in
\siref{supp_moments}.

\fref{sfig_lacI_titration} and \fref{sfig_IPTG_titration} both suggest that
since the dynamic multi-promoter model can reproduce the results of the
equilibrium model at the first moment level it must then also be able to
reproduce the results of the single-promoter model at this level (See
\siref{supp_param_inference}). The interesting comparison comes with higher
moments. A useful metric to consider for gene expression variability is the
noise in gene expression \cite{Shahrezaei2008}. This quantity, defined as the
standard deviation divided by the mean, is a dimensionless metric of how much
variability there is with respect to the mean of a distribution. As we will
show below this quantity differs from the also commonly used metric known as
the Fano factor (variance / mean) in the sense that for experimentally
determined expression levels in fluorescent arbitrary units, the noise is a
dimensionless quantity while the Fano factor is not.

\fref{sfig_noise_comparison} shows the comparison of the predicted protein
noise between  the single- (dashed lines) and the multi-promoter model (solid
lines) for different operators and repressor copy numbers. A striking
difference between both is that the single-promoter model predicts that as the
inducer concentration increases, the standard deviation grows much slower than
the mean, giving a very small noise. In comparison the multi-promoter model has
a much higher floor for the lowest value of the noise, reflecting the expected
result that the variability in gene copy number across the cell cycle should
increase the cell-to-cell variability in gene expression \cite{Peterson2015,
Jones2014a}

\begin{figure}[h!]
	\centering \includegraphics
  {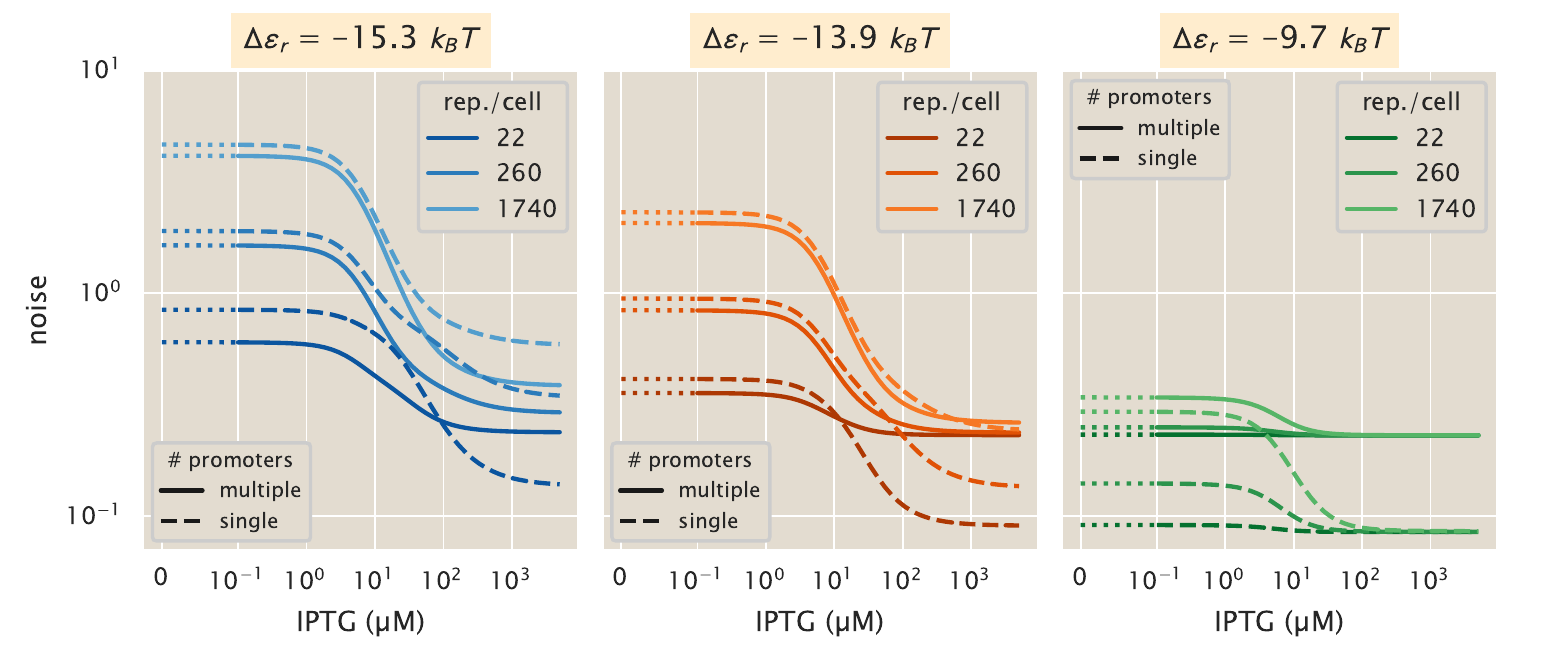}
	\caption{\textbf{Comparison of the predicted protein noise between a single-
	and a multi-promoter kinetic model.} Comparison of the noise
	(standard deviation/mean) between a kinetic model that considers a single
	promoter at all times (dashed line) and the multi-promoter model developed
	in this section (solid line) for different repressor operators. (A) Operator
	O1,  $\eR = -15.3 \; k_BT$, (B) O2, $\eR = -13.9 \; k_BT$, (C) O3, $\eR =
	-9.7 \; k_BT$}
  \label{sfig_noise_comparison}
\end{figure}

\subsection{Comparison with experimental data}\label{supp_theory_vs_data_mom}

Having shown that the kinetic model presented in this section can not only
reproduce the results from the equilibrium picture at the mean level (See
\fref{sfig_lacI_titration} and \fref{sfig_IPTG_titration}), but make predictions
for the cell-to-cell variability as quantified by the noise (See
\fref{sfig_noise_comparison}), we can assess whether or not this model is able
to predict experimental measurements of the noise. For this we take the single
cell intensity measurements (See Methods) to compute the noise at the protein
level.

As mentioned before this metric differs from the Fano factor since for
fluorescent arbitrary units the noise is a dimensionless quantity. To see why
consider that the noise is defined as
\begin{equation}
\text{noise} \equiv \frac{\sqrt{\left\langle p^2 \right\rangle -
                        \left\langle p \right\rangle^2}}
                        {\left\langle p \right\rangle}.
    \label{seq_noise_protein}
\end{equation}
We assume that the intensity level of a cell $I$ is linearly proportional to
the absolute protein count, i.e.
\begin{equation}
I = \alpha p,
\label{seq_calibration_factor}
\end{equation}
where $\alpha$ is the proportionality constant between arbitrary units and
protein absolute number $p$. Substituting this definition on
\eref{seq_noise_protein} gives
\begin{equation}
  \text{noise} = {\sqrt{\ee{(\alpha I)^2} - \ee{\alpha I}^2} \over
                \ee{\alpha I}}.
\end{equation}

Since $\alpha$ is a constant it can be taken out of the average operator
$\ee{\cdot}$, obtaining
\begin{equation}
  \text{noise} = {\sqrt{\alpha^2 \left(\ee{I^2} -
                \ee{I}^2 \right)} \over
                \alpha \ee{I}}
       = {\sqrt{\left(\ee{I^2} - \ee{I}^2 \right)} \over
                \ee{I}}.
\end{equation}

Notice that in \eref{seq_calibration_factor} the linear proportionality between
intensity and protein count has no intercept. This ignores the autofluorescence
that cells without reporter would generate. To account for this, in practice we
compute
\begin{equation}
\text{noise} = {\sqrt{\left(\ee{(I - \ee{I_\text{auto}})^2} -
                    \ee{I - \ee{I_\text{auto}}}^2 \right)} \over
                \ee{I - \ee{I_\text{auto}}}}.
\end{equation}
where $I$ is the intensity of the strain of interest and $\ee{I_\text{auto}}$
is the mean autofluorescence intensity, obtained from a strain that does not
carry the fluorescent reporter gene.

\fref{sfig_noise_delta} shows the comparison between theoretical predictions
and experimental measurements for the unregulated promoter. The reason we split
the data by operator despite the fact that since these are unregulated
promoters, they should in principle have identical expression profiles is to
precisely make sure that this is the case. We have found in the past that
sequences downstream of the RNAP binding site can affect the expression level
of constitutively expressed genes. We can see that both models, the
single-promoter (gray dotted line) and the multi-promoter (black dashed line)
underestimate the experimental noise to different degrees. The single-promoter
model does a worse job at predicting the experimental data since it doesn't
account for the differences in gene dosage during the cell cycle. But still we
can see that accounting for this variability takes us to within a factor of two
of the experimentally determined noise for these unregulated strains.

\begin{figure}[h!]
	\centering \includegraphics
  {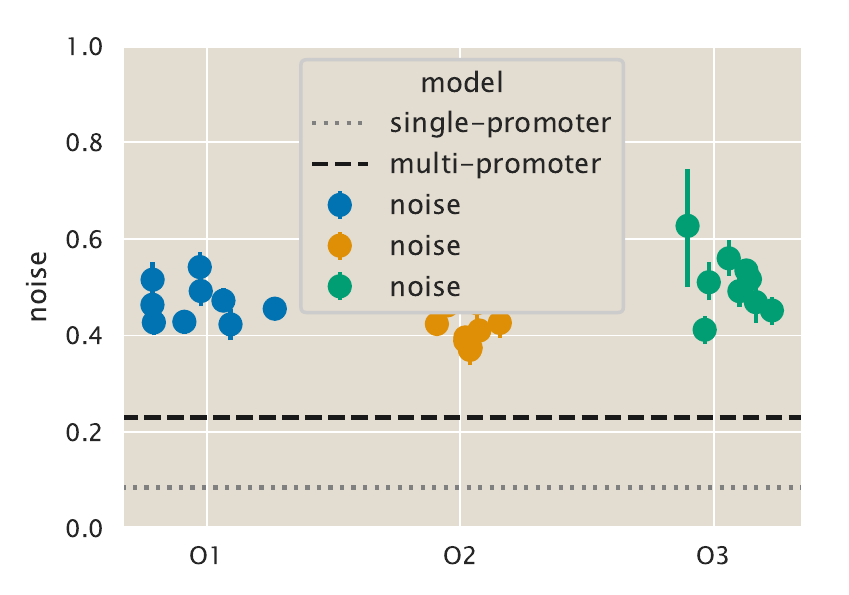}
	\caption{\textbf{Protein noise of the unregulated promoter.} Comparison of
	the experimental noise for different operators with the theoretical
	predictions for the single-promoter (gray dotted line) and the multi-promoter
	model (black dashed line). Each datum represents a single date measurement of
	the corresponding $\Delta lacI$ strain with $\geq 300$ cells. The points
	correspond to the median, and the error bars correspond to the 95\%
	confidence interval as determined by 10,000 bootstrap samples.}
  \label{sfig_noise_delta}
\end{figure}

To further test the model predictive power we compare the predictions for the
three-state regulated promoter. \fref{sfig_noise_reg} shows the theoretical
predictions for the single- and multi-promoter model for varying repressor
copy numbers and repressor-DNA binding affinities as a function of the inducer
concentration. We can see again that our zero-parameter fits systematically
underestimates the noise for all strains and all inducer concentrations. We
highlight that the $y$-axis is shown in a log-scale to emphasize more this
deviation; but, as we will show in the next section, our predictions still fall
within a factor of two from the experimental data.

\begin{figure}[h!]
	\centering \includegraphics
  {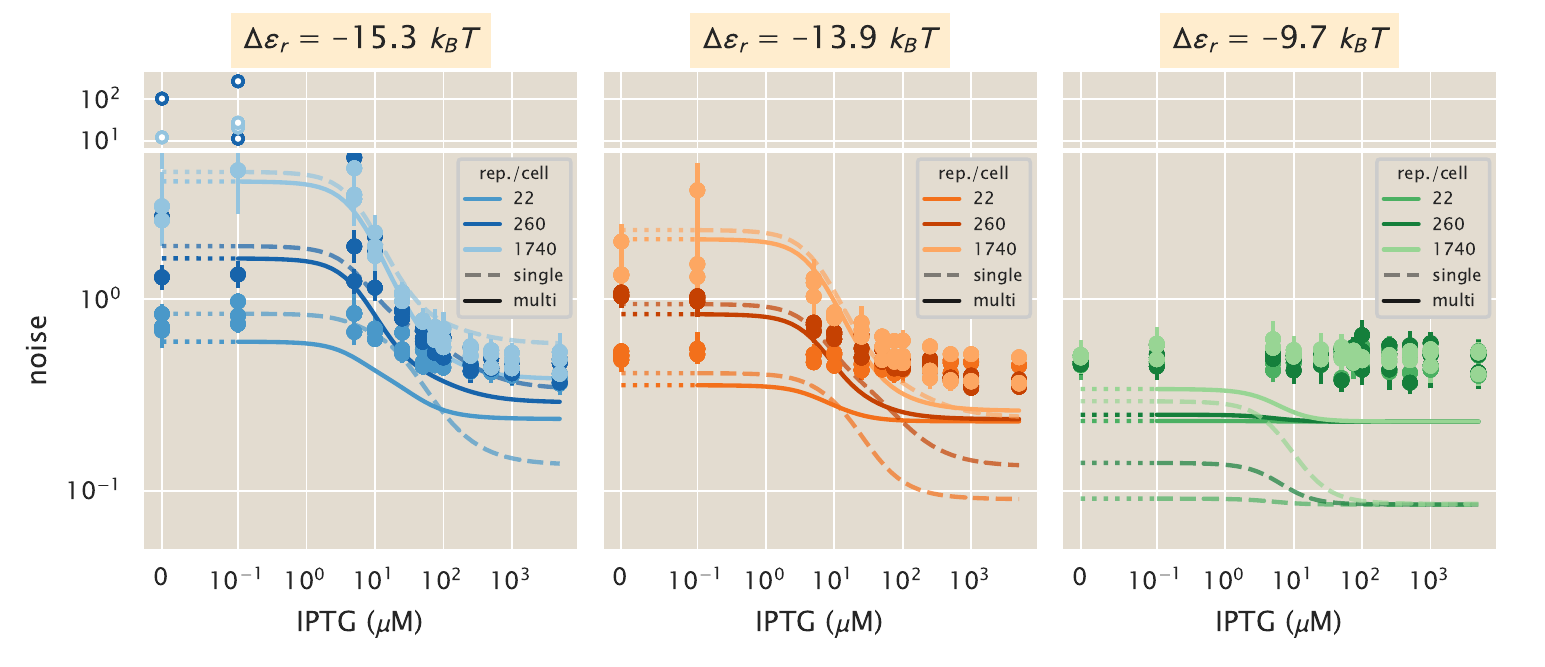}
	\caption{\textbf{Protein noise of the regulated promoter.} Comparison of the
	experimental noise for different operators ((A) O1,  $\eR = -15.3 \; k_BT$,
	(B) O2, $\eR = -13.9 \; k_BT$, (C) O3, $\eR = -9.7 \; k_BT$) with the
	theoretical predictions for the single-promoter (dashed lines) and the
	multi-promoter model (solid lines). Points represent the experimental noise
	as computed from single-cell fluorescence measurements of different {\it E.
	coli} strains under 12 different inducer concentrations. Dotted line
	indicates plot in linear rather than logarithmic scale. Each datum represents
	a single date measurement of the corresponding strain and IPTG concentration
	with $\geq 300$ cells. The points correspond to the median, and the error
	bars correspond to the 95\% confidence interval as determined by 10,000
	bootstrap samples. White-filled dots are plot at a different scale for better
	visualization.}
  \label{sfig_noise_reg}
\end{figure}

\subsubsection{Systematic deviation of the noise in gene expression}

\fref{sfig_noise_delta} and \fref{sfig_noise_reg} highlight that our model
underestimates the cell-to-cell variability as measured by the noise. To
further explore this systematic deviation \fref{sfig_noise_pred_vs_data} shows
the theoretical vs. experimental noise both in linear and log scale. As we can
see the data is systematically above the identity line. The data is colored by
their corresponding experimental fold-change values. The data that has the
largest deviations from the identity line also corresponds to the data with the
largest error bars and the smallest fold-change. This is because measurements
with very small fold-changes correspond to intensities very close to the
autofluorescence background. Therefore minimal changes when computing the noise
are amplified given the ratio of std/mean. In \siref{supp_empirical} we will 
explore empirical ways to improve the agreement between our minimal model and
the experimental data to guide future efforts to improve the minimal.

\begin{figure}[h!]
	\centering \includegraphics
  {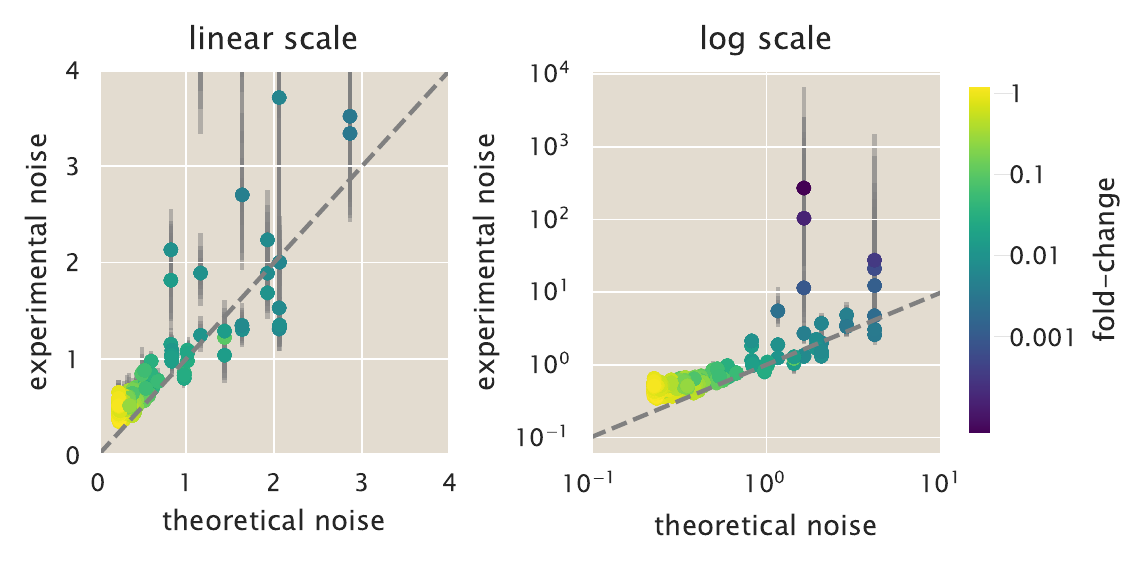}
	\caption{\textbf{Systematic comparison of theoretical vs experimental noise
	in gene expression.} Theoretical vs. experimental noise both in linear
	(left) and log (right) scale. The dashed line shows the identity line of
	slope 1 and intercept zero. All data are colored by the corresponding value
	of the experimental fold-change in gene expression as indicated by the color
	bar. Each datum represents a single date measurement of the corresponding
	strain and IPTG concentration with $\geq 300$ cells. The points correspond
	to the median, and the error bars correspond to the 95\% confidence interval
	as determined by 10,000 bootstrap samples.}
  \label{sfig_noise_pred_vs_data}
\end{figure}

\section{Accounting for the variability in gene copy number during the cell
cycle}\label{supp_multi_gene}

(Note: The Python code used for the calculations presented in this section can
be found in the
\href{https://www.rpgroup.caltech.edu/chann_cap/src/theory/html/moment_dynamics_cell_division.html}{following
link} as an annotated Jupyter notebook)

When growing in rich media, bacteria can double every $\approx$ 20 minutes.
With two replication forks each traveling at $\approx$ 1000 bp per second, and
a genome of $\approx$ 5 Mbp for {\it E. coli} \cite{Moran2010}, a cell would
need $\approx$ 40 minutes to replicate its genome. The apparent paradox  of
growth rates faster than one division per 40 minutes is solved by the fact that
cells have multiple replisomes, i.e. molecular machines that replicate the
genome running in parallel. Cells can have up to 8 copies of the genome being
replicated simultaneously depending on the growth rate \cite{Bremer1996}.

This observation implies that during the cell cycle gene copy number varies.
This variation depends on the growth rate and the relative position of the gene
with respect to the replication origin, having genes close to the replication
origin spending more time with multiple copies compare to genes closer to the
replication termination site. This change in gene dosage has a direct effect on
the cell-to-cell variability in gene expression \cite{Jones2014a,
Peterson2015}.

\subsection{Numerical integration of moment equations}

(Note: The Python code used for the calculations presented in this section can
be found in the
\href{https://www.rpgroup.caltech.edu//chann_cap/software/moment_dynamics_cell_division.html}{following
link} as an annotated Jupyter notebook)

For our specific locus ({\it galK}) and a doubling time of $\approx$ 60 min for
our experimental conditions, cells have on average 1.66 copies of the reporter
gene during the cell cycle \cite{Jones2014a}. What this means is that cells
spend 60\% of the time having one copy of the gene and 40\% of the time with
two copies. To account for this variability in gene copy number across the cell
cycle we numerically integrate the moment equations derived in
\siref{supp_moments} for a time $t = [0, t_s]$ with an mRNA production rate
$r_m$, where $t_s$ is the time point at which the replication fork reaches our
specific locus. For the remaining time before the cell division $t = [t_s,
t_d]$ that the cell spends with two promoters, we assume that the only
parameter that changes is the mRNA production rate from $r_m$ to $2 r_m$. This
simplifying assumption ignores potential changes in protein translation rate
$r_p$ or changes in the repressor copy number that would be reflected in
changes on the repressor on rate $\kron$.

\subsubsection{Computing distribution moments after cell division}

(Note: The Python code used for the calculations presented in this section can
be found in the
\href{https://www.rpgroup.caltech.edu//chann_cap/software/binomial_moments.html}{following
link} as an annotated Jupyter notebook)

We have already solved a general form for the dynamics of the moments of the
distribution, i.e. we wrote differential equations for the moments ${d\ee{m^x
p^y}\over dt}$. Given that we know all parameters for our model we can simply
integrate these equations numerically to compute how the moments of the
distribution evolve as cells progress through their cell cycle. Once the cell
reaches a time $t_d$ when is going to divide the mRNA and proteins that we are
interested in undergo a binomial partitioning between the two daughter cells.
In other words, each molecule flips a coin and decides whether to go to either
daughter. The question then becomes given that we have a value for the moment
$\ee{m^x p^y}_{t_d}$ at a time before the cell division, what would the value
of this moment be after the cell division takes place $\ee{m^x p^y}_{t_o}$?

The probability distribution of mRNA and protein after the cell division
$P_{t_o}(m, p)$ must satisfy
\begin{equation}
  P_{t_o}(m, p) = \sum_{m'=m}^\infty \sum_{p'=p}^\infty 
                  P(m, p \mid m', p') P_{t_d}(m', p'),
\label{eq_dist_post_div}
\end{equation}
where we are summing over all the possibilities of having $m'$ mRNA and $p'$
proteins before cell division. Note that the sums start at $m$ and $p$; this is
because for a cell to have these copy numbers before cell division it is a
requirement that the mother cell had at least such copy number since we are not
assuming that there is any production at the instantaneous cell division time.
Since we assume that the partition of mRNA is independent from the partition of
protein, the conditional probability $P(m, p \mid m', p')$ is simply given by a
product of two binomial distributions, one for the mRNA and one for the
protein, i.e.
\begin{equation}
P(m, p \mid m', p') = {m' \choose m} \left( {1 \over 2} \right)^{m'} \cdot
                      {p' \choose p} \left( {1 \over 2} \right)^{p'}.
\label{eq_binom_prod}
\end{equation}
Because of these product of binomial probabilities are allowed to extend the
sum from
\eref{eq_dist_post_div} to start at $m'=0$ and $p'=0$ as
\begin{equation}
  P_{t_o}(m, p) = \sum_{m'=0}^\infty \sum_{p'=0}^\infty 
                  P(m, p \mid m', p') P_{t_d}(m', p'),
\end{equation}
since the product of the binomial distributions in \eref{eq_binom_prod} is zero
for all $m' < m$ and/or $p' < 0$. So from now on in this section we will assume
that a sum of the form $\sum_x \equiv \sum_{x=0}^\infty$ to simplify notation.

We can then compute the distribution moments after the cell division $\ee{m^x
p^y}_{t_o}$ as
\begin{equation}
\ee{m^x p^y}_{t_o} = \sum_m \sum_p m^x p^y P_{t_o}(m, p),
\end{equation}
for all $x, y \in \mathbb{N}$. Substituting \eref{eq_dist_post_div} results in
\begin{equation}
\ee{m^x p^y}_{t_o} = \sum_m \sum_p m^x p^y
\sum_{m'} \sum_{p'} P(m, p \mid m', p') P_{t_d}(m', p').
\end{equation}
We can rearrange the sums to be 
\begin{equation}
\ee{m^x p^y}_{t_o} = \sum_{m'} \sum_{p'} P_{t_d}(m', p')
                     \sum_m \sum_p m^x p^y P(m, p \mid m', p').
\end{equation}
The fact that \eref{eq_binom_prod} is the product of two independent events
allows us to rewrite the joint probability $P(m, p \mid m', p')$ as
\begin{equation}
P(m, p \mid m', p') = P(m \mid m') \cdot P(p \mid p').
\end{equation}
With this we can then write the moment $\ee{m^x p^y}_{t_o}$ as
\begin{equation}
\ee{m^x p^y}_{t_o} = \sum_{m'} \sum_{p'} P_{t_d}(m', p')
                     \sum_m  m^x  P(m \mid m')
                     \sum_p p^y P(p \mid p').
\end{equation}
Notice that both terms summing over $m$ and over $p$ are the conditional
expected values, i.e.
\begin{equation}
\sum_z  z^x  P(z \mid z') \equiv \ee{z^x \mid z'}, \; 
{\text{ for } z\in \{m, p \}}.
\end{equation}
These conditional expected values are the expected values of a binomial random
variable $z \sim \text{Bin}(z', 1/2)$, which can be easily computed as we will
show later in this section. We then rewrite the expected values after the cell
division in terms of these moments of a binomial distribution
\begin{equation}
\ee{m^x p^y}_{t_o} = \sum_{m'} \sum_{p'} \ee{m^x \mid m'} \ee{p^y \mid p'} 
                     P_{t_d}(m', p').
  \label{eq_general_binom_mom}
\end{equation}

To see how this general formula for the moments after the cell division works
let's compute the mean protein per cell after the cell division $\ee{p}_{t_o}$.
That is setting $x = 0$, and $y = 1$. This results in
\begin{equation}
\ee{p}_{t_o} = \sum_{m'} \sum_{p'} \ee{m^0 \mid m'} \ee{p \mid p'} 
               P_{t_d}(m', p').
\end{equation}
The zeroth moment $\ee{m^0 \mid m'}$ by definition must be one since we have
\begin{equation}
\ee{m^0 \mid m'} = \sum_m m^0 P(m \mid m') = \sum_m P(m \mid m') = 1,
\end{equation}
since the probability distribution must be normalized. This leaves us then with
\begin{equation}
\ee{p}_{t_o} = \sum_{m'} \sum_{p'} P_{t_d}(m', p') \ee{p \mid p'}.
\end{equation}
If we take the sum over $m'$ we simply compute the marginal probability
distribution $\sum_{m'} P_{t_d}(m', p') = P_{t_d}(p')$, then we have
\begin{equation}
\ee{p}_{t_o} = \sum_{p'} \ee{p \mid p'} P_{t_d}(p').
\end{equation}
For the particular case of the first moment of the binomial distribution with
parameters $p'$ and $1/2$ we know that
\begin{equation}
\ee{p \mid p'} = {p' \over 2}.
\end{equation}
Therefore the moment after division is equal to
\begin{equation}
\ee{p}_{t_o} = \sum_{p'} {p' \over 2} P_{t_d}(p')
             = {1 \over 2} \sum_{p'} p' P_{t_d}(p').
\end{equation}
Notice that this is just 1/2 of the expected value of $p'$ averaging over the
distribution prior to cell division, i.e.
\begin{equation}
\ee{p}_{t_o} = {\ee{p'}_{t_d} \over 2},
\end{equation}
where $\ee{\cdot}_{t_d}$ highlights that is the moment of the distribution
prior to the cell division. This result makes perfect sense. What this is
saying is that the mean protein copy number right after the cell divides is
half of the mean protein copy number just before the cell division. That is
exactly we would expect. So in principle to know the first moment of either the
mRNA distribution $\ee{m}_{t_o}$ or the protein distribution $\ee{m}_{t_o}$
right after cell division it suffices to multiply the moments before the cell
division $\ee{m}_{t_d}$ or $\ee{p}_{t_d}$ by 1/2. Let's now explore how this
generalizes to any other moment $\ee{m^x p^y}_{t_o}$.

\subsubsection{Computing the moments of a binomial distribution}

The result from last section was dependent on us knowing the functional form of
the first moment of the binomial distribution. For higher moments we need some
systematic way to compute such moments. Luckily for us we can do so by using
the so-called moment generating function (MGF). The MGF of a random variable
$X$ is defined as
\begin{equation}
M_X(t) = \ee{e^{tX}},
\end{equation}
where $t$ is a dummy variable. Once we know the MGF we can obtain any moment of
the distribution by simply computing
\begin{equation}
  \ee{X^n} = \left. {d^n \over dt^n} M_X(t) \right\vert_{t=0},
  \label{eq_mgf_def}
\end{equation}
i.e. taking the $n$-th derivative of the MGF returns the $n$-th moment of the
distribution. For the particular case of the binomial distribution $X \sim
\text{Bin}(N, q)$ it can be shown that the MGF is of the form
\begin{equation}
M_X(t) = \left[ (1 - q) + qe^{t} \right]^N.
\end{equation}
As an example let's compute the first moment of this binomially distributed
variable. For this, the first derivative of the MGF results in
\begin{equation}
  {d M_X(t) \over dt} = N [(1 - q) + qe^t]^{N - 1} q e^t.
\end{equation}
We just need to follow \eref{eq_mgf_def} and set $t = 0$ to obtain the first
moment
\begin{equation}
  \left. {d M_X(t) \over dt} \right\vert_{t=0} = N q,
  \label{eq_mgf_mean}
\end{equation}
which is exactly the expected value of a binomially distributed random
variable.

So according to \eref{eq_general_binom_mom} to compute any moment $\ee{m^x
p^y}$ after cell division we can just take the $x$-th derivative and the $y$-th
derivative of the binomial MGF to obtain $\ee{m^x \mid m'}$ and $\ee{p^y \mid
p'}$, respectively, and take the expected value of the result. Let's follow on
detail the specific case for the moment $\ee{m p}$. When computing the moment
after cell division $\ee{mp}_{t_o}$ which is of
the form
\begin{equation}
\ee{mp}_{t_o} = \sum_{m'} \sum{p'} \ee{m \mid m'} \ee{p \mid p'} 
                P_{t_d}(m', p'),
\end{equation}
the product $\ee{m \mid m'} \ee{p \mid p'}$ is then
\begin{equation}
\ee{m \mid m'} \ee{p \mid p'} = {m' \over 2} \cdot {p' \over 2},
\end{equation}
where we used the result in \eref{eq_mgf_mean}, substituting $m$ and $p$ for
$X$, respectively, and $q$ for 1/2. Substituting this result into the moment
gives
\begin{equation}
\ee{mp}_{t_o} = \sum_{m'} \sum_{p'} {m' p' \over 4} P_{t_d}(m', p') 
              = {\ee{m' p'}_{t_d} \over 4}.
\end{equation}
Therefore to compute the moment after cell division $\ee{mp}_{t_o}$ we simply
have to divide by 4 the corresponding equivalent moment before the cell
division. 

Not all moments after cell division depend only on the equivalent moment before
cell division. For example if we compute the third moment of the protein
distribution $\ee{p^3}_{t_o}$, we find
\begin{equation}
  \ee{p^3}_{t_o} = {\ee{p^3}_{t_d} \over 8} + {3 \ee{p^2}_{t_d} \over 8}.
\end{equation}
So for this particular case the third moment of the protein distribution
depends on the third moment and the second moment before the cell division. In
general all moments after cell division $\ee{m^x p^y}_{t_o}$ linearly depend on
moments before cell division. Furthermore, there is ``moment closure'' for this
specific case in the sense that all moments after cell division depend on lower
moments before cell division. To generalize these results to all the moments
computed in this work let us then define a vector to collect all moments before
the cell division up the $\ee{m^x p^y}_{t_d}$ moment, i.e.
\begin{equation}
\bb{\ee{m^x p^y}}_{t_d} = \left(
\ee{m^0 p^0}_{t_d}, \ee{m^1}_{t_d}, \ldots , \ee{m^x p^y}_{t_d}
\right).
\end{equation}
Then any moment after cell division $\ee{m^{x'} p^{y'}}_{t_o}$ for $x' \leq x$ and $y' \leq y$ can be computed as
$$
\ee{m^{x'} p^{y'}}_{t_o} = \bb{z}_{x'y'} \cdot \bb{\ee{m^x p^y}}_{t_d},
$$
where we define the vector $\bb{z}_{x'y'}$ as the vector containing all the
coefficients that we obtain with the product of the two binomial distributions.
For example for the case of the third protein moment $\ee{p^3}_{t_o}$ the
vector $\bb{z}_{x'y'}$ would have zeros for all entries except for the
corresponding entry for $\ee{p^2}_{t_d}$ and for $\ee{p^3}_{t_d}$, where it
would have $3/8$ and $1/8$ accordingly.

If we want then to compute all the moments after the cell division up to
$\ee{m^x p^y}_{t_o}$ let us define an equivalent vector
\begin{equation}
\bb{\ee{m^x p^y}}_{t_o} = \left(
\ee{m^0 p^0}_{t_o}, \ee{m^1}_{t_o}, \ldots , \ee{m^x p^y}_{t_o}
\right).
\end{equation}
Then we need to build a square matrix $\bb{Z}$ such that each row of the matrix
contains the corresponding vector $\bb{z}_{x' y'}$ for each of the moments.
Having this matrix we would simply compute the moments after the cell division
as
\begin{equation}
\bb{\ee{m^x p^x}}_{t_o} = \bb{Z} \cdot \bb{\ee{m^x p^x}}_{t_d}.
\end{equation}
In other words, matrix $\bb{Z}$ will contain all the coefficients that we need
to multiply by the moments before the cell division in order to obtain the
moments after cell division. Matrix $\bb{Z}$ was then generated automatically
using Python's analytical math library sympy \cite{sympy}.

\fref{sfig_first_mom_cycles} (adapted from \fref{fig3_cell_cycle}(B)) shows how
the first moment of both mRNA and protein changes over several cell cycles. The
mRNA quickly relaxes to the steady state corresponding to the parameters for
both a single and two promoter copies. This is expected since the parameters
for the mRNA production were determined in the first place under this
assumption (See \siref{supp_model}). We note that there is no apparent delay
before reaching steady state of the mean mRNA count after the cell divides.
This is because the mean mRNA count for the two promoters copies  state is
exactly twice the expected mRNA count for the single promoter state (See
\siref{supp_model}). Therefore once the mean mRNA count is halved after the
cell division, it is already at the steady state value for the single promoter
case. On the other hand, given that the relaxation time to steady state is
determined by the degradation rate, the mean protein count does not reach its
corresponding steady state value for either promoter copy number state.
Interestingly once a couple of cell cycles have passed the first moment has a
repetitive trajectory over cell cycles. We have observed this experimentally by
tracking cells as they grow under the microscope. Comparing cells at the
beginning of the cell cycle with the daughter cells that appear after cell
division shown that on average all cells have the same amount of protein at the
beginning of the cell cycle (See Fig. 18 of \cite{Phillips2019}), suggesting
that these dynamical steady state takes place \textit{in vivo}.

\begin{figure}[h!]
	\centering \includegraphics
  {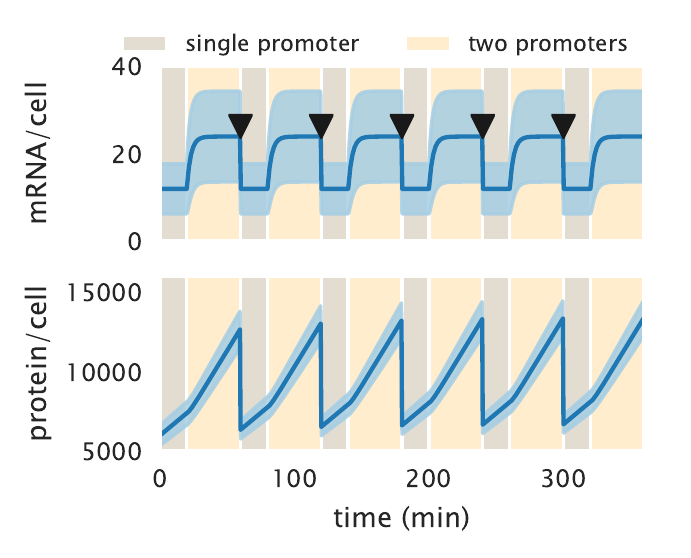}
	\caption{\textbf{First and second moment dynamics over cell the cell cycle.}
	Mean $\pm$ standard deviation mRNA (upper panel) and mean $\pm$ standard
	deviation protein copy number (lower panel) as the cell cycle progresses. The
	dark shaded region delimits the fraction of the cell cycle that cells spend
	with  a single copy of the promoter. The light shaded region delimits the
	fraction of the cell cycle that cells spend with two copies of the promoter.
	For a 100 min doubling time at the {\it galK} locus cells spend 60\% of the
	time with one copy of the promoter and the rest with two copies.}
  \label{sfig_first_mom_cycles}
\end{figure}

In principle when measuring gene expression levels experimentally from an
asynchronous culture, cells are sampled from any time point across their
individual cell cycles. This means that the moments determined experimentally
correspond to an average over the cell cycle. In the following section we
discuss how to account for the fact that cells are not uniformly distributed
across the cell cycle in order to compute these averages.

\subsection{Exponentially distributed ages}

As mentioned in \siref{supp_param_inference}, cells in exponential growth have
exponentially distributed ages across the cell cycle, having more young cells
compared to old ones. Specifically the probability of a cell being at any time
point in the cell cycle is given by \cite{Powell1956}
\begin{equation}
  P(a) = (\ln 2) \cdot 2^{1 - a},
  \label{seq_age_prob}
\end{equation}
where $a \in [0, 1]$ is the stage of the cell cycle, with $a = 0$ being the
start of the cycle and $a = 1$ being the cell division. In
\siref{supp_cell_age_dist} we reproduce this derivation. It is a surprising
result, but can be intuitively thought as follows: If the culture is growing
exponentially, that means that all the time there is an increasing number of
cells. That means for example that if in a time interval $\Delta t$ $N$ ``old''
cells divided, these produced $2N$ ``young'' cells. So at any point there is
always more younger than older cells.

Our numerical integration of the moment equations gave us a time evolution of
the moments as cells progress through the cell cycle. Since experimentally we
sample asynchronous cells that follow \eref{seq_age_prob}, each time point
along the moment dynamic must be weighted by the probability of having sampled
a cell at such specific time point of the cell cycle. Without loss of
generality let's focus on the first mRNA moment $\ee{m(t)}$ (the same can be
applied to all other moments). As mentioned before, in order to calculate the
first moment across the entire cell cycle we must weigh each time point by the
corresponding probability that a cell is found in such point of its cell cycle.
This translates to computing the integral
\begin{equation}
  \ee{m}_c = \int_{\text{beginning cell cycle}}^{\text{end cell cycle}}
                       \ee{m(t)} P(t) dt,
\end{equation}
where $\ee{m}_c$ is the mean mRNA copy number averaged over the entire cell
cycle trajectory, and $P(t)$ is the probability of a cell being at a time $t$ of
its cell cycle.

If we set the time in units of the cell cycle length we can use
\eref{seq_age_prob} and compute instead
\begin{equation}
  \ee{m} = \int_0^1 \ee{m(a)} P(a) da,
  \label{seq_moment_avg}
\end{equation}
where $P(a)$ is given by \eref{seq_age_prob}.

What \eref{seq_moment_avg} implies is that in order to compute the first moment
(or any moment of the distribution) we must weigh each point in the moment
dynamics by the corresponding probability of a cell being at that point along
its cell cycle. That is why when computing a moment we take the time trajectory
of a single cell cycle as the ones shown in \fref{sfig_first_mom_cycles} and
compute the average using \eref{seq_age_prob} to weigh each time point. We
perform this integral numerically for all moments using Simpson's rule.

\subsection{Reproducing the equilibrium picture}

Given the large variability of the first moments depicted in
\fref{sfig_first_mom_cycles} it is worth considering why a simplistic
equilibrium picture has shown to be very successful in predicting the mean
expression level under diverse conditions \cite{Garcia2011c, Brewster2014,
Barnes2019, Razo-Mejia2018}. In this section we compare the simple repression
thermodynamic model with this dynamical picture of the cell cycle. But before
diving into this comparison, it is worth recapping the assumptions that go into
the equilibrium model.

\subsubsection{Steady state under the thermodynamic model}

Given the construction of the thermodynamic model of gene regulation for which
the probability of the promoter microstates rather than the probability of mRNA
or protein counts is accounted for,  we are only allowed to describe the
dynamics of the first moment using this theoretical framework
\cite{Phillips2015}. Again let's only focus on the mRNA first moment $\ee{m}$.
The same principles apply if we consider the protein first moment. We can write
a dynamical system of the form
\begin{equation}
  \dt{\ee{m}} = r_m \cdot \pbound - \gm \ee{m},
\end{equation}
where as before $r_m$ and $\gm$ are the mRNA production and degradation rates
respectively, and $\pbound$ is the probability of finding the RNAP bound to the
promoter \cite{Bintu2005a}. This dynamical system is predicted to have a single
stable fixed point that we can find by computing the steady state. When we
solve for the mean mRNA copy number at steady state $\ee{m}_{ss}$ we find
\begin{equation}
  \ee{m}_{ss} = {r_m \over \gm} \pbound.
\end{equation}

Since we assume that the only effect that the repressor has over the regulation
of the promoter is exclusion of the RNAP from binding to the promoter, we
assume that only $\pbound$ depends on the repressor copy number $R$. Therefore
when computing the fold-change in gene expression we  are left with
\begin{equation}
  \foldchange = {\ee{m (R \neq 0)}_{ss} \over \ee{m (R = 0)}_{ss}}
              = {\pbound (R \neq 0) \over \pbound (R = 0)}.
\end{equation}
As derived in \cite{Garcia2011c} this can be written in the language of
equilibrium statistical mechanics as
\begin{equation}
  \foldchange = \left(1 + {R \over \Nns}e^{-\beta \eR}  \right)^{-1},
  \label{seq_fold_change_thermo}
\end{equation}
where $\beta \equiv (k_BT)^{-1}$, $\eR$ is the repressor-DNA binding energy,
and $\Nns$ is the number of non-specific binding sites where the repressor can
bind.

To arrive at \eref{seq_fold_change_thermo} we ignore the physiological changes
that occur during the cell cycle; one of the most important being the
variability in gene copy number that we are exploring in this section. It is
therefore worth thinking about whether or not the dynamical picture exemplified
in \fref{sfig_first_mom_cycles} can be reconciled with the predictions made by
\eref{seq_fold_change_thermo} both at the mRNA and protein level.

\fref{sfig_lacI_titration} compares the predictions of both theoretical
frameworks for varying repressor copy numbers and repressor-DNA affinities. The
solid lines are directly computed from \eref{seq_fold_change_thermo}. The
hollow triangles and the solid circles, represent the fold-change in mRNA and
protein respectively as computed from the moment dynamics. To compute the
fold-change from the kinetic picture we first numerically integrate the moment
dynamics for both the two- and the three-state promoter (See
\fref{sfig_first_mom_cycles} for the unregulated case) and then average the
time series accounting for the probability of cells being sampled at each stage
of the cell cycle as defined in \eref{seq_moment_avg}. The small systematic
deviations between both models come partly from the simplifying assumption that
the repressor copy number, and therefore the repressor on rate $\kron$ remains
constant during the cell cycle. In principle the gene producing the repressor
protein itself is also subjected to the same duplication during the cell cycle,
changing therefore the mean repressor copy number for both stages.

\begin{figure}[h!]
	\centering \includegraphics
  {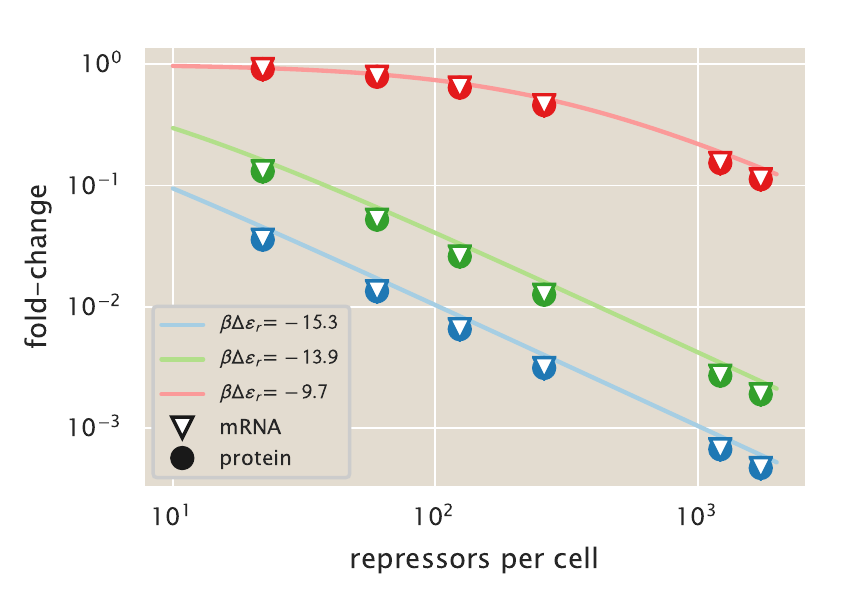}
	\caption{\textbf{Comparison of the equilibrium and kinetic reressor titration
	predictions.} The equilibrium model (solid lines) and the kinetic model with
	variation over the cell cycle (solid circles and white triangles) predictions
	are compared for varying repressor copy numbers and operator binding energy.
	The equilibrium model is directly computed from \eref{seq_fold_change_thermo}
	while the kinetic model is computed by numerically integrating the moment
	equations over several cell cycles, and then averaging over the extent of the
	cell cycle as defined in \eref{seq_moment_avg}.}
  \label{sfig_lacI_titration}
\end{figure}

For completeness \fref{sfig_IPTG_titration} compares the kinetic and
equilibrium models for the extended model of \cite{Razo-Mejia2018} in which the
inducer concentration enters into the equation. The solid line is directly
computed from Eq. 5 of \cite{Razo-Mejia2018}. The hollow triangles and solid
points follow the same procedure as for \fref{sfig_lacI_titration}, where the
only effect that the inducer is assume to have in the kinetics is an effective
change in the number of active repressors, affecting therefore $\kron$.

\begin{figure}[h!]
	\centering \includegraphics
  {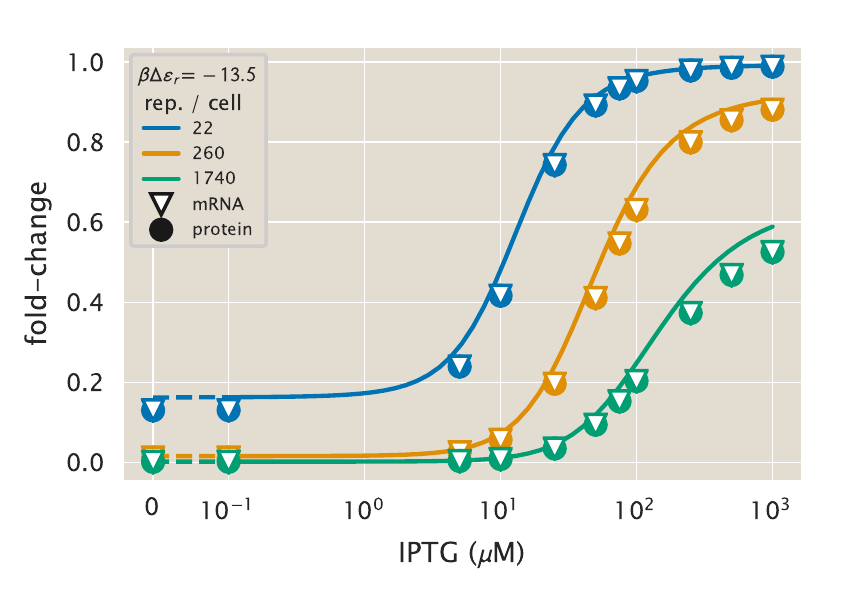}
	\caption{\textbf{Comparison of the equilibrium and kinetic inducer titration
	predictions.} The equilibrium model (solid lines) and the kinetic model with
	variation over the cell cycle (solid circles and white triangles)
	predictions are compared for varying repressor copy numbers and inducer
	concentrations. The equilibrium model is directly computed as Eq. 5 of
	reference \cite{Razo-Mejia2018} with repressor-DNA binding energy $\eR =
	-13.5 \; k_BT$ while the kinetic model is computed by numerically
	integrating the moment dynamics over several cell cycles, and then averaging
	over the extent of a single cell cycle as defined in \eref{seq_moment_avg}.}
  \label{sfig_IPTG_titration}
\end{figure}

\subsection{Comparison between single- and multi-promoter kinetic model}

After these calculations it is worth questioning whether the inclusion of this
change in gene dosage is drastically different with respect to
the simpler picture of a kinetic model that ignores the gene copy number
variability during the cell cycle. To this end we systematically computed the
average moments for varying repressor copy number and repressor-DNA affinities.
We then compare these results with the moments obtained from a single-promoter
model and their corresponding parameters. The derivation of the steady-state
moments of the distribution for the single-promoter model are detailed in
\siref{supp_moments}.

\fref{sfig_lacI_titration} and \fref{sfig_IPTG_titration} both suggest that
since the dynamic multi-promoter model can reproduce the results of the
equilibrium model at the first moment level it must then also be able to
reproduce the results of the single-promoter model at this level (See
\siref{supp_param_inference}). The interesting comparison comes with higher
moments. A useful metric to consider for gene expression variability is the
noise in gene expression \cite{Shahrezaei2008}. This quantity, defined as the
standard deviation divided by the mean, is a dimensionless metric of how much
variability there is with respect to the mean of a distribution. As we will
show below this quantity differs from the also commonly used metric known as
the Fano factor (variance / mean) in the sense that for experimentally
determined expression levels in fluorescent arbitrary units, the noise is a
dimensionless quantity while the Fano factor is not.

\fref{sfig_noise_comparison} shows the comparison of the predicted protein
noise between  the single- (dashed lines) and the multi-promoter model (solid
lines) for different operators and repressor copy numbers. A striking
difference between both is that the single-promoter model predicts that as the
inducer concentration increases, the standard deviation grows much slower than
the mean, giving a very small noise. In comparison the multi-promoter model has
a much higher floor for the lowest value of the noise, reflecting the expected
result that the variability in gene copy number across the cell cycle should
increase the cell-to-cell variability in gene expression \cite{Peterson2015,
Jones2014a}

\begin{figure}[h!]
	\centering \includegraphics
  {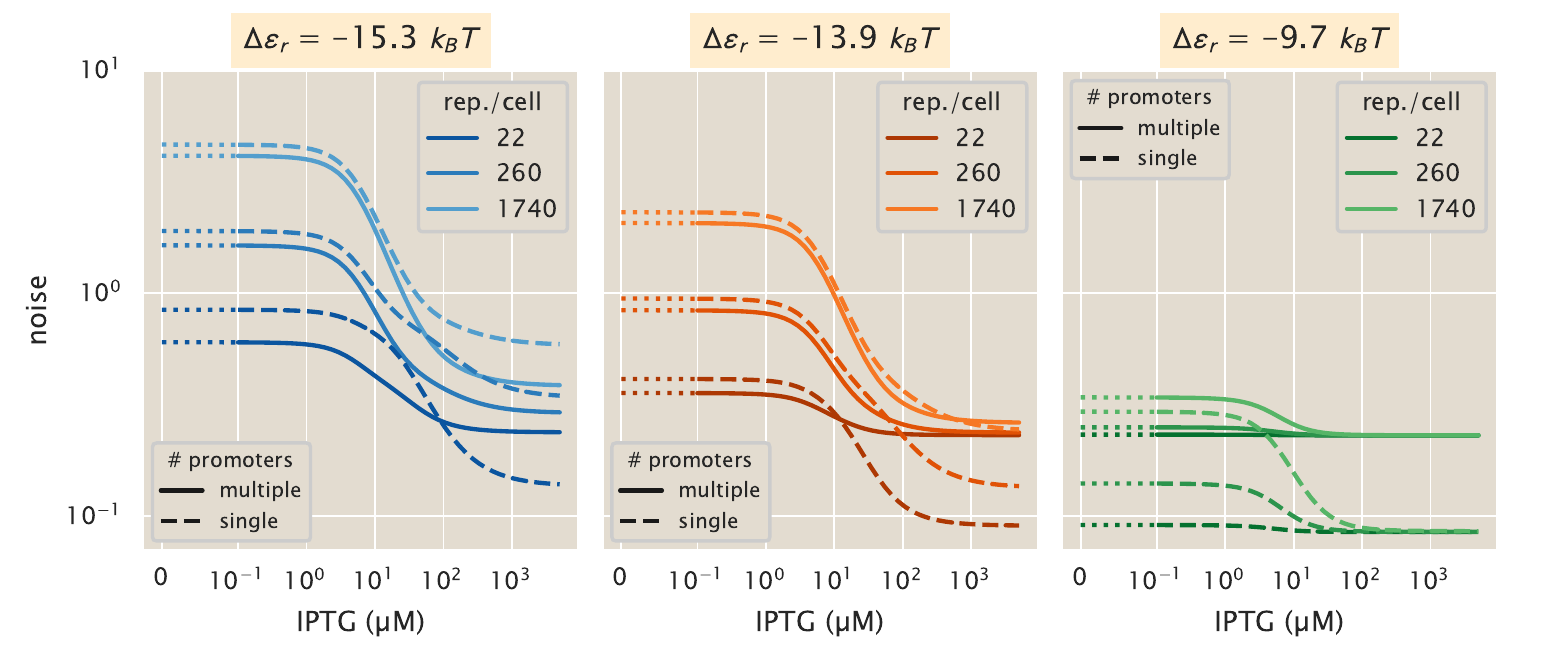}
	\caption{\textbf{Comparison of the predicted protein noise between a single-
	and a multi-promoter kinetic model.} Comparison of the noise
	(standard deviation/mean) between a kinetic model that considers a single
	promoter at all times (dashed line) and the multi-promoter model developed
	in this section (solid line) for different repressor operators. (A) Operator
	O1,  $\eR = -15.3 \; k_BT$, (B) O2, $\eR = -13.9 \; k_BT$, (C) O3, $\eR =
	-9.7 \; k_BT$}
  \label{sfig_noise_comparison}
\end{figure}

\subsection{Comparison with experimental data}\label{supp_theory_vs_data_mom}

Having shown that the kinetic model presented in this section can not only
reproduce the results from the equilibrium picture at the mean level (See
\fref{sfig_lacI_titration} and \fref{sfig_IPTG_titration}), but make predictions
for the cell-to-cell variability as quantified by the noise (See
\fref{sfig_noise_comparison}), we can assess whether or not this model is able
to predict experimental measurements of the noise. For this we take the single
cell intensity measurements (See Methods) to compute the noise at the protein
level.

As mentioned before this metric differs from the Fano factor since for
fluorescent arbitrary units the noise is a dimensionless quantity. To see why
consider that the noise is defined as
\begin{equation}
\text{noise} \equiv \frac{\sqrt{\left\langle p^2 \right\rangle -
                        \left\langle p \right\rangle^2}}
                        {\left\langle p \right\rangle}.
    \label{seq_noise_protein}
\end{equation}
We assume that the intensity level of a cell $I$ is linearly proportional to
the absolute protein count, i.e.
\begin{equation}
I = \alpha p,
\label{seq_calibration_factor}
\end{equation}
where $\alpha$ is the proportionality constant between arbitrary units and
protein absolute number $p$. Substituting this definition on
\eref{seq_noise_protein} gives
\begin{equation}
  \text{noise} = {\sqrt{\ee{(\alpha I)^2} - \ee{\alpha I}^2} \over
                \ee{\alpha I}}.
\end{equation}

Since $\alpha$ is a constant it can be taken out of the average operator
$\ee{\cdot}$, obtaining
\begin{equation}
  \text{noise} = {\sqrt{\alpha^2 \left(\ee{I^2} -
                \ee{I}^2 \right)} \over
                \alpha \ee{I}}
       = {\sqrt{\left(\ee{I^2} - \ee{I}^2 \right)} \over
                \ee{I}}.
\end{equation}

Notice that in \eref{seq_calibration_factor} the linear proportionality between
intensity and protein count has no intercept. This ignores the autofluorescence
that cells without reporter would generate. To account for this, in practice we
compute
\begin{equation}
\text{noise} = {\sqrt{\left(\ee{(I - \ee{I_\text{auto}})^2} -
                    \ee{I - \ee{I_\text{auto}}}^2 \right)} \over
                \ee{I - \ee{I_\text{auto}}}}.
\end{equation}
where $I$ is the intensity of the strain of interest and $\ee{I_\text{auto}}$
is the mean autofluorescence intensity, obtained from a strain that does not
carry the fluorescent reporter gene.

\fref{sfig_noise_delta} shows the comparison between theoretical predictions
and experimental measurements for the unregulated promoter. The reason we split
the data by operator despite the fact that since these are unregulated
promoters, they should in principle have identical expression profiles is to
precisely make sure that this is the case. We have found in the past that
sequences downstream of the RNAP binding site can affect the expression level
of constitutively expressed genes. We can see that both models, the
single-promoter (gray dotted line) and the multi-promoter (black dashed line)
underestimate the experimental noise to different degrees. The single-promoter
model does a worse job at predicting the experimental data since it doesn't
account for the differences in gene dosage during the cell cycle. But still we
can see that accounting for this variability takes us to within a factor of two
of the experimentally determined noise for these unregulated strains.

\begin{figure}[h!]
	\centering \includegraphics
  {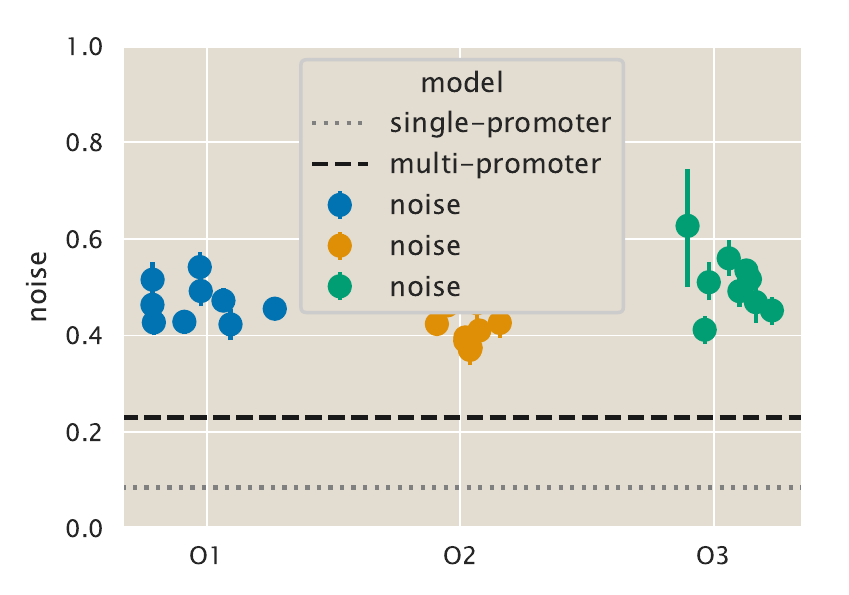}
	\caption{\textbf{Protein noise of the unregulated promoter.} Comparison of
	the experimental noise for different operators with the theoretical
	predictions for the single-promoter (gray dotted line) and the multi-promoter
	model (black dashed line). Each datum represents a single date measurement of
	the corresponding $\Delta lacI$ strain with $\geq 300$ cells. The points
	correspond to the median, and the error bars correspond to the 95\%
	confidence interval as determined by 10,000 bootstrap samples.}
  \label{sfig_noise_delta}
\end{figure}

To further test the model predictive power we compare the predictions for the
three-state regulated promoter. \fref{sfig_noise_reg} shows the theoretical
predictions for the single- and multi-promoter model for varying repressor
copy numbers and repressor-DNA binding affinities as a function of the inducer
concentration. We can see again that our zero-parameter fits systematically
underestimates the noise for all strains and all inducer concentrations. We
highlight that the $y$-axis is shown in a log-scale to emphasize more this
deviation; but, as we will show in the next section, our predictions still fall
within a factor of two from the experimental data.

\begin{figure}[h!]
	\centering \includegraphics
  {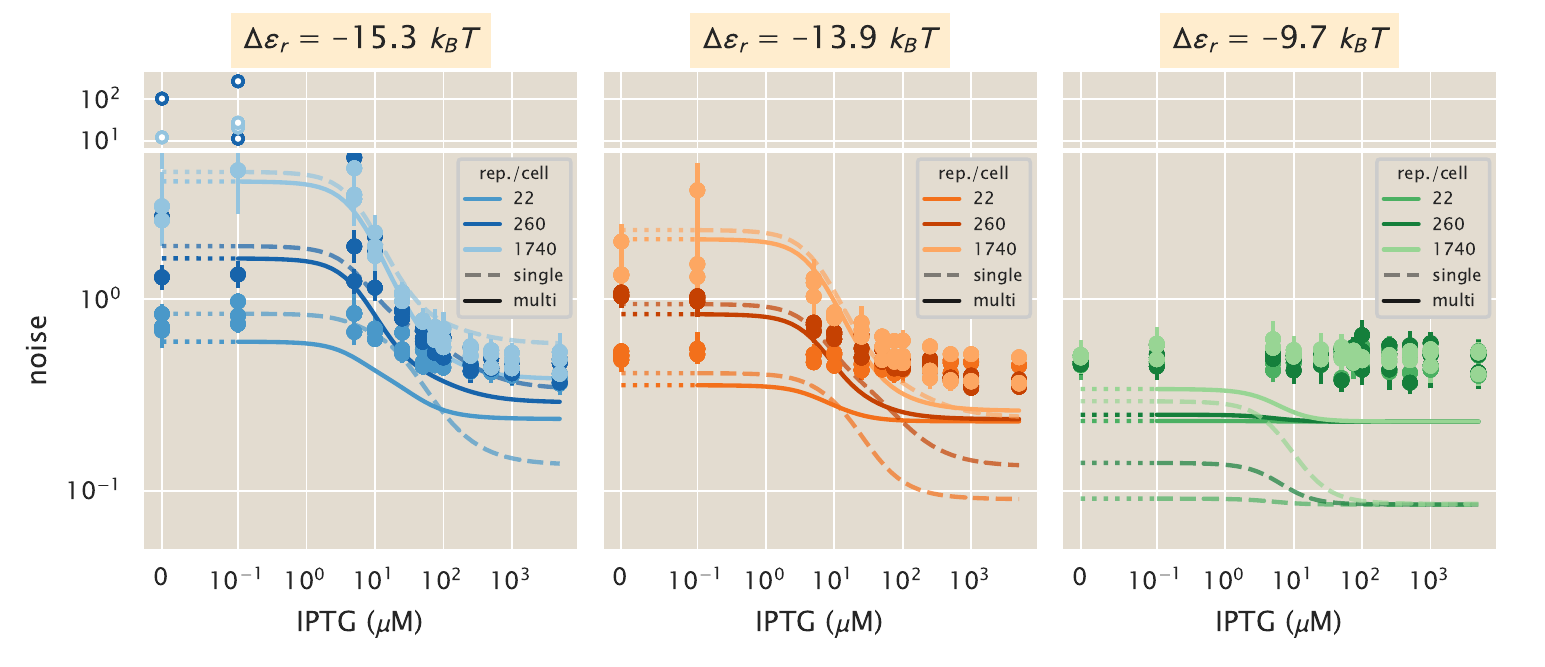}
	\caption{\textbf{Protein noise of the regulated promoter.} Comparison of the
	experimental noise for different operators ((A) O1,  $\eR = -15.3 \; k_BT$,
	(B) O2, $\eR = -13.9 \; k_BT$, (C) O3, $\eR = -9.7 \; k_BT$) with the
	theoretical predictions for the single-promoter (dashed lines) and the
	multi-promoter model (solid lines). Points represent the experimental noise
	as computed from single-cell fluorescence measurements of different {\it E.
	coli} strains under 12 different inducer concentrations. Dotted line
	indicates plot in linear rather than logarithmic scale. Each datum represents
	a single date measurement of the corresponding strain and IPTG concentration
	with $\geq 300$ cells. The points correspond to the median, and the error
	bars correspond to the 95\% confidence interval as determined by 10,000
	bootstrap samples. White-filled dots are plot at a different scale for better
	visualization.}
  \label{sfig_noise_reg}
\end{figure}

\subsubsection{Systematic deviation of the noise in gene expression}

\fref{sfig_noise_delta} and \fref{sfig_noise_reg} highlight that our model
underestimates the cell-to-cell variability as measured by the noise. To
further explore this systematic deviation \fref{sfig_noise_pred_vs_data} shows
the theoretical vs. experimental noise both in linear and log scale. As we can
see the data is systematically above the identity line. The data is colored by
their corresponding experimental fold-change values. The data that has the
largest deviations from the identity line also corresponds to the data with the
largest error bars and the smallest fold-change. This is because measurements
with very small fold-changes correspond to intensities very close to the
autofluorescence background. Therefore minimal changes when computing the noise
are amplified given the ratio of std/mean. In \siref{supp_empirical} we will 
explore empirical ways to improve the agreement between our minimal model and
the experimental data to guide future efforts to improve the minimal.

\begin{figure}[h!]
	\centering \includegraphics
  {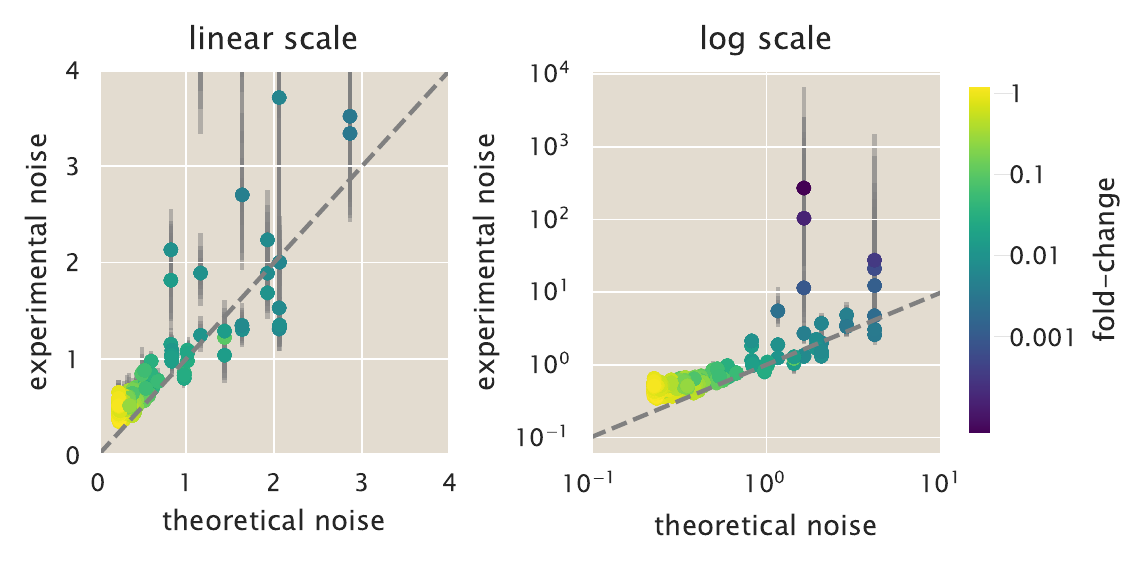}
	\caption{\textbf{Systematic comparison of theoretical vs experimental noise
	in gene expression.} Theoretical vs. experimental noise both in linear
	(left) and log (right) scale. The dashed line shows the identity line of
	slope 1 and intercept zero. All data are colored by the corresponding value
	of the experimental fold-change in gene expression as indicated by the color
	bar. Each datum represents a single date measurement of the corresponding
	strain and IPTG concentration with $\geq 300$ cells. The points correspond
	to the median, and the error bars correspond to the 95\% confidence interval
	as determined by 10,000 bootstrap samples.}
  \label{sfig_noise_pred_vs_data}
\end{figure}

\section{Maximum entropy approximation of distributions}\label{supp_maxent}

(Note: The Python code used for the calculations presented in this section can
be found in the
\href{https://www.rpgroup.caltech.edu//chann_cap/software/MaxEnt_approx_joint.html}{following
link} as an annotated Jupyter notebook)

On the one hand the solution of chemical master equations like the one in
\secref{sec_model} represent a hard mathematical challenge. As presented in
\siref{supp_param_inference} Peccoud and Ycart derived a closed-form solution
for the two-state promoter at the mRNA level \cite{Peccoud1995}. In an
impressive display of mathematical skills, Shahrezaei and Swain were able to
derive an approximate solution for the one- (not considered in this work) and
two-state promoter master equation at the protein level \cite{Shahrezaei2008}.
Nevertheless both of these solutions do not give instantaneous insights about
the distributions as they involve complicated terms such as confluent
hypergeometric functions.

On the other hand there has been a great deal of work to generate methods that
can approximate the solution of these discrete state Markovian models
\cite{Ale2013, Andreychenko2017, Frohlich2016, Schnoerr2017, Smadbeck2013}. In
particular for master equations like the one that concerns us here whose
moments can be easily computed, the moment expansion method provides a simple
method to approximate the full joint distribution of mRNA and protein
\cite{Smadbeck2013}. In this section we will explain the principles behind this
method and show the implementation for our particular case study.

\subsection{The MaxEnt principle}

The principle of maximum entropy (MaxEnt) first proposed by E. T. Jaynes in
1957 tackles the question of given limited information what is the least biased
inference one can make about a particular probability distribution
\cite{Jaynes1957}. In particular Jaynes used this principle to show the
correspondence between statistical mechanics and information theory,
demonstrating, for example, that the Boltzmann distribution is the probability
distribution that maximizes Shannon's entropy subject to a constraint that the
average energy of the system is fixed.

To illustrate the principle let us focus on a univariate distribution $P_X(x)$.
The $n^{\text{th}}$ moment of the distribution for a discrete set of possible
values of $x$ is given by
\begin{equation}
  \ee{x^n} \equiv \sum_x x^n P_X(x).
  \label{eq_mom_ref}
\end{equation}

Now assume that we have knowledge of the first $m$ moments $\bb{\ee{x}}_m = (
\ee{x}, \ee{x^2}, \ldots, \ee{x^m} )$. The question is then how can we use this
information to build an estimator $P_H(x \mid \bb{\ee{x}}_m)$ of the
distribution
such that
\begin{equation}
  \lim_{m \rightarrow \infty} P_H(x \mid \bb{\ee{x}}_m) \rightarrow P_X(x),
\end{equation}
i.e. that the more moments we add to our approximation, the more the estimator
distribution converges to the real distribution.

The MaxEnt principle tells us that our best guess for this estimator is to
build it on the base of maximizing the Shannon entropy, constrained by the
information we have about these $m$ moments. The maximization of Shannon's
entropy guarantees that we are the least committed possible to information that
we do not posses. The Shannon entropy for an univariate discrete distribution
is given by \cite{Shannon1948}
\begin{equation}
  H(x) \equiv - \sum_x P_X(x) \log P_X(x).
\end{equation}

For an optimization problem subject to constraints we make use of the method of
the Lagrange multipliers. For this we define the constraint equation
$\mathcal{L}(x)$ as
\begin{equation}
  \mathcal{L}(x) \equiv H(x) - \sum_{i=0}^m
  \left[ \lambda_i \left( \ee{x^i} - \sum_x x^i P_X(x) \right) \right],
  \label{seq_constraint_eq}
\end{equation}
where $\lambda_i$ is the Lagrange multiplier associated with the $i\th$ moment.
The inclusion of the zeroth moment is an additional constraint to guarantee the
normalization of the resulting distribution. Since $P_X(x)$ has a finite set of
discrete values, when taking the derivative of the constraint equation with
respect to $P_X(x)$, we chose a particular value of $X = x$. Therefore from the
sum over all possible $x$ values only a single term survives. With this in mind
we take the derivative of the constraint equation obtaining
\begin{equation}
  {d\mathcal{L} \over d P_X(x)} = -\log P_X(x) - 1 -
  \sum_{i=0}^m \lambda_i x^i.
\end{equation}

Equating this derivative to zero and solving for the distribution (that we now
start calling $P_H(x)$, our MaxEnt estimator) gives
\begin{equation}
  P_H(x) = \exp \left(- 1 - \sum_{i=0}^m \lambda_i x^i \right)
         ={1 \over \mathcal{Z}}
         \exp \left( - \sum_{i=1}^m \lambda_i x^i \right),
  \label{eq_maxEnt}
\end{equation}
where $\mathcal{Z}$ is the normalization constant that can be obtained by
substituting this solution into the normalization constraint. This results in
\begin{equation}
  \mathcal{Z} \equiv \exp\left( 1 + \lambda_0 \right) =
  \sum_x \exp \left( - \sum_{i=1}^m \lambda_i x^i \right).
\end{equation}

\eref{eq_maxEnt} is the general form of the MaxEnt distribution for a
univariate distribution. The computational challenge then consists in finding
numerical values for the Lagrange multipliers $\{ \lambda_i \}$ such that
$P_H(x)$ satisfies our constraints. In other words, the Lagrange multipliers
weight the contribution of each term in the exponent such that when computing
any of the moments we recover the value of our constraint. Mathematically what
this means is that $P_H(x)$ must satisfy
\begin{equation}
  \sum_x x^n P_H(x) =
  \sum_x {x^n \over \mathcal{Z}}
  \exp \left( - \sum_{i=1}^m \lambda_i x^i \right) = \ee{x^n}.
\end{equation}

As an example of how to apply the MaxEnt principle let us use the classic
problem of a six-face die. If we are only told that after a large number of die
rolls the mean value of the face is $\ee{x} = 4.5$ (note that a fair die has a
mean of $3.5$), what would the least biased guess for the distribution look
like? The MaxEnt principle tells us that our best guess would be of the form
\begin{equation}
  P_H(x) = {1 \over \mathcal{Z}} \exp \left( \lambda x \right).
\end{equation}
Using any numerical minimization package we can easily find the value of the
Lagrange multiplier $\lambda$  that satisfies our constraint.
\fref{fig_maxent_die} shows two two examples of distributions that satisfy the
constraint. Panel (A) shows a distribution consistent with the 4.5 average
where both 4 and 5 are equally likely. Nevertheless in the information we got
about the nature of the die it was never stated that some of the faces were
forbidden. In that sense the distribution is committing to information about
the process that we do not posses. Panel (B) by contrast shows the MaxEnt
distribution that satisfies this constraint. Since this distribution maximizes
Shannon's entropy it is guaranteed to be the least biased distribution given
the available information.

\begin{figure}[h!]
	\centering \includegraphics
  {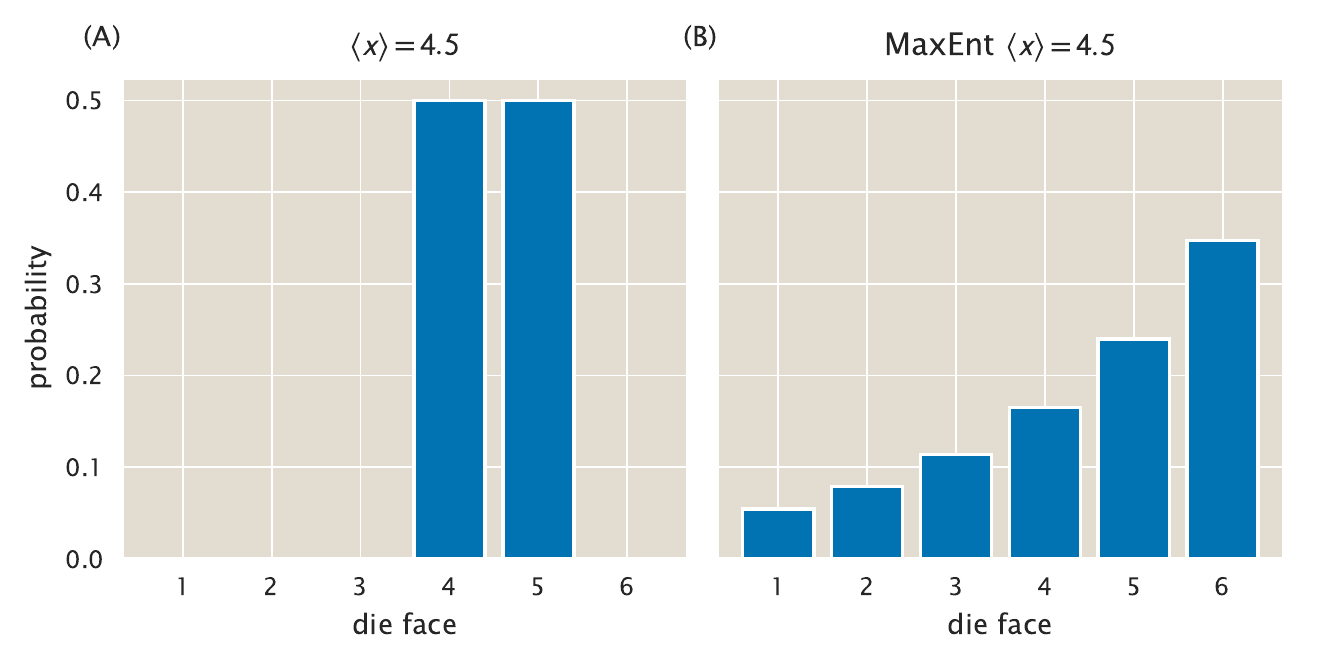}
	\caption{\textbf{Maximum entropy distribution of six-face die.} (A)biased
  distribution consistent with the constraint $\ee{x} = 4.5$. (B) MaxEnt
  distribution also consistent with the constraint.}
  \label{fig_maxent_die}
\end{figure}

\subsubsection{The mRNA and protein joint distribution}

The MaxEnt principle can easily be extended to multivariate distributions. For
our particular case we are interested in the mRNA and protein joint
distribution
$P(m, p)$. The definition of a moment $\ee{m^x p^y}$ is a natural extension of
\eref{eq_mom_ref} of the form
\begin{equation}
  \ee{m^x p^y} = \sum_m \sum_p m^x p^y P(m, p).
\end{equation}

As a consequence the MaxEnt joint distribution $P_H(m, p)$ is of the form
\begin{equation}
  P_H(m, p) = {1 \over \mathcal{Z}}
              \exp \left( - \sum_{(x,y)} \lambda_{(x,y)} m^x p^y \right),
  \label{seq_maxEnt_joint}
\end{equation}
where $\lambda_{(x,y)}$ is the Lagrange multiplier associated with the moment
$\ee{m^x p^y}$, and again $\mathcal{Z}$ is the normalization constant given by
\begin{equation}
  \mathcal{Z} = \sum_m \sum_p
              \exp \left( - \sum_{(x, y)} \lambda_{(x, y)} m^x p^y \right).
\end{equation}
Note that the sum in the exponent is taken over all available $(x, y)$ pairs
that define the moment constraints for the distribution.

\subsection{The Bretthorst rescaling algorithm}

The determination of the Lagrange multipliers suffer from a numerical under and
overflow problem due to the difference in magnitude between the constraints.
This becomes a problem when higher moments are taken into account. The
resulting numerical values for the Lagrange multipliers end up being separated
by several orders of magnitude. For routines such as Newton-Raphson or other
minimization algorithms that can be used to find these Lagrange multipliers
these different scales become problematic.

To get around this problem we implemented a variation to the algorithm due to
G. Larry Bretthorst, E.T. Jaynes' last student. With a very simple argument we
can show that linearly rescaling the constraints, the Lagrange multipliers and
the ``rules'' for how to compute each of the moments, i.e. each of the
individual products that go into the moment calculation, should converge to the
same MaxEnt distribution. In order to see this let's consider again a
univariate distribution $P_X(x)$ that we are trying to reconstruct given the
first two moments $\ee{x}$, and $\ee{x^2}$. The MaxEnt distribution can be
written as
\begin{equation}
  P_H(x) = {1 \over \mathcal{Z}}
  \exp \left(- \lambda_1 x - \lambda_2 x^2 \right) =
  {1 \over \mathcal{Z}}
  \exp \left(- \lambda_1 x \right) \exp \left( - \lambda_2 x^2 \right).
\end{equation}
We can always rescale the terms in any way and obtain the same result. Let's
say that for some reason we want to rescale the quadratic terms by a factor
$a$. We can define a new Lagrange multiplier $\lambda_2' \equiv {\lambda_2
\over a}$ that compensates for the rescaling of the terms, obtaining
\begin{equation}
  P_H(x) = {1 \over \mathcal{Z}}
  \exp \left(- \lambda_1 x \right) \exp \left( - \lambda_2' ax^2 \right).
\end{equation}
Computationally it might be more efficient to find the numerical value of
$\lambda_2'$ rather than $\lambda_2$ maybe because it is of the same order of
magnitude as $\lambda_1$. Then we can always multiply $\lambda_2'$ by $a$ to
obtain back the constraint for our quadratic term. What this means is that that
we can always rescale the MaxEnt problem to make it numerically more stable,
then we can rescale back to obtain the value of the Lagrange multipliers. The
key to the Bretthorst algorithm lies in the selection of what rescaling factor
to choose in order to make the numerical inference more efficient.

Bretthorst's algorithm goes even further by further transforming the
constraints and the variables to make the constraints orthogonal, making the
computation much more effective. We now explain the implementation of the
algorithm for our joint distribution of interest $P(m, p)$.

\subsubsection{Algorithm implementation}

Let the $M \times N$ matrix $\bb{A}$ contain all the factors used to compute
the moments that serve as constraints, where each entry is of the form
\begin{equation}
  A_{ij} = m_i^{x_j} \cdot p_i^{y_j}.
  \label{seq_maxent_rules}
\end{equation}
In other words, recall that to obtain any moment $\ee{m^x p^y}$ we compute
\begin{equation}
  \ee{m^x p^y} = \sum_m \sum_p m^x p^y P(m, x).
\end{equation}
If we have $M$ possible $(m, p)$ pairs in our truncated sample space (because
we can't include the sample space up to infinity) $\{(m, p)_1, (m, p)_2, \ldots
(m, p)_N \}$, and we have $N$ exponent pairs $(x, y)$ corresponding to the $N$
moments used to constraint the maximum entropy distribution $\{(x, y)_1, (x,
y)_2, \ldots, (x, y)_N \}$, then matrix $\bb{A}$ contains all the possible $M$
by $N$ terms of the form described in \eref{seq_maxent_rules}. Let also
$\bb{v}$ be a vector of length $N$ containing all the constraints with each
entry of the form
\begin{equation}
  v_j = \ee{m^{x_j} p^{y_j}},
\end{equation}
i.e. the information that we have about the distribution. That means that the
constraint equation $\mathcal{L}$ to be used for this problem takes the form
\begin{equation}
  \mathcal{L} = -\sum_i P_i \ln P_i + \lambda_0 \left( 1 - \sum_i P_i \right)
  + \sum_{j>0} \lambda_j \left( v_j - \sum_i A_{ij} P_i \right),
\end{equation}
where $\lambda_0$ is the Lagrange multiplier associated with the normalization
constraint, and $\lambda_j$ is the Lagrange multiplier associated with the
$j\th$ constraint. This constraint equation is equivalent to
\eref{seq_constraint_eq}, but now all the details of how to compute the moments
are specified in matrix $\bb{A}$.

With this notation in hand we now proceed to rescale the problem. The first
step consists of rescaling the terms to compute the entries of matrix $\bb{A}$.
As mentioned before, this is the key feature of the Bretthorst algorithm; the
particular choice of rescaling factor used in the algorithm empirically
promotes that the rescaled Lagrange multipliers are of the same order of
magnitude. The rescaling takes the form
\begin{equation}
  A_{ij}' = {A_{ij} \over G_j},
\end{equation}
where $G_j$ serves to rescale the moments, providing numerical stability to the
inference problem. Bretthorst proposes an empirical rescaling that satisfies
\begin{equation}
G_j^2 = \sum_i A_{ij}^2,
\end{equation}
or in terms of our particular problem
\begin{equation}
G_j^2 = \sum_m \sum_p \left( m^{x_j} p^{y_j} \right)^2.
\end{equation}
What this indicates is that each pair $m_i^{x_j} p_i^{y_j}$ is normalized by
the square root of the sum of the all pairs of the same form squared.

Since we rescale the factors involved in computing the constraints, the
constraints must also be rescaled simply as
\begin{equation}
v_j' = \ee{m^{x_j} p^{y_j}}' = {\ee{m^{x_j} p^{y_j}} \over G_j}.
\end{equation}
The Lagrange multipliers must compensate this rescaling since at the end of the
day the probability must add up to the same value. Therefore we rescale the
$\lambda_j$ terms as
\begin{equation}
\lambda_j' = \lambda_j G_j,
\end{equation}
such that any $\lambda_j A_{ij} = \lambda_j' A_{ij}'$. If this empirical value
for the rescaling factor makes the rescaled Lagrange multipliers $\lambda_j'$
be of the same order of magnitude, this by itself would already improve the
algorithm convergence. Bretthorst proposes another linear transformation to
make the optimization routine even more efficient. For this we generate
orthogonal constraints that make Newton-Raphson and similar algorithms converge
faster. The transformation is as follows
\begin{equation}
  A_{ik}'' = \sum_j {e}_{jk} A_{ij}',
\end{equation}
for the entires of matrix $\bb{A}$, and
\begin{equation}
  v_k'' = \sum_j {e}_{jk} u_j',
\end{equation}
for entires of the constraint vector $\bb{v}$, finally
\begin{equation}
  \lambda_k'' = \sum_j {e}_{jk} \beta_j,
\end{equation}
for the Lagrange multipliers. Here ${e}_{jk}$ is the $j\th$ component
of the $k\th$ eigenvector of the matrix $\bb{E}$ with entries
\begin{equation}
  {E}_{kj} = \sum_i {A}_{ik}' {A}_{ij}'.
\end{equation}
This transformation guarantees that the matrix $\bb{A}''$ has the property
\begin{equation}
  \sum_i A_{ij}'' A_{jk}'' = \beta_j \delta_{jk},
\end{equation}
where $\beta_j$ is the $j\th$ eigenvalue of the matrix $\bb{E}$ and
$\delta_{jk}$ is the Kronecker delta function. What this means is that, as
desired, the constraints are orthogonal to each other, improving the algorithm
convergence speed.

\subsection{Predicting distributions for simple repression constructs}

Having explained the theoretical background along with the practical
difficulties and a workaround strategy proposed by Bretthorst, we implemented
the inference using the moments obtained from averaging over the variability
along the cell cycle (See \siref{supp_multi_gene}). \fref{fig_pmf_mRNA} and
\fref{fig_pmf_protein} present these inferences for both mRNA and protein
levels respectively for different values of the repressor-DNA binding energy
and repressor copy numbers per cell. From these plots we can easily appreciate
that despite the fact that the mean of each distribution changes as the
induction level changes, there is a lot of overlap between distributions. This
as a consequence means that at the single-cell level cells cannot perfectly
resolve between different inputs.

\begin{figure}[h!]
	\centering \includegraphics
  {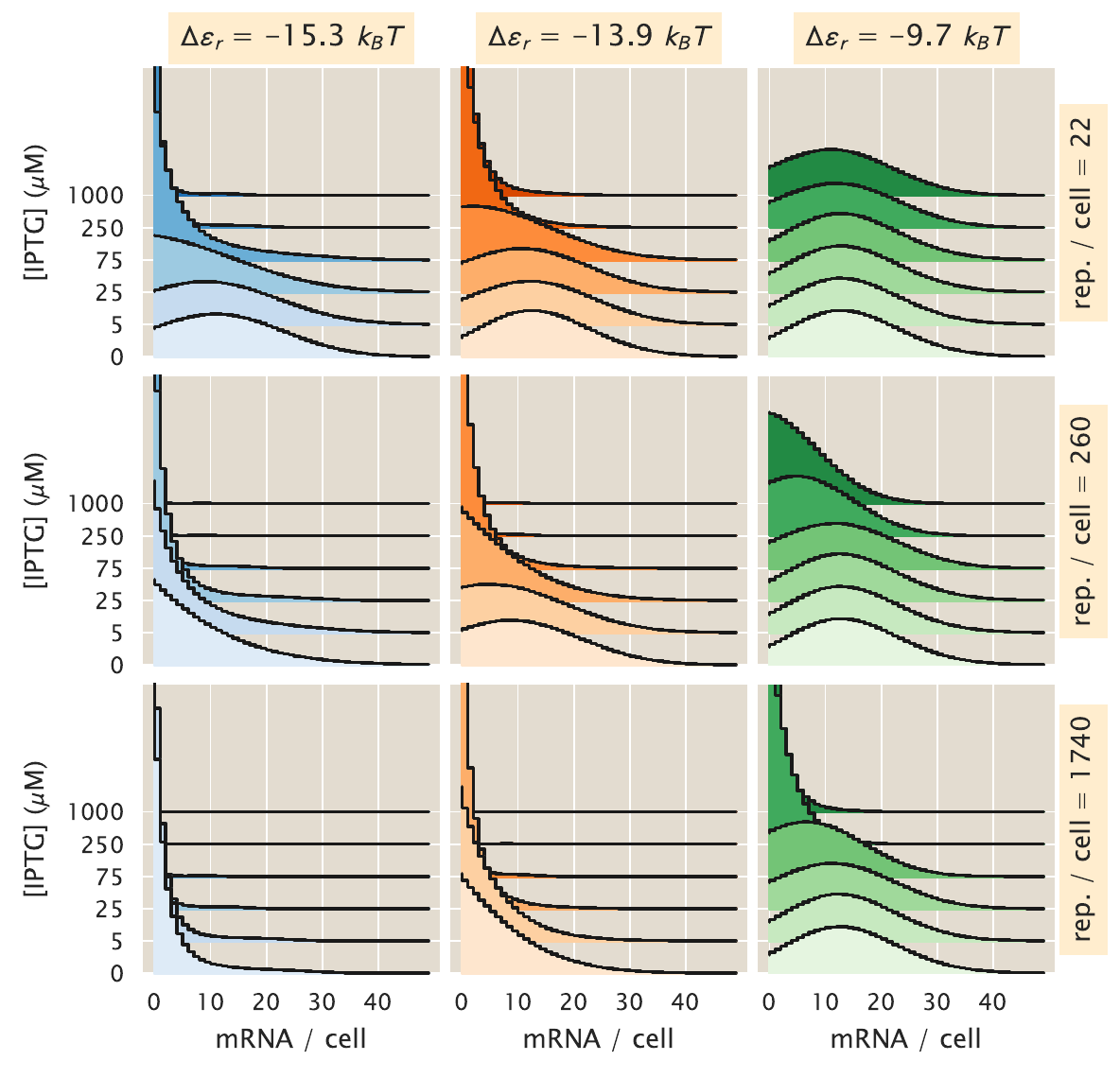}
	\caption{\textbf{Maximum entropy mRNA distributions for simple repression
	constructs.} mRNA distributions for different biophysical parameters. From
	left to right the repressor-DNA affinity decreases as defined by the three
	lacI operators O1 ($-15.3 \; k_BT$), O2 ($ -13.9 \; k_BT$), and O3 ($-9.7 \;
	k_BT$). From top to bottom the mean repressor copy number per cell increases.
	The curves on each plot represent different IPTG concentrations. Each
	distribution was fitted using the first three moments of the mRNA
	distribution.}
  \label{fig_pmf_mRNA}
\end{figure}

\begin{figure}[h!]
	\centering \includegraphics
  {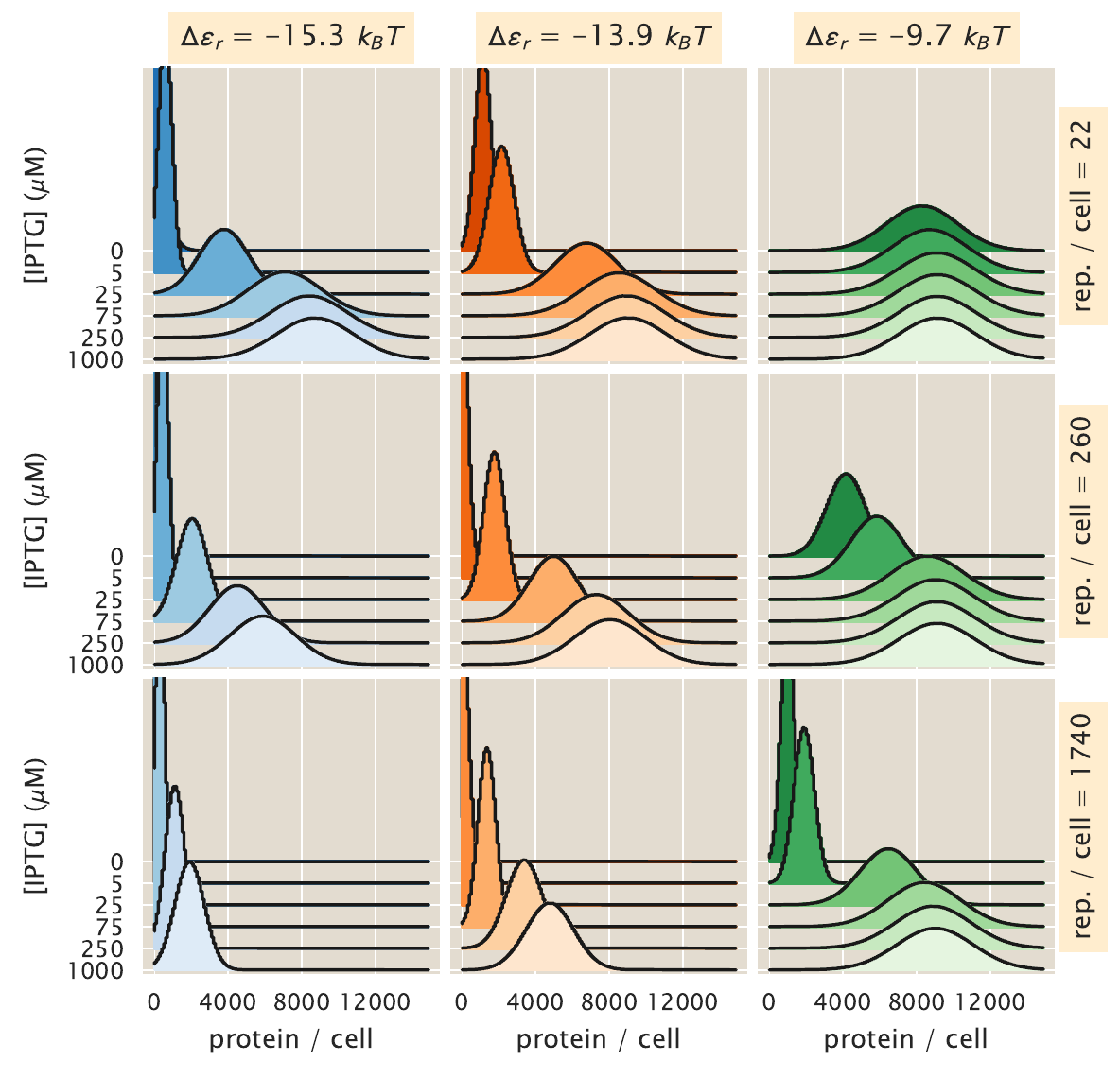}
	\caption{\textbf{Maximum entropy protein distributions for simple repression
	constructs.} Protein distributions for different biophysical parameters. From
	left to right the repressor-DNA affinity decreases as defined by the three
	lacI operators O1 ($-15.3 \; k_BT$), O2 ($ -13.9 \; k_BT$), and O3 ($-9.7 \;
	k_BT$). From top to bottom the mean repressor copy number per cell increases.
	The curves on each plot represent different IPTG concentrations. Each
	distribution was fitted using the first six moments of the protein
	distribution.}
  \label{fig_pmf_protein}
\end{figure}

\subsection{Comparison with experimental data}

Now that we have reconstructed an approximation of the probability distribution
$P(m, p)$ we can compare this with our experimental measurements. But just as
detailed in \siref{supp_theory_vs_data_mom} the single-cell microscopy
measurements are given in arbitrary units of fluorescence. Therefore we cannot
compare directly our predicted protein distributions with these values. To get
around this issue we use the fact that the fold-change in gene expression that
we defined as the ratio of the gene expression level in the presence of the
repressor and the expression level of a knockout strain is a non-dimensional
quantity. Therefore we normalize all of our single-cell measurements by the
mean fluorescence value of the $\Delta lacI$ strain with the proper background
fluorescence subtracted as explained in \siref{supp_theory_vs_data_mom} for the
noise measurements. In the case of the theoretical predictions of the protein
distribution we also normalize each protein value by the predicted mean protein
level $\ee{p}$, having now non-dimensional scales that can be directly
compared. \fref{sfig_cdf_delta} shows the experimental (color curves) and
theoretical (dark dashed line) cumulative distribution functions for the three
$\Delta lacI$ strains. As in \fref{sfig_noise_delta}, we do not expect
differences between the operators, but we explicitly plot them separately to
make sure that this is the case. We can see right away that as we would expect
given the limitations of the model to accurately predict the noise and skewness
of the distribution, the model doesn't accurately predict the data. Our model
predicts a narrower distribution compared to what we measured with single-cell
microscopy. 

\begin{figure}[h!]
	\centering \includegraphics
  {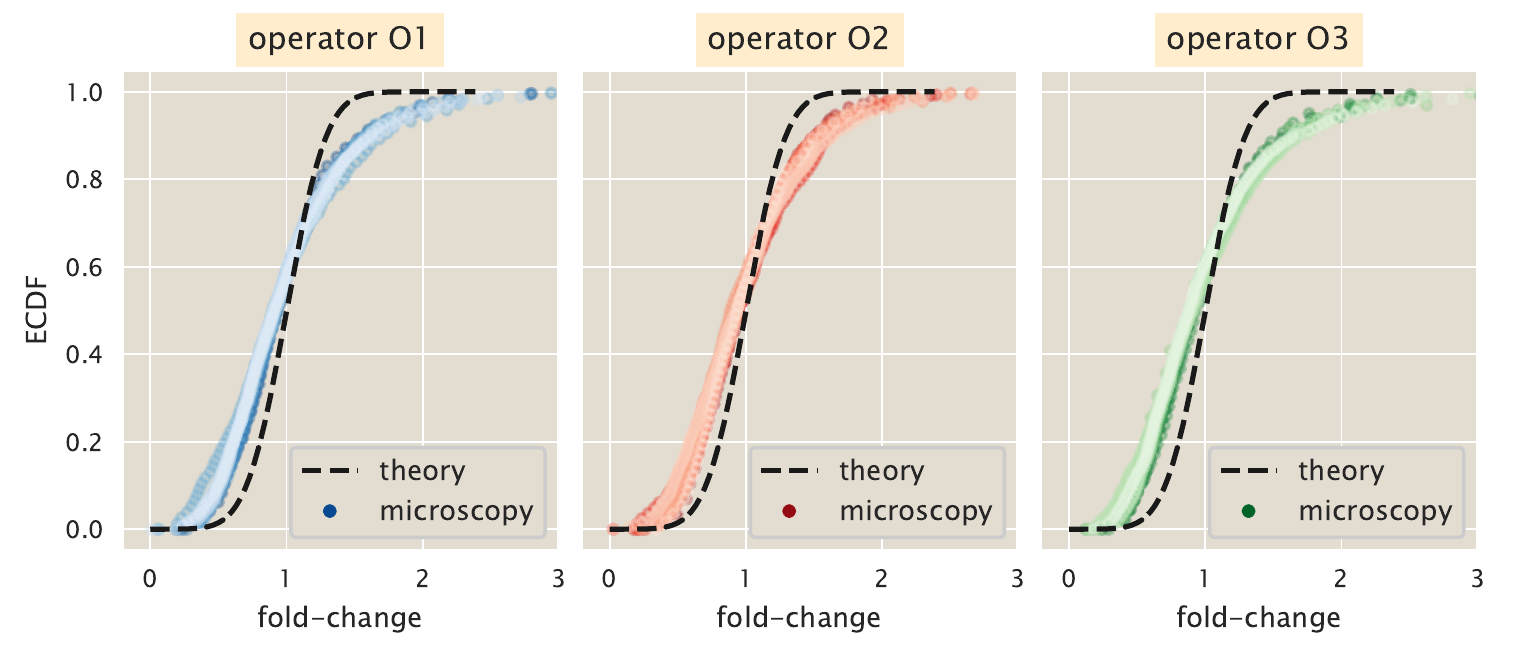}
	\caption{\textbf{Experiment vs. theory comparison for $\Delta lacI$ strain.}
  Example fold-change empirical cumulative distribution functions (ECDF) for
  strains with no repressors and different operators. The color curves
  represent single-cell microscopy measurements while the dashed black lines
  represent the theoretical distributions as reconstructed by the maximum
  entropy principle. The theoretical distributions were fitted using the first
  six moments of the protein distribution.}
  \label{sfig_cdf_delta}
\end{figure}

The same narrower prediction applies to the regulated promoters.
\fref{sfig_cdf_reg}, shows the theory-experiment comparison of the cumulative
distribution functions for different repressor binding sites (different
figures), repressor copy numbers (rows), and inducer concentrations (columns).
In general the predictions are systematically narrower compared to the actual
experimental data.

\begin{figure}[h!]
	\centering \includegraphics
  {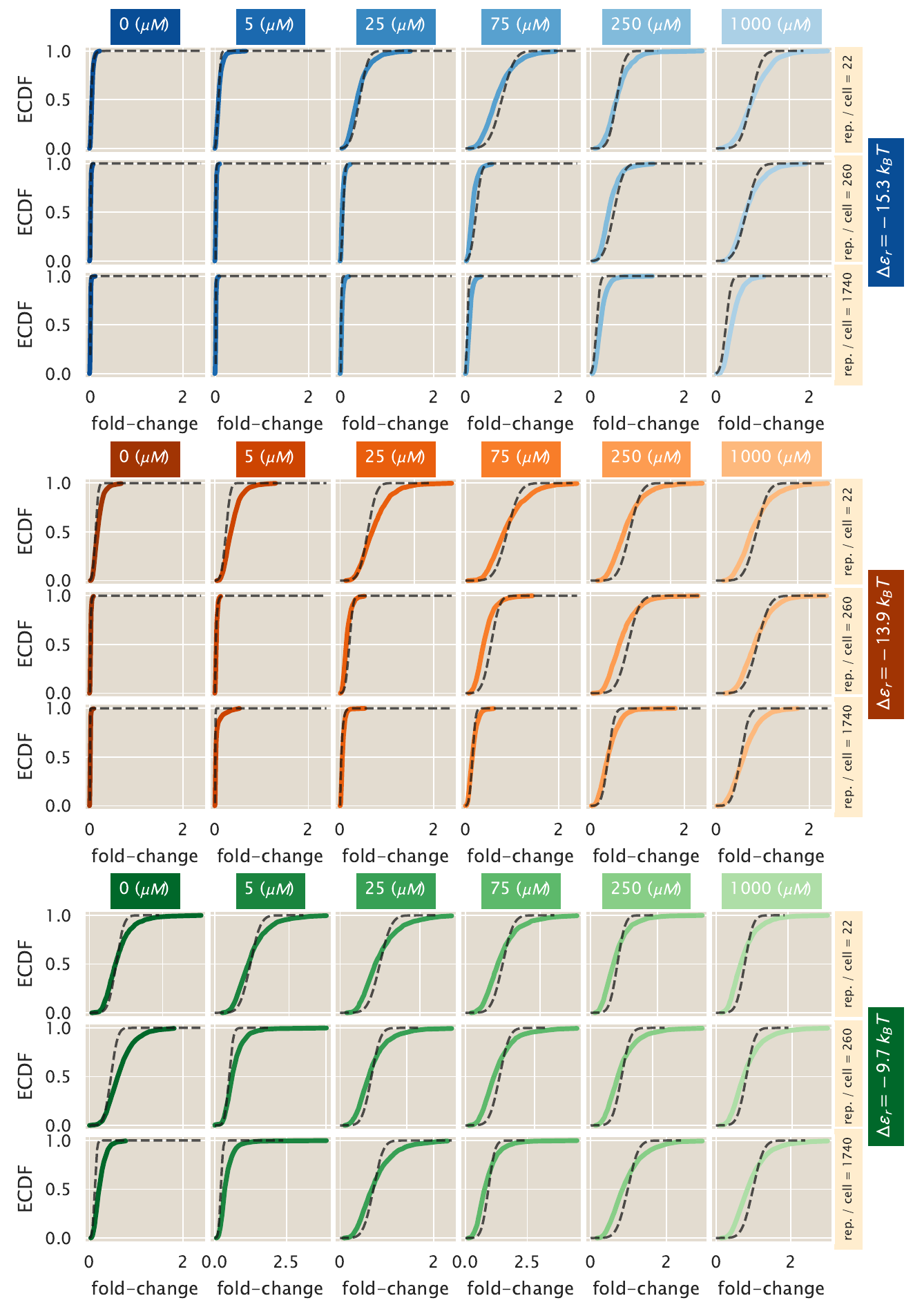}
	\caption{\textbf{Experiment vs. theory comparison for regulated promoters.}
  Example fold-change empirical cumulative distribution functions (ECDF) for
  regulated strains with the three operators (different colors) as a function
  of repressor copy numbers (rows) and inducer concentrations (columns). The
  color curves represent single-cell microscopy measurements while the dashed
  black lines represent the theoretical distributions as reconstructed by the
  maximum entropy principle. The theoretical distributions were fitted using
  the first six moments of the protein distribution}
  \label{sfig_cdf_reg}
\end{figure}

\section{Gillespie simulation of master equation}\label{supp_gillespie}

(Note: The Python code used for the calculations presented in this section can
be found in the
\href{https://www.rpgroup.caltech.edu//chann_cap/software/gillespie_simulation.html}{following
link} as an annotated Jupyter notebook)

So far we have generated a way to compute an approximated form of the joint
distribution of protein and mRNA $P(m, p)$ as a function of the moments of the
distribution $\ee{m^x p^y}$. This is a non-conventional form to work with the
resulting distribution of the master equation. A more conventional approach to
work with master equations whose closed-form solutions are not known or not
computable is to use stochastic simulations commonly known as Gillespie
simulations. To benchmark the performance of our approach based on distribution
moments and maximum entropy we implemented the Gillespie algorithm. Our
implementation as detailed in the corresponding Jupyter notebook makes use of
just-in-time compilation as implemented with the Python package
\href{http://numba.pydata.org}{numba}.

\subsection{mRNA distribution with Gillespie simulations}

To confirm that the implementation of the Gillespie simulation was correct we
perform the simulation at the mRNA level for which the closed-form solution of
the steady-state distribution is known as detailed in
\siref{supp_param_inference}. \fref{sfig_gillespie_mRNA} shows example
trajectories of mRNA counts. Each of these trajectories were computed over
several cell cyles, where the cell division was implemented generating a
binomially distributed random variable that depended on the last mRNA count
before the division event.

\begin{figure}[h!]
	\centering \includegraphics
  {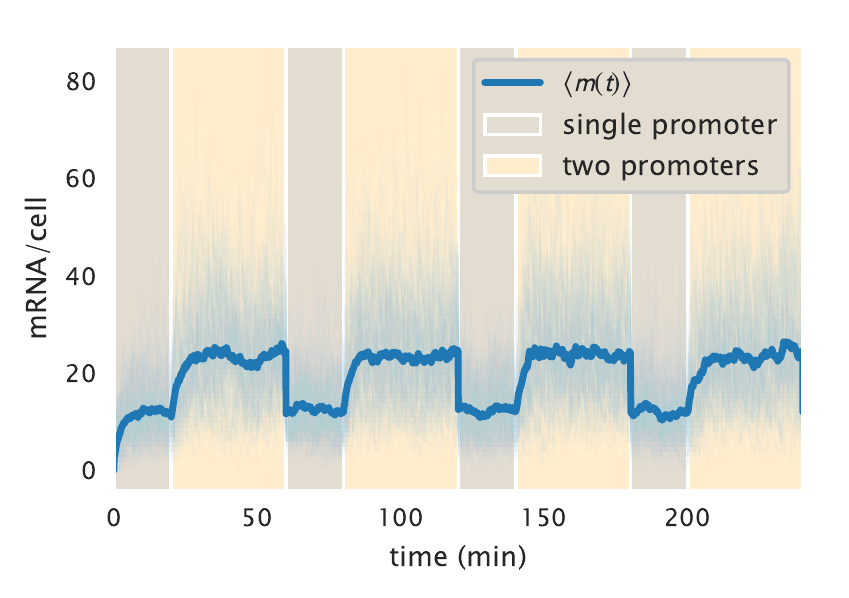}
	\caption{\textbf{Stochastic trajectories of mRNA counts.} 100 stochastic
	trajectories generated with the Gillespie algorithm for mRNA counts over
	time for a two-state unregulated promoter. Cells spend a fraction of the
	cell cycle with a single copy of the promoter (light brown) and the rest of
	the cell cycle with two copies (light yellow). When trajectories reach a
	new cell cycle, the mRNA counts undergo a binomial partitioning to simulate
	the cell division.}
  \label{sfig_gillespie_mRNA}
\end{figure}

To check the implementation of our stochastic algorithm we generated several of
these stochastic trajectories in order to reconstruct the mRNA steady-state
distribution. These reconstructed distributions for a single- and double-copy
of the promoter can be compared with \eref{seq_two_state_mRNA} - the
steady-state distribution for the two-state promoter.
\fref{sfig_gillespie_mRNA_dist} shows the great agreement between the
stochastic simulation and the analytical result, confirming that our
implementation of the Gillespie simulation is correct.

\begin{figure}[h!]
	\centering \includegraphics
  {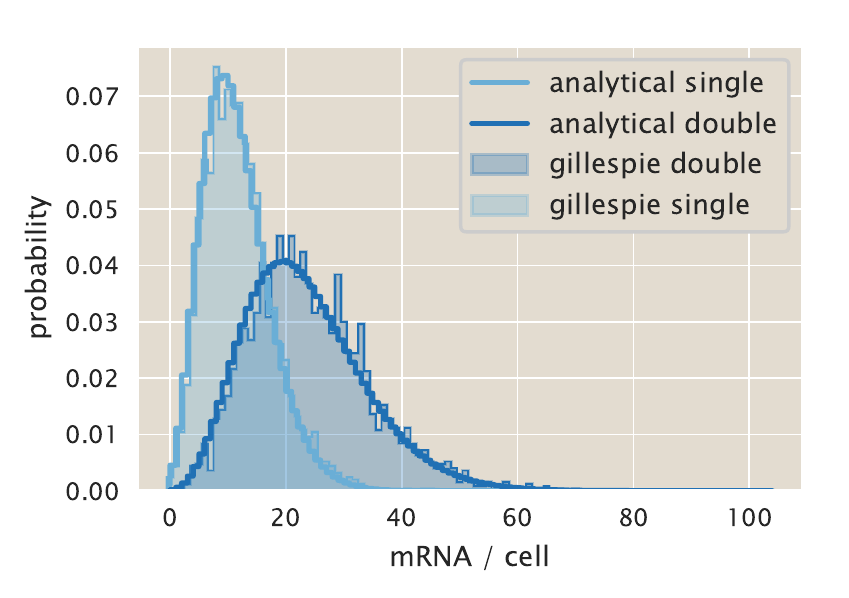}
	\caption{\textbf{Comparison of analytical and simulated mRNA distribution.}
	Solid lines show the steady-state mRNA distributions for one copy (light
	blue) and two copies of the promoter (dark blue) as defined by
    \eref{seq_two_state_mRNA}. Shaded regions represent the corresponding 
    distribution obtained using 2500 stochastic mRNA trajectories and taking 
    the last cell-cyle to approximate the distribution.}
  \label{sfig_gillespie_mRNA_dist}
\end{figure}

\subsection{Protein distribution with Gillespie simulations}

Having confirmed that our implementation of the Gillespie algorithm that
includes the binomial partitioning of molecules reproduces analytical results
we extended the implementation to include protein counts.
\fref{sfig_gillespie_proteins} shows representative trajectories for both mRNA
and protein counts over several cell cycles. Specially for the protein we can
see that it takes several cell cycles for counts to converge to the dynamical
steady-state observed with the deterministic moment equations. Once this
steady-state is reached, the ensemble of trajectories between cell cycles look
very similar.

\begin{figure}[h!]
	\centering \includegraphics
  {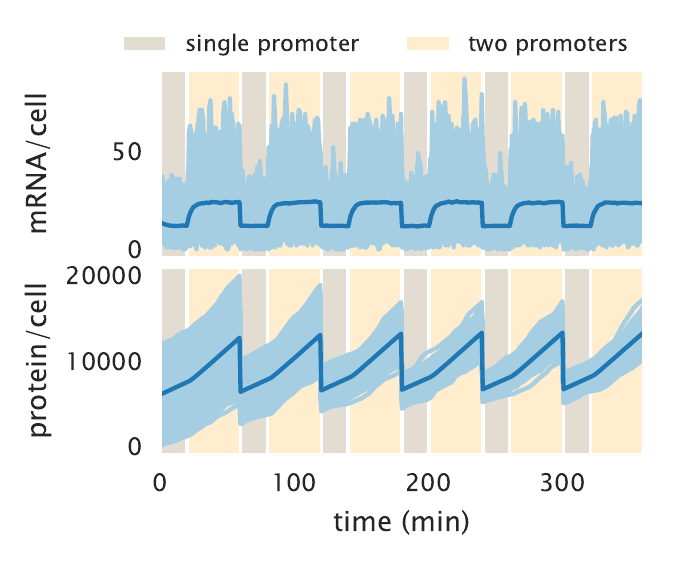}
	\caption{\textbf{Stochastic trajectories of mRNA and protein counts.} 2500
	protein counts over time for a two-state unregulated promoter. Cells spend
	a fraction of the cell cycle with a single copy of the promoter (light
	brown) and the rest of the cell cycle with two copies (light yellow). When
	trajectories reach a new cell cycle, the molecule counts undergo a binomial
	partitioning to simulate the cell division.}
  \label{sfig_gillespie_proteins}
\end{figure}

From these trajectories we can compute the protein steady-state distribution,
taking into account the cell-age distribution as detailed in
\siref{supp_maxent}. \fref{sfig_gillespie_proteins_dist} shows the comparison
between this distribution and the one generated using the maximum entropy
algorithm. Despite the notorious differences between the distributions, the
Gillespie simulation and the maximum entropy results are indistinguishable in
terms of the mean, variance, and skewness of the distribution. We remind the
reader that the maximum entropy is an approximation of the distribution that
gets better the more moments we add. We therefore claim that the approximation
works sufficiently well for our purpose. The enormous advantage of the maximum
entropy approach comes from the computation time. for the number of
distributions that were needed for our calculations the Gillespie algorithm
proved to be a very inefficient method given the large sample space. Our
maximum entropy approach reduces the computation time by several orders of
magnitude, allowing us to extensively explore different parameters of the
regulatory model.

\begin{figure}[h!]
	\centering \includegraphics
  {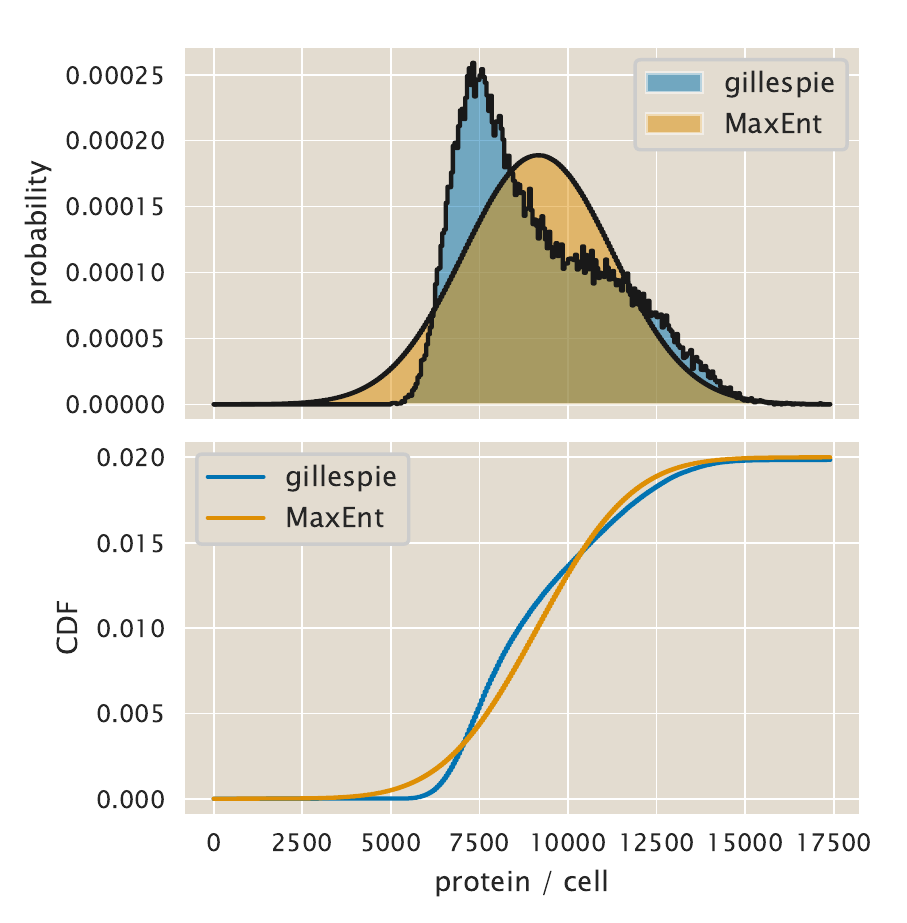}
	\caption{\textbf{Comparison of protein distributions.} Comparison of the
	protein distribution generated with Gillespie stochastic simulations (blue
	curve) and the maximum entropy approach presented in \siref{supp_maxent}
	(orange curve). The upper panel shows the probability mass function. The
	lower panel compares the cumulative distribution functions.}
  \label{sfig_gillespie_proteins_dist}
\end{figure}

\section{Computational determination of the channel capacity}
\label{supp_channcap}

(Note: The Python code used for the calculations presented in this section can
be found in the
\href{https://www.rpgroup.caltech.edu//chann_cap/software/blahut_algorithm_channel_capacity.html}{following
link} as an annotated Jupyter notebook)

In this section we detail the computation of the channel capacity of the simple
genetic circuit shown in \fref{fig5_channcap}. As detailed in
\secref{sec_channcap} the channel capacity is defined as the mutual information
between input $c$ and output $p$ maximized over all possible input
distributions $P(c)$ \cite{Shannon1948}. In principle there are an infinite
number of input distributions, so the task of finding $\hat{P}(c)$, the input
distribution at channel capacity, requires an algorithmic approach that
guarantees the convergence to this distribution. Tkačik, Callan and Bialek
developed a clever analytical approximation to find the $\hat{P}(c)$
distribution \cite{Tkacik2008a}. The validity of their so-called small noise
approximation requires the standard deviation of the output distribution $P(p
\mid c)$ to be much smaller than the domain of the distribution. For our
particular case such condition is not satisfied given the spread of the
inferred protein distributions shown in \fref{fig4_maxent}.

Fortunately there exists a numerical algorithm to approximate $\hat{P}(c)$ for
discrete distributions. In 1972 Richard Blahut and Suguru Arimoto independently
came up with an algorithm mathematically shown to converge to $\hat{P}(c)$
\cite{Blahut1972}. To compute both the theoretical and the experimental channel
capacity shown in \fref{fig5_channcap}, we implemented Blahut's algorithm. In
the following section we detail the definitions needed for the algorithm. Then
we detail how to compute the experimental channel capacity when the bins of the
distribution are not clear given the intrinsic arbitrary nature of microscopy
fluorescence measurements.

\subsection{Blahut's algorithm}

Following \cite{Blahut1972} we implemented the algorithm to compute the channel
capacity. We define $\bb{p_c}$ to be an array containing the probability of
each of the input inducer concentrations (twelve concentrations, See Methods).
Each entry $j$ of the array is then of the form
\begin{equation}
  p_c^{(j)} = P(c = c_j),
\end{equation}
with $j \in \{1, 2, \ldots, 12 \}$. The objective of the algorithm is to find
the entries $p_c^{(j)}$ that maximize the mutual information between inputs and
outputs. We also define $\bb{Q}$ to be a $\vert \bb{p_c} \vert$ by
$\vert \bb{p_{p \mid c}} \vert$ matrix, where $\vert \cdot \vert$ specifies the
length of the array, and $\bb{p_{p \mid c}}$ is an array containing the
probability distribution of an output given a specific value of the input. In
other words, the matrix $\bb{Q}$ recollects all of the individual output
distribution arrays $\bb{p_{p \mid c}}$ into a single object. Then each entry
of the matrix $\bb{Q}$ is of the form
\begin{equation}
  Q^{(i, j)} = P(p = p_i \mid c = c_j).
\end{equation}

For the case of the theoretical predictions of the channel capacity (Solid
lines in \fref{fig5_channcap}) the entries of matrix $\bb{Q}$ are given by the
inferred maximum entropy distributions as shown in \fref{fig4_maxent}. In the
next section we will discuss how to define this matrix for the case of the
single-cell fluorescence measurements. Having defined these matrices we proceed
to implement the algorithm shown in Figure 1 of \cite{Blahut1972}.

\subsection{Channel capacity from arbitrary units of fluorescence}

A difficulty when computing the channel capacity between inputs and outputs
from experimental data is that ideally we would like to compute
\begin{equation}
C(g; c) \equiv \sup_{P(c)} I(g; c),
\end{equation}
where $g$ is the gene expression level, and $c$ is the inducer concentration.
But in reality we are computing
\begin{equation}
C(f(g); c) \equiv \sup_{P(c)} I(f(g); c),
\end{equation}
where $f(g)$ is a function of gene expression that has to do with our mapping
from the YFP copy number to some arbitrary fluorescent value as computed from
the images taken with the microscope. The data processing inequality, as
derived by Shannon himself, tells us that for a Markov chain of the form $c
\rightarrow
g \rightarrow f(g)$ it must be true that \cite{Shannon1948}
\begin{equation}
I(g; c) \geq I(f(g); c),
\end{equation}
meaning that information can only be lost when mapping from the real
relationship between gene expression and inducer concentration to a
fluorescence value.

On top of that, given the limited number of samples that we have access to when
computing the channel capacity, there is a bias in our estimate given this
undersampling. The definition of accurate unbiased descriptors of the mutual
information is still an area of active research. For our purposes we will use
the method described in \cite{Cheong2011a}. The basic idea of the method is to
write the mutual information as a series expansion in terms of inverse powers
of the sample size, i.e.
\begin{equation}
I_{\text{biased}} = I_\infty + \frac{a_1}{N} + \frac{a_2}{N^2} + \cdots,
\end{equation}
where $I_{\text{biased}}$ is the biased estimate of the mutual information as
computed from experimental data, $I_\infty$ is the quantity we would like to
estimate, being the unbiased mutual information when having access to infinity
number of experimental samples, and the coefficients $a_i$ depend on the
underlying distribution of the signal and the response. This is an empirical
choice to be tested. Intuitively this choice satisfies the limit that as the
number of samples from the distribution grows, the empirical estimate of the
mutual information $I_{\text{biased}}$ should get closer to the actual value
$I_\infty$.

In principle for a good number of data points the terms of higher order become
negligible. So we can write the mutual information as
\begin{equation}
I_{\text{biased}} \approx I_\infty + \frac{a_1}{N} + \mathcal{O}(N^{-2}).
\label{seq_mutual_biased}
\end{equation}
This means that if this particular arbitrary choice of functional form is  a
good approximation, when computing the mutual information for varying number of
samples - by taking subsamples of the experimental data - we expect to find a
linear relationship as a function of the inverse of these number of data
points. From this linear relationship the intercept is a bias-corrected
estimate of the mutual information. We can therefore bootstrap the data by
taking different sample sizes and then use the Blahut-Arimoto algorithm we
implemented earlier to estimate the biased channel capacity. We can then fit a
line and extrapolate for when $1/N = 0$ which corresponds to our unbiased
estimate of the channel capacity.

Let's go through each of the steps to illustrate the method.
\fref{sfig_fluor_dist} show a typical data set for a strain with an O2 binding
site ($\eR = -13.9 \; k_BT$) and $R = 260$ repressors per cell. Each of the
distributions in arbitrary units is binned into a specified number of bins to
build matrix $\bb{Q}$.

\begin{figure}[h!]
	\centering \includegraphics
  {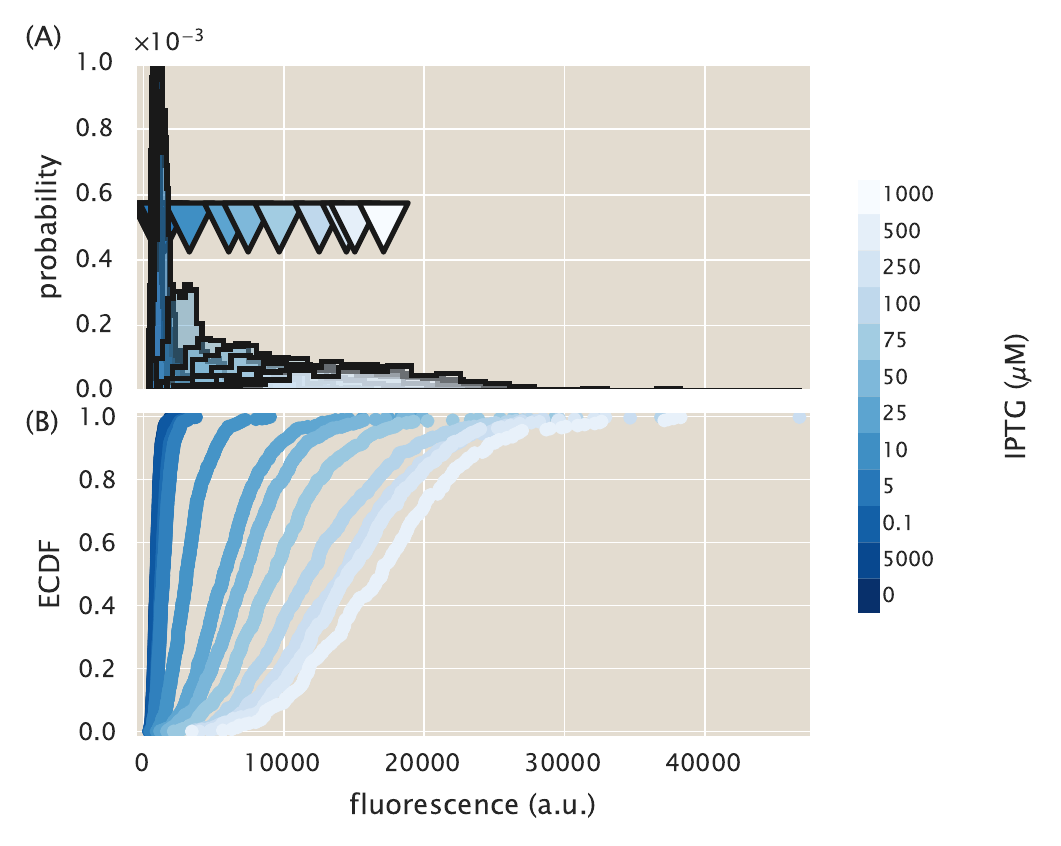}
	\caption{\textbf{Single cell fluorescence distributions for different inducer
	concentrations.} Fluorescence distribution histogram (A) and cumulative
	distribution function (B) for a strain with 260 repressors per cell and a
	binding site with binding energy $\eR = -13.9\; k_BT$. The different curves
	show the single cell fluorescence distributions under the 12 different IPTG
	concentrations used throughout this work. The triangles in (A) show the mean
	of each of the distributions.}
  \label{sfig_fluor_dist}
\end{figure}

Given a specific number of bins used to construct $\bb{Q}$, we subsample a
fraction of the data and compute the channel capacity for such matrix using the
Blahut-Arimoto algorithm. \fref{sfig_channcap_bootstrap} shows an example where
50\% of the data on each distribution from \fref{sfig_fluor_dist} was sampled
and binned into 100 equal bins. The counts on each of these bins are then
normalized and used to build matrix $\bb{Q}$ that is then fed to the
Blahut-Arimoto algorithm. We can see that for these 200 bootstrap samples the
channel capacity varies by $\approx$ 0.1 bits. Not a significant variability,
nevertheless  we consider that it is important to bootstrap the data multiple
times to get a better estimate of the channel capacity.

\begin{figure}[h!]
	\centering \includegraphics
  {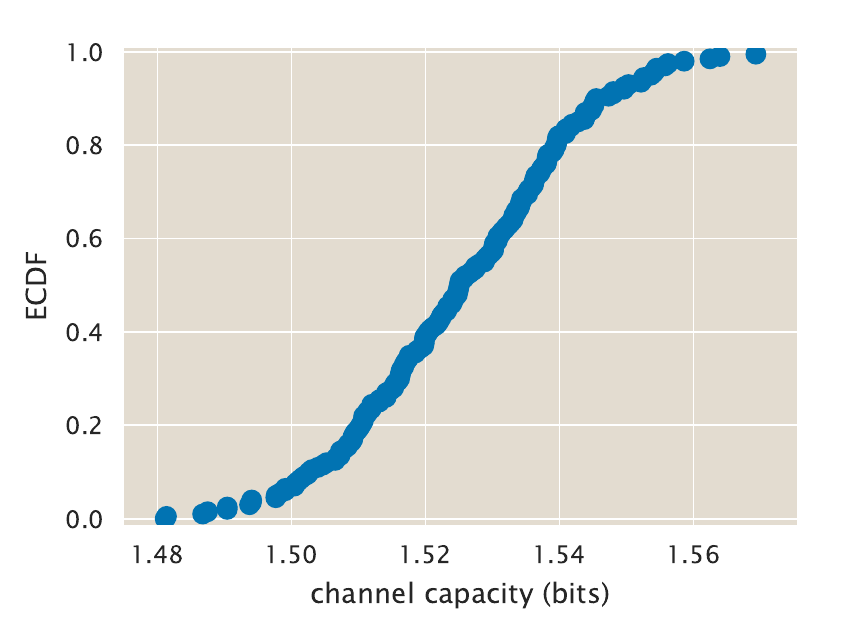}
	\caption{\textbf{Channel capacity bootstrap for experimental data.}
	Cumulative distribution function of the resulting channel capacity estimates
	obtained by subsampling 200 times 50\% of each distribution shown in
	\fref{sfig_fluor_dist}, binning it into 100 bins, and feeding the resulting
	$\bb{Q}$ matrix to  the Blahut-Arimoto algorithm.}
  \label{sfig_channcap_bootstrap}
\end{figure}

\eref{seq_mutual_biased} tells us that if we subsample each of the
distributions from \fref{sfig_fluor_dist} at different fractions, and plot them
as a function of the inverse sample size we will find a linear relationship if
the expansion of the mutual information is valid. To test this idea we repeated
the bootstrap estimate of \fref{sfig_channcap_bootstrap} sampling 10\%, 20\%,
and so on until taking 100\% of the data. We repeated this for different number
of bins since a priori for arbitrary units of fluorescence we do not have a way
to select the optimal number of bins. \fref{sfig_channcap_lin_reg} shows the
result of these estimates. We can see that the linear relationship proposed in
\eref{seq_mutual_biased} holds true for all number of bins selected. We also
note that the value of the intercept of the linear regression varies depending
on the number of bins.

\begin{figure}[h!]
	\centering \includegraphics
  {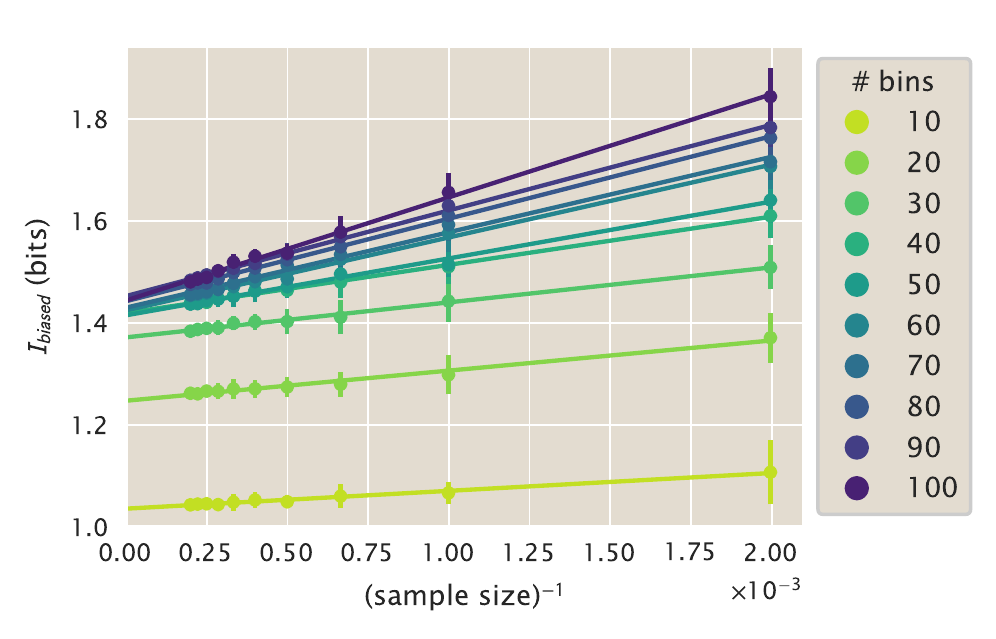}
	\caption{\textbf{Inverse sample size vs channel capacity.} As indicated in
	\eref{seq_mutual_biased} if the channel capacity obtained for different
	subsample sizes of the data is plotted against the inverse sample size there
	must exist a linear relationship between these variables. Here we perform 15
	bootstrap samples of the data from \fref{sfig_fluor_dist}, bin these samples
	using  different number of bins, and perform a linear regression (solid
	lines) between the bootstrap channel capacity estimates, and the inverse
	sample size.}
  \label{sfig_channcap_lin_reg}
\end{figure}

To address the variability in the estimates of the unbiased channel capacity
$I_\infty$ we again follow the methodology suggested in \cite{Cheong2011a}. We
perform the data subsampling and computation of the channel capacity for a
varying number of bins. As a control we perform the same procedure with
shuffled data, where the structure that connects the fluorescence distribution
to the inducer concentration input is lost. The expectation is that this
control should give a channel capacity of zero if the data is not
``over-binned.'' Once the number of bins is too high, we would expect some
structure to emerge in the data that would cause the Blahut-Arimoto algorithm
to return non-zero channel capacity estimates.

\fref{sfig_bins_channcap} shows the result of the unbiased channel capacity
estimates obtained for the data shown in \fref{sfig_fluor_dist}. For the blue
curve we can distinguish three phases:
\begin{enumerate}
  \item A rapid increment from 0 bits to about 1.5 bits as the number of bins
  increases.
  \item A flat region between $\approx$ 50 and 1000 bins.
  \item A second rapid increment for large number of bins.
\end{enumerate}
We can see that the randomized data presents two phases only:
\begin{enumerate}
  \item A flat region where there is, as expected no information being
  processed since the structure of the data was lost when the data was
  shuffled.
  \item A region with fast growth of the channel capacity as the over-binning
  generates separated peaks on the distribution, making it look like there is
  structure in the data.
\end{enumerate}
We take the flat region of the experimental data ($\approx$ 100 bins) to be our
best unbiased estimate of the channel capacity from this experimental dataset.

\begin{figure}[h!]
	\centering \includegraphics
  {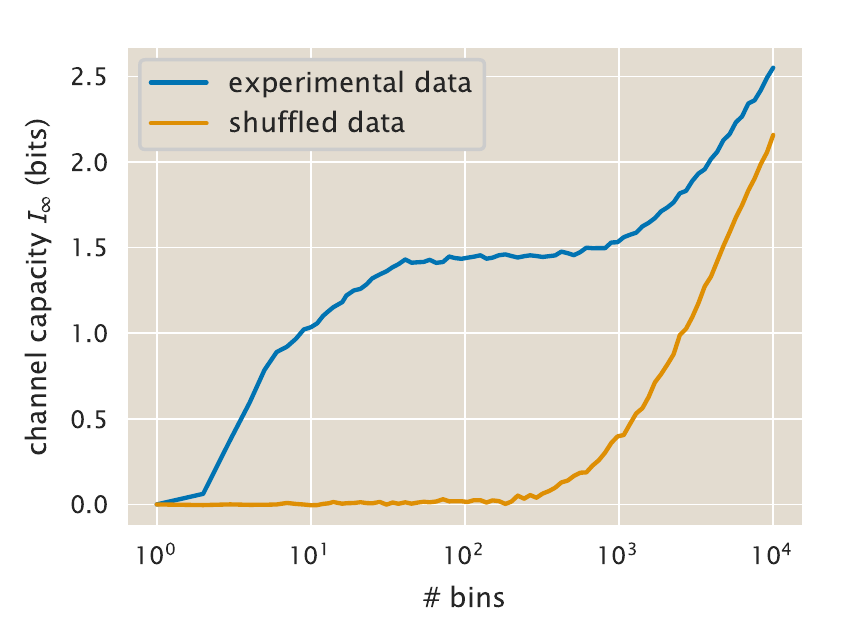}
  \caption{\textbf{Channel capacity as a function of the number of bins.}
  Unbiased channel capacity estimates obtained from linear regressions as in
  \fref{sfig_channcap_lin_reg}. The blue curve show the estimates obtained from
  the data shown in \fref{sfig_fluor_dist}. The orange curve is generated from
  estimates where the same data is shuffled, loosing the relationship between
  fluorescence distributions and inducer concentration.}
  \label{sfig_bins_channcap}
\end{figure}

\subsection{Assumptions involved in the computation of the channel capacity}

An interesting suggestion by Professor Gasper Tkacik was to dissect the
different physical assumptions that went into the construction of the
input-output function $P(p \mid c)$, and their relevance when comparing the
theoretical channel capacities with the experimental inferences. In what
follows we describe the relevance of four important aspects that all affect the
predictions of the information processing capacity.

\paragraph{(i) Cell cycle variability.}
We think that the inclusion of the gene copy number variability during the cell
cycle and the non-Poissoninan protein degradation is a key component to our
estimation of the input-output functions and as a consequence of the channel
capacity. This variability in gene copy number is an additional source of noise
that systematically decreases the ability of the system to resolve different
inputs. The absence of the effects that the gene copy number variability and
the protein partition has on the information processing capacity leads to an
overestimate of the channel capacity as shown in \fref{sfig_channcap_compare}.
Only when these noise sources are included in our inferences is that we get to
capture the experimental channel capacities with no further fit parameters.

\begin{figure}[h!]
	\centering \includegraphics
  {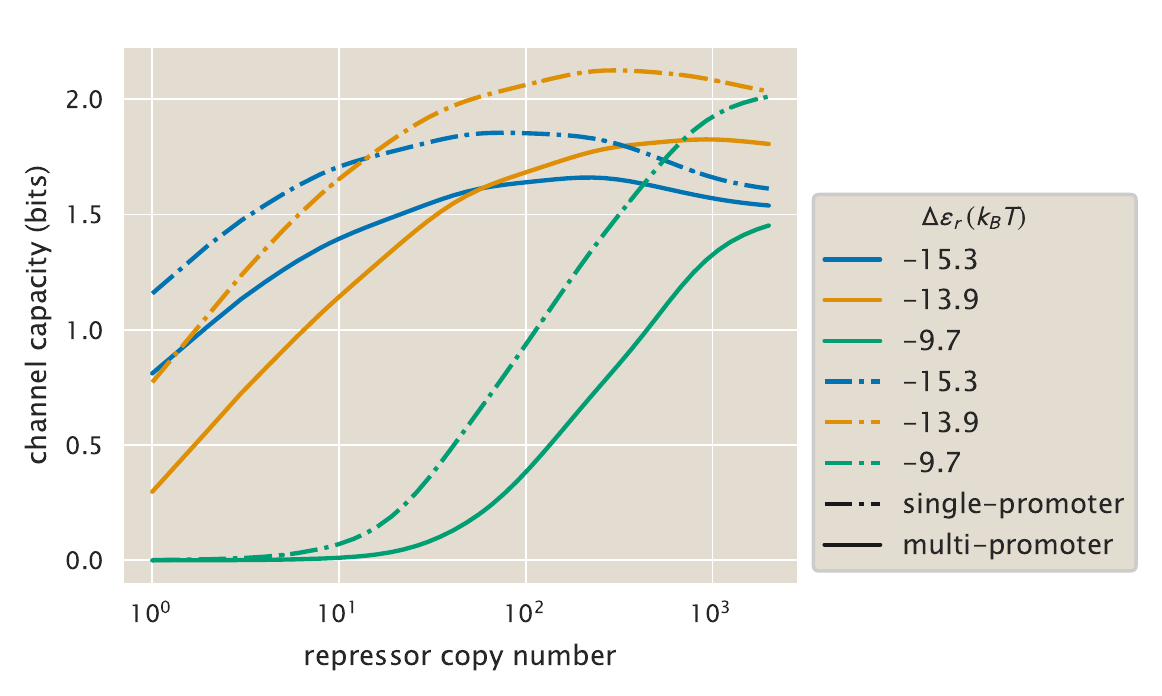}
  \caption{\textbf{Comparison of channel capacity predictions for single- and
  multi-promoter models.} Channel capacity for the multi-promoter model (solid
  lines) vs. the single-promoter steady state model (dot-dashed lines) as a
  function of repressor copy numbers for different repressor-DNA binding
  energies. The single-promoter model assumes Poissonian protein degradation
  ($\gp > 0$) and steady state, while the multi-promoter model accounts for
  gene copy number variability and during the cell cycle and has protein
  degradation as an effect due to dilution as cells grow and divide.}
  \label{sfig_channcap_compare}
\end{figure}

\paragraph{(ii) Non-Gaussian noise distributions.}
For the construction of the probability distributions used in the main text
(\fref{fig4_maxent}) we utilized the first 6 moments of the protein
distribution. The maximum entropy formalism tells us that the more constraints
we include in the inference, the closer the maximum entropy distribution will
be to the real distribution. But \textit{a priori} there is no way of knowing
how many moments should be included in order to capture the essence of the
distribution. In principle two moments could suffice to describe the entire
distribution as happens with the Gaussian distribution. To compare the effect
that including more or less constraints on the maximum entropy inference we
constructed maximum entropy distributions using an increasing number of moments
from 2 to 6. We then computed the Kullback-Leibler divergence $D_{KL}$ of the
form
\begin{equation}
	D_{KL}(P_6(p \mid c) || P_i(p \mid c)) =
	\sum_p P_6(p \mid c) \log_2 {P_6(p \mid c) \over P_i(p \mid c)},
\end{equation}
where $P_i(p \mid c)$ is the maximum entropy distribution constructed with the
first $i$ moments, $i \in \{2, 3, 4, 5, 6\}$. Since the Kullback-Leibler
divergence $D_{KL}(P || Q)$ can be interpreted as the amount of information
lost by assuming the incorrect distribution $Q$ when the correct distribution
is $P$, we used this metric as a way of how much information we would have lost
by using less constraints compared to the six moments used in the main text.

\fref{sfig_dkl} shows this comparison for different operators and repressor
copy numbers. We can see from here that using less moments as constraints gives
basically the same result. This is because most of the values of the
Kullback-Leibler divergence are significantly smaller than 0.1 bits, and the
entropy of these distributions is in general $> 10$ bits, so we would lose less
than 1\% of the information contained in these distributions by utilizing only
two moments as constraints. Therefore the use of non-Gaussian noise is not an
important feature for our inferences.

\begin{figure}[h!]
	\centering \includegraphics
  {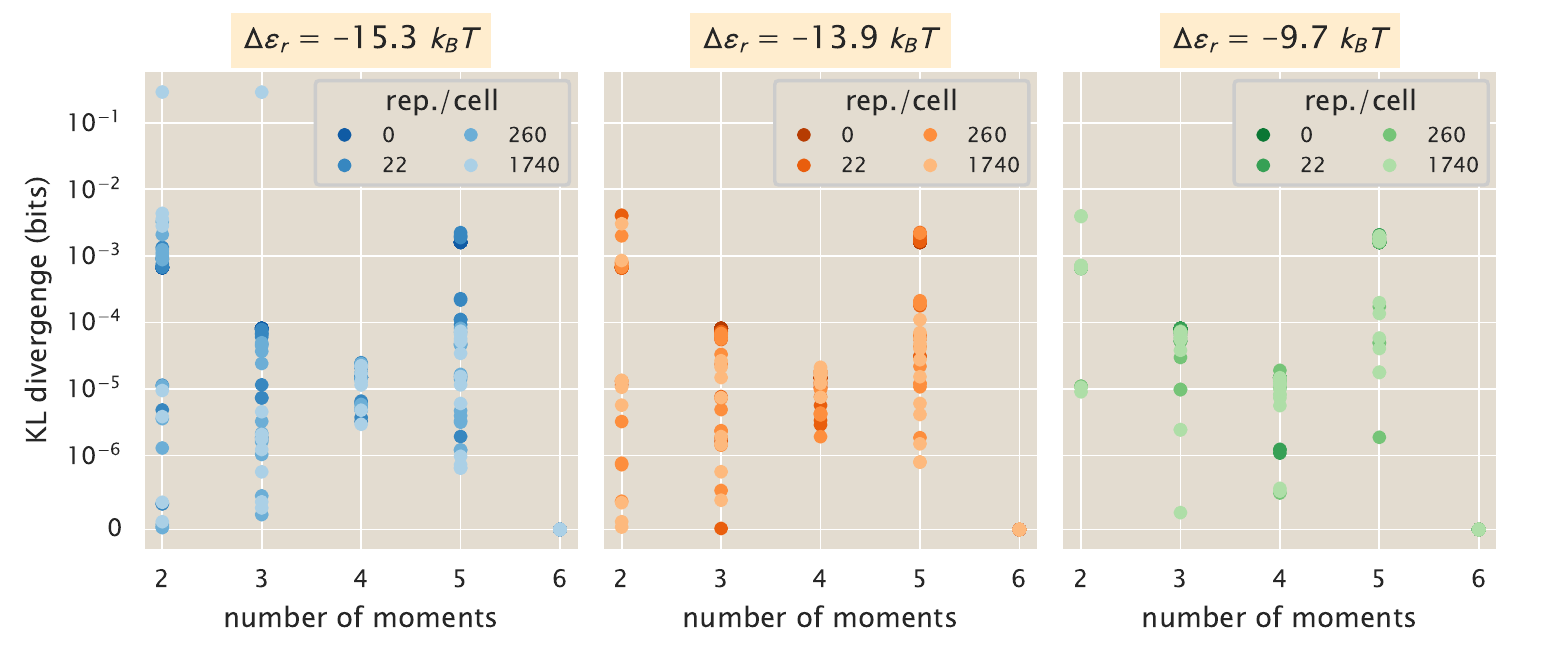}
	\caption{\textbf{Measuring the loss of information by using different number
	of constraints.} The Kullback-Leibler divergence was computed between the
	maximum entropy distribution constructed using the first 6 moments of the
	distribution and a variable number of moments.}
  \label{sfig_dkl}
\end{figure}

\paragraph{(iii) Multi-state promoter.}
This particular point is something that we are still exploring from a
theoretical perspective. We have shown that in order to capture the
single-molecule mRNA FISH data a single-state promoter wouldn't suffice. This
model predicts a Poisson distribution as the steady-state and the data
definitely shows super Poissonian noise. Given the bursty nature of gene
expression we opt to use a two-state promoter where the states reflect
effective transcriptionally ``active'' and ``inactive'' states. We are
currently exploring alternative formulations of this model to turn it into a
single state with a geometrically distributed burst-size.

\paragraph{(iv) Optimal vs Log-flat Distributions.}
The relevance of having use the Blahut-Arimoto algorithm to predict the maximum
mutual information between input and outputs was just to understand the best
case scenario. We show the comparison between theoretical and experimental
input-output functions $P(p \mid c)$ in \fref{sfig_cdf_reg}. Given the good
agreement between these distributions we could compute the mutual information
$I(c; p)$ for any arbitrary input distribution $P(c)$ and obtain a good
agreement with the corresponding experimental mutual information.

The reason we opted to specifically report the mutual information at channel
capacity was to put the results in a context. By reporting the upper bound in
performance of these genetic circuits we can start to dissect how different
molecular parameters such as repressor-DNA binding affinity or repressor copy
number affect the ability of this genetic circuit to extract information from
the environmental state.

\section{Empirical fits to noise predictions} \label{supp_empirical}

(Note: The Python code used for the calculations presented in this section can
be found in the
\href{https://www.rpgroup.caltech.edu/chann_cap/src/theory/html/empirical_constants.html}{following
link} as an annotated Jupyter notebook)

In \fref{fig3_cell_cycle}(C) in the main text we show that our minimal model has
a systematic deviation on the gene expression noise predictions compared to the
experimental data. This systematics will need to be addressed on an improved
version of the minimal model presented in this work. To guide the insights into
the origins of this systematic deviation in this appendix we will explore
empirical modifications of the model to improve the agreement between theory and
experiment.

\subsection{Multiplicative factor for the noise}
\label{supp_mult_factor_noise}

The first option we will explore is to modify our noise predictions by a
constant multiplicative factor. This means that we assume the relationship
between our minimal model predictions and the data for noise in gene expression
are of the from
\begin{equation}
    \text{noise}_{\text{exp}} = \alpha \cdot \text{noise}_{\text{theory}},
\end{equation}
where $\alpha$ is a dimensionless constant to be fit from the data. The data,
especially in \fref{sfig_noise_delta} suggests that our predictions are within a
factor of $\approx$ two from the experimental data. To further check that
intuition we performed a weighted linear regression between the experimental and
theoretical noise measurements. The weight for each datum was taken to be
proportional to the bootstrap errors in the noise estimate, this to have poorly
determined noises weigh less during the regression. The result of this
regression with no intercept shows exactly that a factor of two systematically
improves the theoretical vs. experimental predictions.
\fref{sfig_noise_mult_factor} shows the improved agreement when the theoretical
predictions for the noise are multiplied by $\approx 1.5$.

\begin{figure}[h!]
	\centering \includegraphics
  {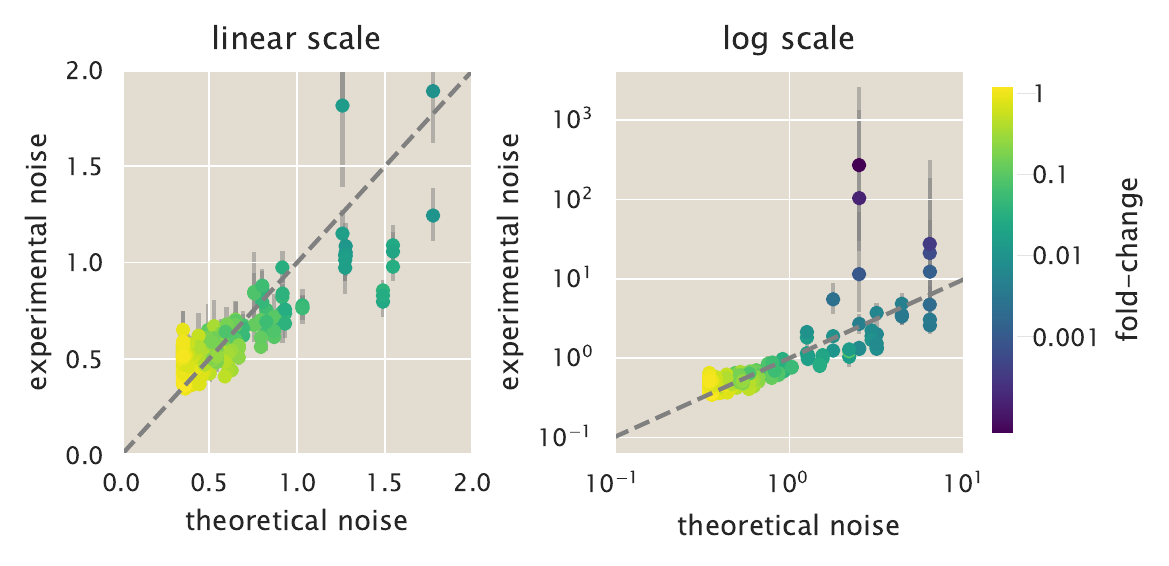}
	\caption{\textbf{Multiplicative factor to improve theoretical vs.
	experimental comparison of noise in gene expression.} Theoretical vs.
	experimental noise both in linear (left) and log (right) scale. The dashed
	line shows the identity line of slope 1 and intercept zero. All data are
	colored by the corresponding value of the experimental fold-change in gene
	expression as indicated by the color bar. The $x$-axis was multiplied by a
	factor of $\approx 1.5$ as determined by a linear regression from the data
	in \fref{sfig_noise_comparison}. Each datum represents a single date
	measurement of the corresponding strain and IPTG concentration with $\geq
	300$ cells. The points correspond to the median, and the error bars
	correspond to the 95\% confidence interval as determined by 10,000 bootstrap
	samples.}
  \label{sfig_noise_mult_factor}
\end{figure}

For completeness \fref{sfig_noise_reg_corrected} shows the noise in gene
expression as a function of the inducer concentration including this factor of
$\approx 1.5$. It is clear that overall a simple multiplicative factor improves
the predictive power of the model.

\begin{figure}[h!]
	\centering \includegraphics
  {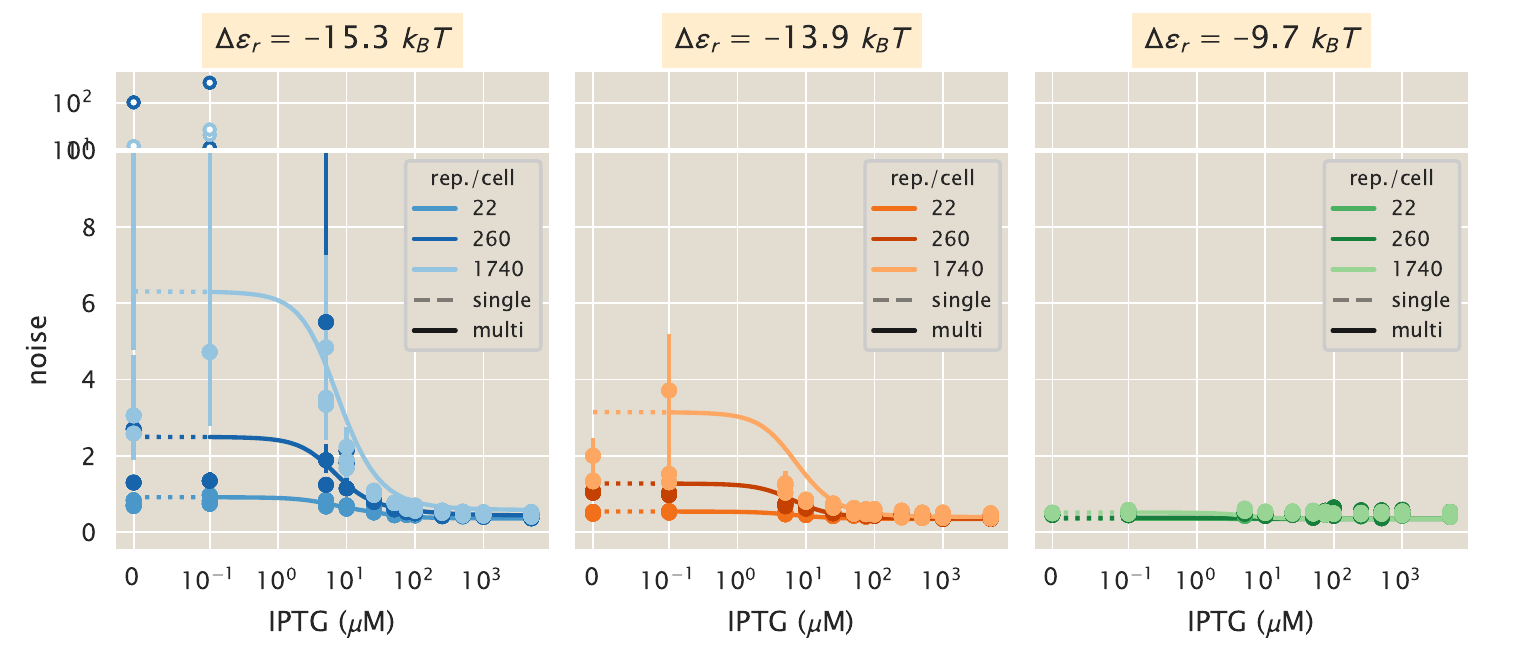}
	\caption{\textbf{Protein noise of the regulated promoter with multiplicative
	factor.} Comparison of the experimental noise for different operators ((A)
	O1,  $\eR = -15.3 \; k_BT$, (B) O2, $\eR = -13.9 \; k_BT$, (C) O3, $\eR =
	-9.7 \; k_BT$) with the theoretical predictions for the  the multi-promoter
	model. A linear regression revealed that multiplying the theoretical noise
	prediction by a factor of $\approx 1.5$ would improve agreement between
	theory and data. Points represent the experimental noise as computed from
	single-cell fluorescence measurements of different {\it E. coli} strains
	under 12 different inducer concentrations. Dotted line indicates plot in
	linear rather than logarithmic scale. Each datum represents a single date
	measurement of the corresponding strain and IPTG concentration with $\geq
	300$ cells. The points correspond to the median, and the error bars
	correspond to the 95\% confidence interval as determined by 10,000 bootstrap
	samples. White-filled dots are plot at a different scale for better
	visualization.}
  \label{sfig_noise_reg_corrected}
\end{figure}

\subsection{Additive factor for the noise}
\label{supp_add_factor_noise}

As an alternative way to empirically improve the predictions of our model we will now test the idea of an additive constant. What this means is that our minimal model underestimates the noise in gene expression as
\begin{equation}
    \text{noise}_{\text{exp}} = \beta + \text{noise}_{\text{theory}},
\end{equation}
where $\beta$ is an additive constant to be determined from the data. As with
the multiplicative constant we performed a regression to determine this
empirical additive constant comparing experimental and theoretical gene
expression noise values. We use the error in the 95\% bootstrap confidence
interval as a weight for the linear regression. \fref{sfig_noise_add_factor}
shows the resulting theoretical vs. experimental noise where $\beta \approx
0.2$. We can see a great improvement in the agreement between theory and 
experiment with this additive constant

\begin{figure}[h!]
	\centering \includegraphics
  {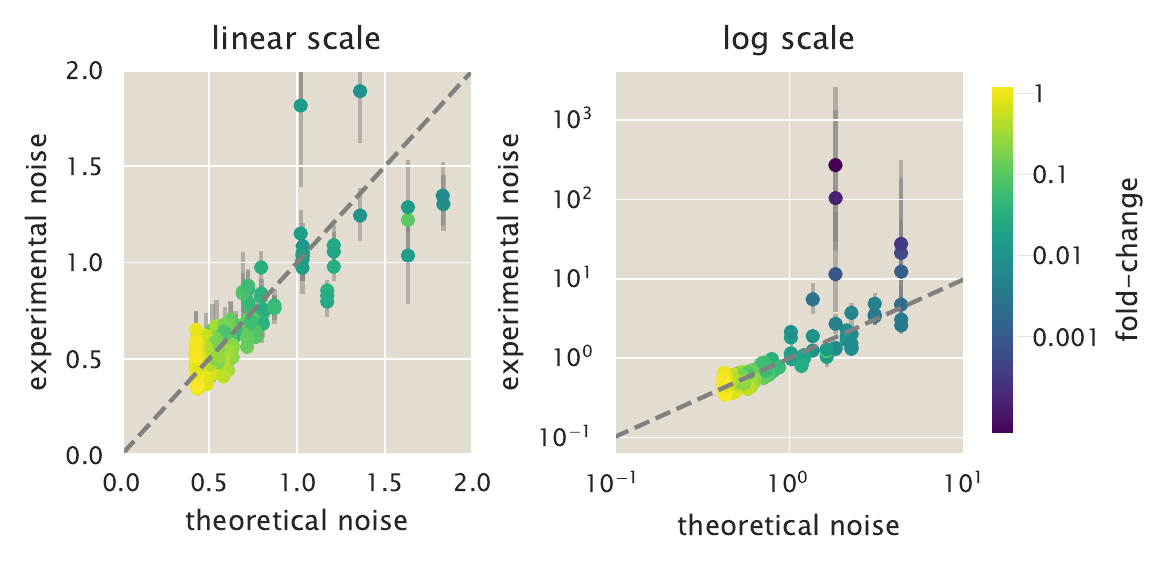}
	\caption{\textbf{Additive factor to improve theoretical vs. experimental
	comparison of noise in gene expression.} Theoretical vs. experimental noise
	both in linear (left) and log (right) scale. The dashed line shows the
	identity line of slope 1 and intercept zero. All data are colored by the
	corresponding value of the experimental fold-change in gene expression as
	indicated by the color bar. A value of $\approx 0.2$ was added to all values
	in the $x$-axis as determined by a linear regression from the data in
	\fref{sfig_noise_comparison}. Each datum represents a single date
	measurement of the corresponding strain and IPTG concentration with $\geq
	300$ cells. The points correspond to the median, and the error bars
	correspond to the 95\% confidence interval as determined by 10,000 bootstrap
	samples.}
  \label{sfig_noise_add_factor}
\end{figure}

For completeness \fref{sfig_noise_reg_add} shows the noise in gene expression as
a function of the inducer concentration including this additive factor of $\beta
\approx 0.2$. If anything, the additive factor seems to improve the agreement 
between theory and data even more than the multiplicative factor.

\begin{figure}[h!]
	\centering \includegraphics
  {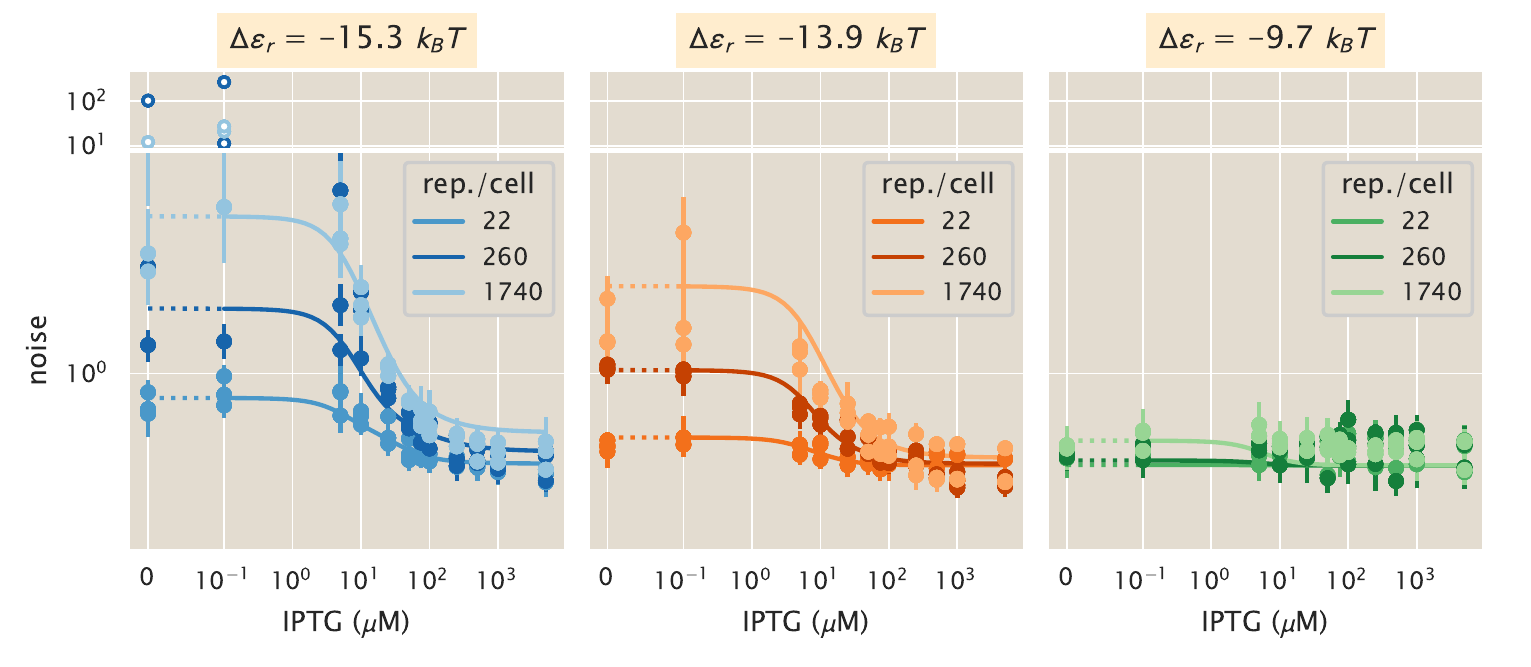}
	\caption{\textbf{Protein noise of the regulated promoter with additive
	factor.} Comparison of the experimental noise for different operators ((A)
	O1,  $\eR = -15.3 \; k_BT$, (B) O2, $\eR = -13.9 \; k_BT$, (C) O3, $\eR =
	-9.7 \; k_BT$) with the theoretical predictions for the  the multi-promoter
	model. A linear regression revealed that an additive factor of $\approx 0.2$
	to the the theoretical noise prediction  would improve agreement between
	theory and data. Points represent the experimental noise as computed from
	single-cell fluorescence measurements of different {\it E. coli} strains
	under 12 different inducer concentrations. Dotted line indicates plot in
	linear rather than logarithmic scale. Each datum represents a single date
	measurement of the corresponding strain and IPTG concentration with $\geq
	300$ cells. The points correspond to the median, and the error bars
	correspond to the 95\% confidence interval as determined by 10,000 bootstrap
	samples. White-filled dots are plot at a different scale for better
	visualization.}
  \label{sfig_noise_reg_add}
\end{figure}

\subsection{Correction factor for channel capacity with multiplicative factor}

As seen in \siref{supp_multi_gene} a constant multiplicative factor can reduce
the discrepancy between the model predictions and the data with respect to the
noise (standard deviation / mean) in protein copy number. To find the
equivalent correction would be for the channel capacity requires gaining
insights from the so-called small noise approximation \cite{Tkacik2008a}. The
small noise approximation assumes that the input-output function can be modeled
as a Gaussian distribution in which the standard deviation is small. Using
these assumptions one can derive a closed-form for the channel capacity.
Although our data and model predictions do not satisfy the requirements for the
small noise approximation, we can gain some intuition for how the channel
capacity would scale given a systematic deviation in the cell-to-cell
variability predictions compared with the data.

Using the small noise approximation one can derive the form of the input
distribution at channel capacity $P^*(c)$. To do this we use the fact that
there is a deterministic relationship between the input inducer concentration
$c$ and the mean output protein value  $\ee{p}$, therefore we can work with
$P(\ee{p})$ rather than $P(c)$ since the deterministic relation allows us to
write
\begin{equation}
  P(c) dc = P(\ee{p}) d\ee{p}.
\end{equation}
Optimizing over all possible distributions $P(\ee{p})$ using calculus of
variations results in a distribution of the form
\begin{equation}
  P^*(\ee{p}) = {1 \over \mathcal{Z}} {1 \over \sigma_p(\ee{p})},
\end{equation}
where $\sigma_p(\ee{p})$ is the standard deviation of the protein distribution
as a function of the mean protein expression, and $\mathcal{Z}$ is a
normalization constant defined as
\begin{equation}
  \mathcal{Z} \equiv \int_{\ee{p(c=0)}}^{\ee{p(c\rightarrow \infty)}}
  {1 \over \sigma_p(\ee{p})} d\ee{p}.
\end{equation}
Under these assumptions the small noise approximation tells us that the channel
capacity is of the form \cite{Tkacik2008a}
\begin{equation}
  I = \log_2 \left( {\mathcal{Z} \over \sqrt{2 \pi e}} \right).
\end{equation}

From the theory-experiment comparison in \siref{supp_multi_gene} we know that
the standard deviation predicted by our model is systematically off by a factor
of two compared to the experimental data, i.e.
\begin{equation}
  \sigma_p^{\exp} = 2 \sigma_p^{\text{theory}}.
\end{equation}
This then implies that the normalization constant $\mathcal{Z}$ between theory
and experiment must follow a relationship of the form
\begin{equation}
  \mathcal{Z}^{\exp} = {1 \over 2} \mathcal{Z}^{\text{theory}}.
\end{equation}
With this relationship the small noise approximation would predict that the
difference between the experimental and theoretical channel capacity should be
of the form
\begin{equation}
  I^{\exp} = \log_2 \left( {\mathcal{Z}^{\exp} \over \sqrt{2 \pi e}} \right)
  = \log_2 \left( {\mathcal{Z}^{\text{theory}} \over \sqrt{2 \pi e}} \right)
  - \log_2(2).
\end{equation}
Therefore under the small noise approximation we would expect our predictions
for the channel capacity to be off by a constant of 1 bit ($\log_2(2)$) of
information. Again, the conditions for the small noise approximation do not
apply to our data given the intrinsic level of cell-to-cell variability in the
system, nevertheless what this analysis tells is is that we expect that an
additive constant should be able to explain the discrepancy between our model
predictions and the experimental channel capacity. To test this hypothesis we
performed a ``linear regression'' between the model predictions and the
experimental channel capacity with a fixed slope of 1. The intercept of this
regression, -0.56 bits, indicates the systematic deviation we expect should
explain the difference between our model and the data.
\fref{sfig_channcap_corr} shows the comparison between the original predictions
shown in \fref{fig5_channcap}(A) and the resulting predictions with this shift.
Other than the data with zero channel capacity, this shift is able to correct
the systematic deviation for all data. We therefore conclude that our model
ends up underestimating the experimentally determined channel capacity by a
constant amount of 0.43 bits.

\begin{figure}[h!]
	\centering \includegraphics
  {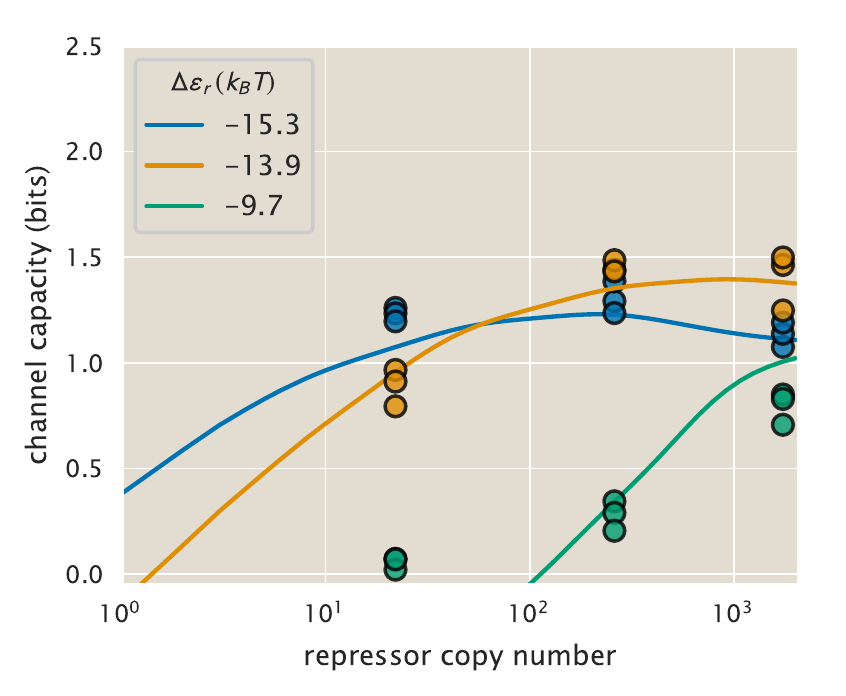}
	\caption{\textbf{Additive correction factor for channel capacity.} Solid
	lines represent the theoretical predictions of the channel capacity shown in
	\fref{fig5_channcap}(A). The dashed lines show the resulting predictions
	with a constant shift of -0.43 bits. Points represent single biological
	replicas of the inferred channel capacity.}
  \label{sfig_channcap_corr}
\end{figure}

\section{Derivation of the cell age distribution}
\label{supp_cell_age_dist}

E. O. Powell first derive in 1956 the distribution of cell age for a cell
population growing steadily in the exponential phase \cite{Powell1956}. This
distribution is of the form
\begin{equation}
  P(a) = \ln(2) \cdot 2^{1 - a},
\end{equation}
where $a \in [0, 1]$ is the fraction of the cell cycle, 0 being the moment right
after the mother cell divides, and 1 being the end of the cell cycle just before
cell division. In this section we will reproduce and expand the details on each
of the steps of the derivation.

For an exponentially growing bacterial culture, the cells satisfy the growth law
\begin{equation}
  {\dt{n}} = \mu n,
  \label{seq_growth_ode}
\end{equation}
where $n$ is the number of cells and $\mu$ is the growth rate in units of
time$^{-1}$. We begin by defining $P(a)$ to be the probability density function
of a cell having age $a$. At time zero of a culture in exponential growth, i.e.
the time when we start considering the growth, not the initial condition of the
culture, there are $NP(a)da$ cells with age range between $[a, a + da]$. In
other words, for $N \gg 1$ and $da \ll a$
\begin{equation}
  N P(a \leq x \leq a + da) \approx N P(a)da.
\end{equation}
We now define
\begin{equation}
  F(\tau) = \int_\tau^\infty f(\xi) d\xi,
\end{equation}
as the fraction of cells whose division time is greater than $\tau$. This is
because in principle not all cells divide exactly after $\tau$ minutes, but
there is a distribution function $f(\tau)$ for the division time after birth.
Empirically it has been observed that a generalize Gamma distribution fits well
to experimental data on cell division time, but we will worry about this
specific point later on.

From the definition of $F(\tau)$ we can see that if a cell reaches an age $a$,
the probability of surviving to an age $a + t$ without dividing is given by
\begin{equation}
  P(\text{age} = (a + t) \mid \text{age} = a) = F(a + t \mid a) =
  {F(a + t) \over F(a)}.
  \label{seq_prob_a+t_give_a}
\end{equation}
This result comes simply from the definition of conditional probability. Since
$F(a)$ is the probability of surviving $a$ or more minutes without dividing, by
the definition of conditional probability we have that
\begin{equation}
  F(a + t \mid a) = {F(a, a + t) \over F(a)},
\end{equation}
where $F(a, a + t)$ is the joint probability of surviving $a$ minutes and $a +
t$ minutes. But the probability of surviving $a + t$ minutes or more implies
that the cell already survived $a$ minutes, therefore the information is
redundant and we have
\begin{equation}
  F(a, a + t) = F(a + t).
\end{equation}
This explains \eref{seq_prob_a+t_give_a}. From this equation we can find that
out of the $N P(a)da$ cells with age $a$ only a fraction
\begin{equation}
  \left[ NP(a)da \right] F(a + t \mid a) = NP(a) {F(a + t)\over F(a)} da
\end{equation}
will survive without dividing until time $a + t$. During that time interval $t$
the culture has passed from $N$ cells to $N e^{\mu t}$ cells given the
assumption that they are growing exponentially. The survivors $NP(a)F(a + t \mid
a)da$ then represent a fraction of the total number of cells
\begin{equation}
  {\text{\# survivors} \over \text{\# total cells}} =
  {\left[ NP(a)da \right] F(a + t \mid a) \over Ne^{\mu t}} =
  P(a){F(a + t)\over F(a)}da {1 \over e^{\mu t}},
\end{equation}
and their ages lie in the range $[a+t, a+t+da]$. Since we assume that the
culture is in steady state then it follows that the fraction of cells that
transitioned from age $a$ to age $a + t$ must be $P(a + t)da$. Therefore we have
a difference equation - the discrete analogous of a differential equation - of
the form
\begin{equation}
  P(a + t) da = P(a) {F(a + t) \over F(a)}e^{-\mu t} da.
  \label{seq_difference_eq}
\end{equation}
What this equation shows is a relationship that connects the probability of
having a life time of $a + t$ with a probability of having a shorter life time
$a$ and the growth of the population. If we take $t$ to be very small,
specifically if we assume $t \ll \mu^{-1}$ we can Taylor expand around $a$ the
following terms:
\begin{equation}
  F(a + t) \approx F(a) + {dF \over da} t,
\end{equation}
\begin{equation}
  P(a + t) \approx P(a) + {dP \over da} t,
\end{equation}
and
\begin{equation}
  e^{-\mu t} \approx 1 - \mu t.
\end{equation}
Substituting these equations into \eref{seq_difference_eq} gives
\begin{equation}
  P(a) + {dP \over da} t = P(a) \left( {F(a) + {dF \over da}t \over
  F(a)} \right) (1 - \mu t).
\end{equation}
This can be rewritten as
\begin{equation}
  {1 \over P(a)} {dP \over da} =
  {1 \over F(a)} {dF \over da} - \mu - {\mu t \over F(a)} {dF \over da}.
\end{equation}
Since we assumed $t \ll \mu^{-1}$ we then approximate the last term to be close
to zero. We can then simplify this result into
\begin{equation}
  {1 \over P(a)} {dP \over da} = {1 \over F(a)} {dF \over da} - \mu.
\end{equation}
Integrating both sides of the equation with respect to $a$ gives
\begin{equation}
  \ln P(a) = \ln F(a) - \mu a + C,
\end{equation}
where $C$ is the integration constant. Exponentiating both sides gives
\begin{equation}
  P(a) = C' F(a)e^{-\mu a}.
\end{equation}
Where $C' \equiv e^C$. To obtain the value of the unknown constant we recall
that $F(0) = 1$ since the probability of having a life equal or longer than zero
must add up to one, therefore we have that $P(0) = C'$. This gives then
\begin{equation}
  P(a) = P(0) e^{-\mu a} F(a).
  \label{seq_Pa_result}
\end{equation}
Substituting the definition of $F(a)$ gives
\begin{equation}
  P(a) = P(0) e^{-\mu a} \int_a^\infty f(\xi) d\xi.
  \label{seq_Pa_result_int}
\end{equation}
The last step of the derivation involves writing $P(0)$ and the growth rate
$\mu$ in terms of the cell cycle length distribution $f(\tau)$.

The growth rate of the population cell number (not the growth of cell mass) is
defined as the number of cell doublings per unit time divided by the number of
cells. This is more clear to see if we write \eref{seq_growth_ode} as a finite
difference
\begin{equation}
  {N(t + \Delta t) - N(t) \over \Delta t} = \mu N(t).
\end{equation}
If the time $\Delta t$ is the time interval it takes to go from $N$ to $2N$
cells we have
\begin{equation}
  {2N - N \over \Delta t} = \mu N.
\end{equation}
Solving for $\mu$ gives
\begin{equation}
  \mu = \overbrace{{2N - N \over \Delta t}}
  ^{\text{\# doubling events per unit time}}
  \overbrace{{1 \over N}}^{{1 \over \text{population size}}}.
\end{equation}
We defined $F(a)$ to be the probability of a cell reaching an age $a$ or
greater. For a cell to reach an age $a + da$ we can then write
\begin{equation}
  F(a + da) = \int_{a + da}^{\infty} f(\xi) d\xi
  = \int_a^{\infty} f(\xi) d\xi - \int_a^{a + da} f(\xi) d\xi.
\end{equation}
We can approximate the second term on the right hand side to be
\begin{equation}
  \int_a^{a + da} f(\xi) d\xi \approx f(a) da,
\end{equation}
for $da \ll a$, obtaining
\begin{equation}
  F(a + da) \approx F(a) - f(a)da.
\end{equation}
What this means is that from the original fraction of cells $F(a)$ with age $a$
or greater a fraction $f(a)da / F(a)$ will not reach age $(a + da)$ because they
will divide. So out of the $NP(a)$ cells that reached exactly age $a$, the
number of doubling events on a time interval $da$ is given by
\begin{equation}
  {\text{\# doublings of cells of age } a {\text{ on interval } da}} =
  \overbrace{NP(a)}^{\text{\# cells of age }a}
  \overbrace{f(a) da \over F(a)}^{\text{fraction of doublings per unit time}}.
\end{equation}
The growth rate then is just the sum (integral) of each age contribution to the
total number of doublings. This is
\begin{equation}
  \mu = {1 \over N} \int_0^\infty NP(a) {f(a)da \over F(a)}.
\end{equation}
Substituting \eref{seq_Pa_result} gives
\begin{equation}
  \mu = \int_0^\infty [P(0) e^{-\mu a} F(a)] {f(a)da \over F(a)}
  = \int_0^\infty P(0) e^{-\mu a} f(a)da.
  \label{seq_mu_integral}
\end{equation}
We now have the growth rate $\mu$ written in terms of the cell cycle length
probability distribution $f(a)$ and the probability $P(0)$. Since $P(a)$ is a
probability distribution it must be normalized, i.e.
\begin{equation}
  \int_0^\infty P(a) da = 1.
\end{equation}
Substituting \eref{seq_Pa_result} into this normalization constraint gives
\begin{equation}
  \int_0^\infty P(0) e^{-\mu a} F(a) da = 1.
\end{equation}
From here we can integrate the left hand side by parts. We note that given the
definition of $F(a)$, the derivative with respect to $a$ is $-f(a)$ rather than
$f(a)$. This is because if we write the derivative of $F(a)$ we have
\begin{equation}
  {dF(a) \over da} \equiv \lim_{da \rightarrow 0}
  {F(a + da) - F(a) \over da}.
\end{equation}
Substituting the definition of $F(a)$ gives
\begin{equation}
  {dF(a) \over  da} = \lim_{da \rightarrow 0} {1 \over da}
  \left[\int_{a + da}^\infty f(\xi) d\xi - \int_a^\infty f(\xi) d\xi \right].
\end{equation}
This difference in the integrals can be simplified to
\begin{equation}
  \lim_{da \rightarrow 0} {1 \over da} \left[ \int_{a + da}^\infty f(\xi) d\xi -
  \int_a^\infty f(\xi) d\xi \right]\approx {-f(a)da \over da} = -f(a).
\end{equation}
Taking this into account we now perform the integration by parts obtaining
\begin{equation}
 P(0) \left[ {e^{-\mu t} \over -\mu} F(a) \right]^\infty_0
 - P(0) \int_0^\infty {e^{-\mu a} \over -\mu} (-f(a)) da = 1.
 \label{seq_inte_byparts}
\end{equation}
On the first term on the left hand side we have that as $a \rightarrow \infty$,
both terms $e^{-\mu a}$ and $F(a)$ go to zero. We also have that $e^{\mu 0} = 1$
and $F(0) = 1$. This results in
\begin{equation}
  {P(0) \over \mu} - P(0) \int_0^\infty {e^{-\mu a} \over \mu} f(a) da = 1.
\end{equation}
The second term on the left hand side is equal to \eref{seq_mu_integral} since
\begin{equation}
  \mu = \int_0^\infty P(0) e^{-\mu a} f(a)da \Rightarrow
  1 = \int_0^\infty P(0) {e^{-\mu a} \over \mu} f(a)da.
\end{equation}
This implies that on \eref{seq_inte_byparts} we have
\begin{equation}
  {P(0) \over \mu} - 1 = 1 \Rightarrow P(0) = 2 \mu.
\end{equation}
With this result in hand we can rewrite \eref{seq_Pa_result_int} as
\begin{equation}
  P(a) = 2\mu e^{-\mu a} \int_a^\infty f(\xi) d\xi.
  \label{seq_Pa_Posubs}
\end{equation}
Also we can rewrite the result for the growth rate $\mu$ on
\eref{seq_mu_integral} as
\begin{equation}
  \mu = 2 \mu \int_0^\infty e^{-\mu a} f(a) da \Rightarrow
  2 \int_0^\infty e^{-\mu a} f(a) da = 1.
  \label{seq_fa_int}
\end{equation}

As mentioned before the distribution $f(a)$ has been empirically fit to a
generalize Gamma distribution. But if we assume that our distribution has almost
negligible dispersion around the mean average doubling time $a = \tau_d$, we can
approximate $f(a)$ as
\begin{equation}
  f(a) = \delta(a - \tau_d),
\end{equation}
a Dirac delta function. Applying this to \eref{seq_fa_int} results in
\begin{equation}
  2 \int_0^\infty e^{-\mu a} \delta(a - \tau_a) da = 1
  \Rightarrow 2 e^{-\mu \tau_d} = 1.
\end{equation}
Solving for $\mu$ gives
\begin{equation}
  \mu = {\ln 2 \over \tau_d}.
\end{equation}
This delta function approximation for $f(a)$ has as a consequence that
\begin{equation}
  F(a) =
  \begin{cases}
    1 \text{ for } a \in [0, \tau_d],\\
    0 \text{ for } a > \tau_d.
  \end{cases}
\end{equation}
Fianlly we can rewrite \eref{seq_Pa_Posubs} as
\begin{equation}
  P(a) = 2 \left( {\ln 2 \over \tau_d} \right)
  e^{- {\ln 2 \over \tau_d} a} \int_a^\infty \delta(\xi - \tau_d) d\xi
  \Rightarrow = 2 \ln 2 \cdot 2^{-a \over \tau_d}.
\end{equation}
Simplifying this we obtain
\begin{equation}
  P(a) =
  \begin{cases}
    \ln 2 \cdot 2^{1 - {a \over \tau_d}} \text{ for } a \in [0, \tau_d],\\
    0 \text{ otherwise}.
  \end{cases}
\end{equation}
This is the equation we aimed to derive. The distribution of cell ages over
the cell cycle.

		\printbibliography[title={Supplemental References},
		segment=\therefsegment, filter=notother]
	\end{refsegment}

}{} %
\end{document}